\documentclass[10pt]{article}

\usepackage{latexsym,amsmath,amssymb,amsbsy,amstext,amscd,amsfonts,appendix}
\usepackage{graphics,graphicx,epstopdf}
\usepackage{tabularx}
\usepackage{multirow,caption,subcaption}
\usepackage{booktabs}
\usepackage{soul,color}
\usepackage{listings}
\usepackage{float}
\usepackage{comment}
\usepackage{subfig,latexsym,amsmath,amssymb,amsbsy,amstext,amscd,amsfonts,appendix,color,multirow,graphicx}
\usepackage{natbib}

\usepackage{latexsym,amsmath,amssymb,amsbsy,amstext,amscd,amsfonts,appendix}
\usepackage{graphics,graphicx}
\usepackage{tabularx}
\usepackage{multirow}
\usepackage{booktabs}
\usepackage{soul,color}
\usepackage{listings}
\usepackage{float}
 \usepackage{subfig,latexsym,amsmath,amssymb,amsbsy,amstext,amscd,amsfonts,appendix,color,multirow,graphicx}

\date{\today}

\title{Remarks on application of different variables for the PKN model of hydrofracturing. Various fluid-flow regimes.}

\author{P. Kusmierczyk$^{1,2}$, G. Mishuris$^{2}$ and M. Wrobel$^{1,2}$,
\\[2mm]
{\it
$^{1}$EUROTECH Sp. z o.o., ul. Wojska Polskiego 3, }
{\it 39-300 Mielec, Poland }
\\[2mm]
{\it
$^{2}$IMAPS, Aberystwyth University, }
{\it Ceredigion SY23 3BZ, Wales U.K.}
}

\begin{document}

\maketitle

\begin{abstract}
The problem of hydraulic fracture for the PKN model is considered within the framework presented recently by Linkov (2011).
The modified formulation is further enhanced by employing an improved regularized boundary condition near the crack tip.
This increases solution accuracy especially for singular leak-off regimes.
A new dependent variable having clear physical sense is introduced.
A comprehensive analysis of numerical algorithms based on various dependent variables is provided. Comparison with know numerical results has been given.
\end{abstract}
\\
\noindent \textbf{PACS:} PACS 91.55.Fg, 91.55.Jk, 91.60.Ba
\\
\\
\\
\thanks{This work has been done in the framework of the EU FP7 PEOPLE project under contract
number PIAP-GA-2009-251475-HYDROFRAC.}

\section{Introduction}

In its broadest definition, hydraulic fracturing refers to a
problem of a fluid driven fracture propagating in a brittle medium.
The process has been known for at least fifty years
(\citet{Crittendon}, \citet{Desroches}, \citet{Desroches_1994},
\citet{Detournay}, \citet{Geertsma}, \citet{Harrison},
\citet{Hubbert}, \citet{Khristianovic}). The
respective technology has been utilized in the petroleum industry to
intensify the extraction of hydrocarbons for decades.

Recently, due
to economical reasons, it has been revived for exploiting
non-conventional hydrocarbon deposits. The process has many other
technological applications (e.g. disposal of waste drill cuttings
underground \citep{Moschovidis}, geothermal reservoirs exploitation
\citep{Pine} or $\it{in}$ $\it{situ}$ stress measurements
\citep{Desroches}. Hydrofracturing also appears in nature (e.g.
geological processes, like magma-driven dy\-kes \citep{Rubin},
\citep{Lister} or a subglacial drainage of water \citep{Rice}).
Due to the complexity of this multiphysical phenomenon, mathematical
and numerical simulation of the process still represents a
challenging task, in spite of the fact that immense progress has
been made since the first algorithms were developed.

The mathematical model of the problem should account for coupled
mechanisms driving the process, which are: i) solid mechanics
equations, describing the deformation of the rock induced by the
fluid pressure; ii) equations for the fluid flow within the fracture
and the leak-off to the rock formation; iii) fracture mechanics
criteria defining the conditions for fracture propagation. Further
development of the model involves incorporation of mass transport
for the proppant movement, fluid diffusion to account for the rock
saturation by the leak-off flux, thermal effects
affecting rheological properties of the fluid and others.

The computational challenges of the hydrofracturing models result
from several factors: (i) strong non-linearity introduced by the
Poiseuille equation describing the fluid flow; (ii) in the general case,
a non-local relationship between the fracture opening and the net
fluid pressure; (iii) moving boundaries of the fluid front and the
fracture contour; (iv) degeneration of the governing PDE at the fracture front; (v) possible lag between the crack tip and the
fluid front.

The first simplified mathematical descriptions of hydraulic fracture
were summarized in the following three main classical
models. The so-called PKN model was considered in \cite{PK}, where
the authors adopted Sneddon's solution \citep{Sneddon} which was
further enhanced by \citet{Nordgren} to account for the fluid loss
effect and fracture volume change. As a result, the crack length was
determined as a part of the solution. The so-called KGD (plain
strain) model was developed independently by \citet{Khristianovic},
and \citet{Geertsma}. Finally, the radial or penny-shaped model was
introduced by \citet{Sneddon_1} with constant fluid pressure and
was extended for the general case by \citet{S$S}.

Different variations of the aforementioned models were used for
treatment designs for decades, despite the fact that each of them is valid
only under very specific assumptions (like elliptic cross-section of
the fracture, and fracture half-length much greater than its
constant height for the PKN model). Recently, the classical models
have been largely replaced by the pseudo 3D models \citep{Mack}. A
comprehensive review of the history and techniques of
hydrofracturing simulation can be found in \citet{Adachi_2007}.

Although the classical models have been superseded in most of the
practical applications, they still play a crucial role when
developing and analyzing new computational algorithms. The models
enable one to understand the nature of, and reasons for, computational
difficulties, find the remedies for them and to extend these ideas
to the general case.

Thus, in the pioneering work by \citet{Nordgren}, main peculiarities of the model were analyzed and a numerical algorithm
was proposed to deal with the problem. Asymptotic analysis of the solution near the crack tip for impermeable rock model was presented in \citet{Kemp}
and an approximate solution for the zero leak-off
case, with an accuracy to the first four leading asymptotic terms, was given
\footnote{Extension of this solution to the full series representation was given in \citet{Kovalyshen}, while other form, in terms of fast converging series, was obtained in \citet{Linkov_3}.}.
Here, probably for the first time, the speed equation was efficiently implemented in the model.
The fourth degree of the crack opening was considered as the proper variable and used in the numerical computations.
Finally, \citet{Kemp} suggested to use a special tip element, compatible with the asymptotic behaviour of the solution, within the Finite Volume (FV) scheme.

For the leak-off function defined by the Carter law the two leading terms of asymptotics can be found in \citet{Kovalyshen}, where the PKN
problem was revisited to take into account the multi-scale arguments in the spirit of \cite{Garagash et al}.
The authors also used the FV scheme with a special tip element to tackle the transient regime.

Recently,  the classical
models of Nordgren and Spen\-ce $\&$ Sharp have been revisited again
in \citet{Linkov_1} -- \citet{Linkov_4}. The author discovered that
in some formulations the hyd\-rau\-lic fracture problem may exhibit
ill-posed properties. To
eliminate the difficulties resulting from these facts, a number of
measures were proposed: i) speed equation to trace the crack front
instead of the usually applied total flux balance condition\footnote{Probably, for
the first time, this idea was recalled in indirect way in \citet{S$S} and
utilized by \cite{Kemp}, but later was abandoned as the particle velocity
at the crack tip is difficult to compute numerically};
ii) the
so-called $\varepsilon$-regularization technique which consists of
imposing the computational domain boundary at a small distance
behind the crack tip; iii) new boundary condition to be imposed in
the regularized formulation; iv) new dependent variables: the
particle velocity and the crack opening taken in a degree to exploit
the asymptotic behaviour of the solution; v) the spatial coordinates
moving with the crack front and evaluation of the temporal
derivative under fixed values of these coordinates.

The advantages of the modified formulation were cle\-ar\-ly
demonstrated on the basis of de\-ve\-loped analytical benchmarks in
\citet{Linkov_4}. An immense improvement of solution accuracy,
computation efficiency and stability was shown.
In \citet{MWL} a further step in employing the modified formulation
was done by analyzing the stiffness of the system of differential
equations arising after spatial discretization. An efficient
modification of the algorithm has been proposed to trace the
fracture propagation. Finally, the investigation of solution
sensitivity to some process parameters has been considered.

However, the aforementioned analysis is concerned with the case when
the leak-off vanishes near the crack tip and, as a result, the fluid velocity does
not change much in this region.

The primary aim of this paper is to verify the recipes delivered in
\citet{Linkov_1,Linkov_2,Linkov_3,Linkov_4} and \citet{MWL} for an arbitrary leak off regime.
In particular, the analysis includes: (i) Investigation of
performance of numerical algorithms for hydrofracturing based on
different formulations (different dependent and independent
variables); (ii) Utilization of a new dependent variable defined as
an integral of the crack opening. Such a variable has a clear physical
and technological sense and is not related to any specific type of
solution asymptotic behaviour near the crack tip; (iii)
Modification of the way to impose boundary conditions in the
framework of the $\varepsilon$-regularization technique \footnote{Note that the problem regularization is an important issue. It can be done by various techniques. Another type of the direct regularisation is shown in \citet{Wrobel}.}; (iv)
Identification of optimal ranges of the various technique
parameters, accuracy of computations and stiffness of the resulting
dynamical systems; (v) Comparison of the solution for the Carter leak-off model with the numerical results from \citet{Kovalyshen} and discussion on the solution sensitivity.

The structure of this paper is as follows. In the next section we
collect known results for the PKN model in various formulations. We
restrict ourselves only to the information which is absolutely
necessary to understand the paper. Subsection~\ref{second_approach}
contains an alternative formulation in terms of a new proper
dependent variable - fracture volume.
In Section~\ref{numerics} we discuss in details the numerical procedures and present main results of the computations
comparing performances of different solvers under considerations.
The solution obtained for the Carter leak-off model is compared with the available numerical data from \citet{Kovalyshen}.
The main findings of this paper are summarized in the
conclusion section.
Some new results
concerning the asymptotic behaviour of the solutions for different
leak-off regimes are presented also in Appendices A and B.

\section{Problem formulation and preliminary results}
\label{section_2}

\subsection{Physical fundamentals and basic equations}
\label{basic}

Consider a symmetrical crack of the length $2l$ situated in the
plane $x\in(-l,l)$, where the length $l=l(t)$ is one of the solution
components changing as a result of the fluid flow inside the crack. The initial crack length is assumed to
be nonzero: $l(0)=l_*>0$. There are reasonable motivations behind
this assumption. Namely, for the initial (unstable) stage of the
crack propagation the acceleration of the process is too large to
neglect the inertial terms. For this reason, any classical model not
accounting for this effect is not credible. On the other hand, in
many cases the hydrofracturing process is associated with the
so-called 'perforation technique'. The latter consists of the creation
of a number of finger-shaped initial fractures by detonations of
shaped charges spaced along the wellbore \citep{Economides2000}. In this way,
hydrofracturing starts simultaneously from a number of non-zero
length cracks. Moreover, in many rock formations (e.g. shale
reservoirs) cracks already exist but are closed by the confining
stress \citep{Economides2000}. Finally, the uncertainties involved in this complex
multiphysics problem itself do not allow one to make any reliable
modelling of the crack nucleation in the rock formation.

By convention, we assume that the crack is fully filled by a
Newtonian liquid injected at known rate $q_0(t)$ at the crack mouth
$x=0$.

The Poiseulle equation for the Newtonian liquid flow in a narrow
channel is written in the form:
\begin{equation}\label{poiseulle}
q=-\frac{1}{M}w^{3}\frac{\partial p}{\partial x},
\end{equation}
where $q=q(t,x)$ is  the fluid flow rate and $w=w(t,x)$ is the crack
opening, while $p=p(t,x)$ is the net fluid pressure, that is, the
difference between the fluid pressure $p_f$ inside the fracture and the confining stress
$\sigma_0$ ($p=p_f-\sigma _0)$. The constant $M$ involved in the
equation is defined as $M=12\mu$, where $\mu$ stands for the dynamic
viscosity (see for example \citet{Economides2000}).

The continuity equation, accounting for the crack expansion and the
leak-off of the fluid, may be expressed as:
\begin{equation}\label{continuity}
\frac{\partial w}{\partial t}+\frac{\partial q}{\partial x}+q_l=0,\quad t>0,\quad 0<x<l(t),
\end{equation}
where $q_l=q_l(t,x)$ is the volume rate of fluid loss to formation
in the direction perpendicular to the crack surfaces per unit length
of fracture.

Numerical algorithms for the PKN model, with and without leak-off,
have been considered in \citet{Nordgren}, \citet{Kemp}, \citet{Kovalyshen_1} and others.
Improvements based on the speed equation and the $\varepsilon$-re\-gu\-la\-ri\-sa\-tion technique
 have been introduced in
\citet{Linkov_3,Linkov_4,MWL}, for the case when leak-off
vanishes near the crack tip. Below, we discuss the
effectiveness of this approach on the three most popular leak-off
models: Carter law \citep{Carter}, modified law incorporating pressure difference \citep{Clifton}
and bounded leak-off near the crack tip. An extensive discussion on possible
behaviour of the leak-off function can be found in \citet{Kovalyshen_1}.

In our numerical simulations, we utilise one of the following
leak-off variants:
\begin{equation}\label{leak-off}
q_l(t,x)=q_l^{(j)}(t,x)+q_j^*(t,x), \quad j=1,2,3,
\end{equation}
where
\begin{equation}
\begin{array}{l}
\displaystyle q_l^{(1)}=\frac{C_1(t)}{\sqrt{t-\tau (x)}},\quad
q_l^{(2)}=\frac{C_2(t)p}{\sqrt{t-\tau(x)}}, \quad
 \label{carter}
\displaystyle q_l^{(3)}=C_{31}(t)p+C_{32}(t),\quad \quad
0<x<l(t).
\end{array}
\end{equation}
Here $C_1=C_L$ is usually assumed to be a known constant defined
experimentally \citep{Carter}. Recently it was estimated analytically
for a poro-elastic material in \citet{Kovalyshen_1}. The function
$\tau(x)$ contains information on the history of the process. It
defines the time at which the fracture tip reaches the point $x$ and can
be computed as the inverse of the crack length:
\begin{equation}\label{tau}
\tau(x)=l^{-1}(x),\quad x>l_*.
\end{equation}
For $x\le l_*$ we conventionally set $\tau(x)\equiv 0$.
Other constants in \eqref{carter}, $C_j(t)=C_j(t,w,p)$, ($j=2,3$)
may depend on the solution itself but are bounded functions in time.
Finally, we assume that the terms $q_j^*$, ($j=1,2,3$) in \eqref{carter}
are negligible in comparison with $q_l^{(j)}$ near the crack tip.
Note that application of the Carter leak-off law \citep{Carter} which is a simplified
model of established fluid diffusion through the fracture walls, may be not justified at some stages of the process
\citep{Nordgren,Lenoach,Mathias}.

In this paper, we are aiming to build
a general numerical framework for the problem under consideration.
Thus, the collection of possible leak-off representations given in
\eqref{carter} covers the whole
spectrum of possible bahaviours used in the hydrofracturing
simulations \citep{Kovalyshen_1}.

The system of equations \eqref{poiseulle} - \eqref{continuity}
should be supplemented by the elasticity equation. We consider the
simplest relationship used in the PKN formulation
\begin{equation}\label{elasticity}
p=kw,
\end{equation}
with a known proportionality coefficient $k=\frac{2}{\pi
h}\frac{E}{1-\nu^2}$ found from the solution of a plane strain
elasticity problem for an elliptical crack of height
$h$ \citep{Nordgren}.  Constants $E$ and $\nu$ are the Young modulus and the Poisson
ratio, respectively. 
In physical interpretation, this condition refers to the case when the fracture resistance of the solid is so small,
that the energy dissipated by the fracture extension is negligible compared to the energy dissipated in the viscous fluid flow
\citep{Adachi_2002}. However, it turns out that, even in the models where the
toughness dominated regime can be accounted for, it may be of a minor importance.
For example, in \citet{Sav_Det_2002} it has been proven that
\emph{radial hydraulic fractures in impermeable rocks generally
propagate in the viscosity regime, and that the toughness regime is
relevant only in exceptional circumstances} (for the average values
of the field parameters the fracture would remain in the viscosity
dominated regime for many years).

On substitution of the Poiseulle equation \eqref{poiseulle} and
elasticity relationship \eqref{elasticity} into the continuity equation
\eqref{continuity}, one obtains a well known lubrication (Reynolds)
equation defined in the trapezoidal domain ($t>0,\,\,0<x<l(t)$):
\begin{equation}\label{reynold}
\frac{\partial w}{\partial t}-\frac{k}{M}\frac{\partial}{\partial
x}\left(w^{3}\frac{\partial w}{\partial x}\right)+q_l=0.
\end{equation}

Since the system has its natural symmetry with respect to variable
$x$ and the equations are local, it is convenient to consider only
half (symmetrical part) of the interval $[0, l(t)]$ instead of the
full crack length $[-l(t), l(t)]$.

Following the discussion on the initial crack length above, the
initial conditions for the problem are:
\begin{equation}\label{initial}
l(0)=l_*,\quad w(0,x)=w_*(x),\quad x\in(0,l_*).
\end{equation}

The boundary conditions include: known fluid injection rate at
the crack mouth, $q_0$, zero crack opening and zero fluid flux
rate at the crack tip:
\begin{equation}\label{boundary_conditions}
q(t,0)=q_0(t),\quad w(t,l(t))=0, \quad q(t,l(t))=0.
\end{equation}
Note that the problem formulated in this way looks overdetermined as
the governing equation (\ref{reynold}) is of the second order with
respect to spatial variable. This issue shall be discussed later.

Finally, by consecutive integration of equation (\ref{reynold}) over
time and then space, one can also derive the standard formula for
the global fluid balance in the form:
\begin{equation}
\begin{array}{l}
\label{balance_1}
\displaystyle
\int_0^{l(t)}[w(t,x)-w_*(x)]dx-\int_{0}^tq_0(t)dt+\int_0^{l(t)}\int_{0}^tq_l(t,x)dtdx=0,
\end{array}
\end{equation}
where it is accepted that $w_*(x)=0$ when $x>l_*$ and $l'(t)\ge0$.

As has been shown in \cite{Linkov_4}, the crucial role in the analysis
of the problem plays the particle velocity defined in the following
manner:
\begin{equation}\label{velocity_1}
V(t,x)=\frac{q}{w},\quad t>0,\quad 0\le x \le l(t),
\end{equation}
which indicates the average velocity of fluid flow through the cross-sections of the fracture.

Under the assumption that the crack is fully filled by the fluid and
sucking, ejection or discharge through the front can be neglected,
the fluid velocity defines the crack propagation speed and the
following speed equation is valid \citep{Kemp,Linkov_1,Linkov_2}\footnote{In fact, the speed equation in this form is valid only under the assumption of zero spurt loss at the crack tip \citep{Nordgren,Clifton,Adachi_2007}}:
\begin{equation}\label{velocity_2}
l'(t)=V\left(t,l(t)\right),\quad t>0.
\end{equation}
Moreover, for physical reasons, one can assume that the fluid
velocity at the crack tip is finite
\begin{equation}\label{velocity_b}
0\le V\left(t,x\right)<\infty,\quad t>0, \quad x \le l(t)
\end{equation}
Note that, allowing the crack propagation speed to be infinite, one
has to simultaneously include the inertia term in the equations. Thus,
the estimate (\ref{velocity_b}) is a direct consequence of
neglecting the acceleration terms.

\subsection{Asymptotic behaviour of the solution and its consequences}
\label{asymptotics}

As was mentioned in \citet{S$S}, the fact that both $w$ and $q$ are present in \eqref{velocity_1}, creates
serious difficulties when trying to use the fluid velocity as a
variable. However, as shown in \citet{Linkov_1,Linkov_2,Linkov_3,Linkov_4}, proper usage of
fluid velocity may be extremely beneficial. First, it allows one to
replace two boundary conditions at the crack tip
(\ref{boundary_conditions})$_{2,3}$ with a single one additionally incorporating
 information from the speed equation (\ref{velocity_2}),
(\ref{velocity_b}).

Indeed, the boundary conditions (\ref{boundary_conditions})$_{2,3}$
in view of (\ref{poiseulle}) and (\ref{elasticity}) lead to the
estimate
\begin{equation}
\label{w_asym_0}
w(t,x)= o\left((l(t)-x)^{\frac{1}{4}}\right),
\quad x\rightarrow l(t),
\end{equation}
which does not necessarily guarantee (\ref{velocity_b}). However,
further analysis of the problem, for different leak-off functions
(see \citet{Kemp,Kovalyshen} and Appendix B of this paper), shows that the particle velocity is
bounded near the crack tip and the crack opening exhibits the
following asymptotic behaviour:
\begin{equation}
\label{w_asym_1}
\begin{array}{l}
w(t,x)=
w_0(t)\big(l(t)-x\big)^{\frac{1}{3}}+w_1(t)\big(l(t)-x\big)^{\alpha}
+\,o\big((l(t)-x)^{\alpha}\big), \quad \text{as} \quad x\rightarrow
l(t),
\end{array}
\end{equation}
with some $\alpha>1/3$. For the classical PKN model for an impermeable
solid (or when leak-off vanishes near the crack tip
at least as fast as the crack opening) the
exponent $\alpha = 4/3$ was found in \cite{Kemp}. For
the case of the singular Carter's type leak-off, the exponent
$\alpha=1/2$ was determined in \cite{Kovalyshen}.

Note that the asymptotics \eqref{w_asym_1} shows that fluid velocity is indeed bounded near the crack tip. Moreover,
\begin{equation}
\label{V_asym_0}
V(t,x)= V_0(t)+V_1(t)\big(l(t)-x\big)^{\beta}+o\big((l(t)-x)^{\beta}\big),
\end{equation}
as $x\to l(t)$, where $\beta=\alpha-1/3$ and
\begin{equation}\label{V_asym_0_coefs}
V_0=\frac{k}{3M}w_0^3(t),\quad V_1=\frac{k}{M}\left(\alpha+\frac{2}{3}\right)w_0^2(t)w_1(t).
\end{equation}
As follows from Appendix B, $V(t,x)$ may not be so smooth near
the crack tip as one could expect and the exponent $\beta$ in
(\ref{V_asym_0}) plays an important role for this. Indeed, if
$\beta\ge1$ then $V(t,\cdot)\in C^1[0,l(t)]$ and the particle
velocity function is smooth enough near the crack tip. However, this
 happens only in the special case of $\alpha=4/3$ when $V_x(t,x)$ is
bounded near the crack tip. In case of singular leak-off
($0<\beta<1$), the particle velocity near the crack tip is only of
the H\"older type $V(t,\cdot)\in C^1[0,l(t))\bigcap
H^\beta[0,l(t)]$. In Appendix B we present an exact form of the
asymptotic expansion \eqref{w_asym_1}, which yields the
aforementioned smoothness deterioration of $V$ near the crack tip
for the singular leak-off models.

Note that estimate \eqref{w_asym_1} (or
\eqref{V_asym_0}) is equivalent to the condition \eqref{velocity_b}. Thus,
in view of \eqref{velocity_1}, the pair of conditions
\eqref{boundary_conditions}$_2$ and
 \eqref{boundary_conditions}$_3$ is equivalent to  \eqref{boundary_conditions}$_2$ and \eqref{w_asym_1}.
This discussion clearly illustrates why accounting for asymptotic
behaviour of the solution in form
  \eqref{w_asym_1} is of crucial importance for effective numerical realisation
of any algorithm utilised in hydrofracturing \citep{Adachi_2007b,Garagash et al,MKP}. On the other hand, the
fact that the particle velocity function is smooth enough near the
crack tip has been one of the important arguments to use \emph{the
speed equation and proper variable approach} as the basis for
improvement of the existing numerical algorithms \citep{Linkov_1}. It
should be emphasized that behaviour of $V(t,x)$ near the
crack tip may have serious implications when using
$\varepsilon$-regularization technique.

\emph{Therefore, the main aim of this paper is to show that,
regardless of possible smoothness of the particle velocity near the
crack tip, the approach proposed in \citet{Linkov_1} -- \citet{Linkov_4} and \citet{MWL} is still
efficient.}

\subsection{Normalised formulation}

Let us normalize the problem by introducing the following
dimensionless variables:
\[
\tilde x=\frac{x}{l(t)}, \quad \tilde t = \frac{t}{t_n},\quad
t_n=\frac{M}{kl_*},\quad \tilde w_*(\tilde x)= w_*(x),
\]
\[
\tilde w(\tilde t,\tilde x)=\frac{w(t,x)}{l_*}, \quad
\tilde V(\tilde t, \tilde x)=\frac{t_n}{l_*}V(t,x),\quad L(\tilde
t)=\frac{l(t)}{l_*},
\]
\begin{equation}
\label{dimensionless} l^2_{*}\tilde q_0(\tilde t)= t_nq_0(t),\quad
l_*\tilde q_l(\tilde t,\tilde x)=t_nq_l(t,x),
\end{equation}
where $\tilde x\in(0,1)$ and $L(0)=1$.

Using this notation, one
defines the normalised particle velocity as:
\begin{equation}\label{norm_speed}
 \tilde V(\tilde t,\tilde x)=-\frac{\tilde w^2}{L(\tilde t)}\frac{\partial \tilde w}{\partial \tilde
 x}.
 \end{equation}
The conservation law \eqref{continuity} in the normalised domain is
rewritten in the following manner:
\begin{equation}\label{norm_continuity}
\frac{\partial \tilde w}{\partial \tilde t}=
\frac{1}{L(\tilde t)}\left[\left(\tilde x \tilde V(\tilde t,1)-\tilde V(\tilde t,\tilde x)\right)\frac{\partial \tilde w}{\partial \tilde x}-\tilde w\frac{\partial \tilde V}{\partial \tilde x}\right]-\tilde q_l(\tilde t,\tilde x),
\end{equation}
The leading terms of the asymptotic estimate of the leak-off
function from (\ref{carter}) are now:
\begin{equation}
\tilde q_{l}^{(1)}(\tilde t,\tilde x)=\frac{\tilde C_1(\tilde t)D\left(\tilde t\right)}{\sqrt{1-\tilde{x}}},\,\,\,
\tilde q_{l}^{(2)}(\tilde t,\tilde x)=\frac{\tilde C_2(\tilde t)D\left(\tilde t\right)}{\sqrt{1-\tilde{x}}}\tilde w(\tilde t,\tilde x),\quad
\label{norm_leak_off_1}
\tilde q_{l}^{(3)}(\tilde t,\tilde x)=\tilde C_{31}(\tilde t)\tilde w(\tilde t,\tilde x)+\tilde C_{32}(\tilde t).
\end{equation}
Here, the function
\begin{equation}\label{function_D}
D\left(\tilde t\right)=\sqrt{\frac{L'\left(\tilde t\right)}{L\left(\tilde t\right)}},
\end{equation}
is introduced in the Appendix \ref{app:B}, where the remainder
between the normalised total flux and the leading term
(\ref{norm_leak_off_1}) has been effectively estimated. Thus the
normalised term $\tilde q^*_j(\tilde t,\tilde x)$ vanishes near the
crack tip faster than the solution itself.

Finally, normalised initial conditions (\ref{initial}) and boundary
conditions (\ref{boundary_conditions}) are:
\begin{equation}\label{norm_initial}
L(0)=1,\quad \tilde w(0,\tilde x)=\tilde w_*(\tilde x),\quad
x\in(0,1),
\end{equation}
and
\begin{equation}\label{norm_boundary}
  -\frac{1}{L(\tilde t)}\tilde w^3\frac{\partial \tilde w}{\partial \tilde x} (\tilde t,0)=\tilde q_0(\tilde t),\quad \tilde w(\tilde t,1)=0.
\end{equation}
The global fluid balance (\ref{balance_1}) can be written in the
form:
\begin{equation}
L(\tilde t)\int_0^{1}\tilde w(\tilde t, x)d x-\int_0^1 \tilde
w(x,0)d x-\int_{0}^{\tilde t} \tilde q_0( t)d t+
\label{balance_2} \int_{0}^{\tilde t} L(t)\int_0^{1}\tilde q_l( t,
x)d x d t=0.
\end{equation}
For convenience, from this point on we will omit the "$\sim$" symbol for
all dependent and independent variables and will only consider the
respective dimensionless values.

Note that the particular representation (\ref{norm_continuity}) of
the Rey\-nolds equation  highlights an essential feature of the problem
- it is singularly perturbed near the crack tip. Indeed, both
coefficients in front of the spatial derivatives on the right-hand
side of the equation (\ref{norm_continuity}) tend to zero at $x=1$.
Thus, the asymptotic behaviour of the solution near the crack tip
($x\to 1$)
\begin{equation}
\label{w_asym_1n}
w= w_0(t)\left(1- x\right)^{\frac{1}{3}}+w_1(t)\left(1-x\right)^{\alpha}+o\left((1-x)^{\alpha}\right),
\end{equation}
\begin{equation}
\label{V_asym_1n}
V= V_0(t)+V_1(t)\left(1- x\right)^{\alpha-\frac{1}{3}}+o\left((1-x)^{\alpha-\frac{1}{3}}\right),
\end{equation}
represents nothing but the boundary layer. Moreover, normalizing
\eqref{V_asym_0_coefs}$_{1}$ one obtains:
\begin{equation}
\label{v_0_w_0_relation}
V_0(t)=\frac{1}{3L(t)} w_0^3(t).
\end{equation}
The terms $w_0$, $w_1$ and $V_0$, $V_1$ in (\ref{w_asym_1n})
and (\ref{V_asym_1n}) are different from those in (\ref{w_asym_1})
and (\ref{V_asym_0}). In fact, the former should be written with
"$\sim$" symbol.

On substitution of \eqref{norm_speed} into \eqref{norm_continuity},
one eliminates the particle velocity function from Reynolds
equation:
\begin{equation}\label{w_system}
\frac{\partial  w}{\partial t}=
\frac{1}{L^2(t)}\left[\frac{1}{3}w_0^3 x\frac{\partial  w}{\partial
x}+3\tilde w^2\left(\frac{\partial  w}{\partial   x}\right)^2+
w^3\frac{\partial^2  w}{\partial x^2}\right] - q_l.
\end{equation}
Here $w_0$ is the multiplier of the first term of the asymptotic
expansion \eqref{w_asym_1}. This form of lubrication equation
exhibits the same degenerative properties as
(\ref{norm_continuity}). Also the coefficients appearing in
front of the leading terms tend to zero near the crack tip.

The speed equation (\ref{velocity_2}) defining the crack propagation
speed is given in the normalised variables as:
\begin{equation}\label{norm_speed_2}
L'(t)= V_0(t),\quad t>0.
\end{equation}
Taking into account (\ref{v_0_w_0_relation}), the latter can be
rewritten in  the following form
\begin{equation}
\label{new_speed} \frac{d }{dt}L^2=\frac{2}{3}w_0^3(t),\quad \quad
t>0.
\end{equation}
This equation serves us to determine the unknown value of the crack
length $L(t)$. As it has been shown in \cite{MWL}, such an approach
has clear advantages over the standard one based on the global fluid
balance equation (\ref{balance_2}).

As a result of the foregoing transformations, one can formulate a
system of PDEs describing the hydrofracturing
process. The system is composed of two operators:
\begin{equation}\label{w_DS}
\frac{d}{d t} w=\mathcal{A}_w( w,L^2),\quad \frac{d}{d t}L^2(t)=\mathcal{B}_w(w),
\end{equation}
where $\mathcal{A}_w$ is defined by the right-hand side of equation
\eqref{w_system} with the boundary conditions
\eqref{norm_boundary}$_{1,2}$, while the second operator
$\mathcal{B}_w$ is given by \eqref{new_speed}. The system is
equipped with the initial conditions:
\begin{equation}\label{norm_initial_DS}
L(0)=1,\quad w(0,x)=w_*(x) ,\quad x \in (0,1).
\end{equation}

\subsubsection{Reformulation of the problem in proper dependent variables. First approach}
\label{first_approach}

In \cite{Linkov_4} and later in \cite{MWL} it has been shown that
the dependent variable
\begin{equation}\label{U}
U(t,x)= w^3(t,x)
\end{equation}
is more favorable for the solution of the system  (\ref{w_DS}),
(\ref{norm_initial_DS}) than the crack opening itself. This idea is
based on the fact that, according to the asymptotics of the solution
near the crack tip, the dependent variable $U$ is much smoother than
$w$. In the case of an impermeable solid, the solution $U$ is analytic in
the closed interval $[0,1]$ (see \cite{Linkov_4}). However, the type
of leak-off function is of significant importance here. Thus,
adopting asymptotic representation (\ref{w_asym_1n}), one can see
that for $x \to 1$
\begin{equation}\label{U_asymp}
U=U_0(t)(1- x)+U_1(t)(1- x)^{\frac{2}{3}+\alpha}+o((1- x)^{\frac{1}{3}+2\alpha}),
\end{equation}
where the coefficients $U_0(t)$ and $U_1(t)$ are directly related to
those appearing in the crack opening formulation:
\begin{equation}\label{U_coefs}
U_0(t)=w_0^3(t),\quad U_1(t)=  w_0^2(t) w_1(t).
\end{equation}
Depending on the type of leak-off described in \eqref{carter}, the
exponent in the second asymptotic term $\frac{2}{3}+\alpha$ may take
value $3/2$, $11/6$ or 2, respectively. Thus in the first two
cases, the transformation (\ref{U}) no longer results in polynomial
representations of asymptotic expansion for $U$. For this reason, the
advantage of the approach using variable $U$ in more general cases, when the leak-off is singular
near the crack tip, should still be confirmed. This is one of the aims of
this paper.
On the other hand, at least two factors work in favor of this formulation.
First, the spatial derivative of $U$ is not singular and, second, the particle velocity
is given by a linear relationship
\begin{equation}\label{V_norm}
 V(t,x)=-\frac{1}{3L(t)}\frac{\partial U}{\partial x}.
\end{equation}
The governing equation \eqref{norm_continuity} in terms of the new
variable can be written in the normalized domain $x\in(0,1)$ as:
\begin{equation}\label{U_continuity}
\frac{\partial U}{\partial t}=
\frac{1}{L(t)}\left[\left( x  V(t,1)- V(t, x)\right)\frac{\partial U}{\partial  x}-
3U\frac{\partial V}{\partial x}\right]-3U^{\frac{2}{3}} q_l,
\end{equation}
Similarly to (\ref{w_system}) the particle velocity function may be
eliminated from the lubrication equation:
\begin{equation}\label{U_system}
\frac{\partial U}{\partial t}=
\frac{1}{3L^2(t)}\left[ x U_0\frac{\partial U}{\partial  x}+\left(\frac{\partial U}{\partial  x}\right)^2+3U\frac{\partial^2 U}{\partial  x^2}\right]-3U^{\frac{2}{3}} q_l.
\end{equation}
Note that equations \eqref{U_continuity}-\eqref{U_system} are of a
very similar structure to those evaluated for the crack opening $w$.
They exhibit the same degenerative nature near the crack tip.

Finally, boundary conditions (\ref{norm_boundary}) transform to:
\begin{equation}\label{U_boundary}
-\frac{\sqrt[3]{U(t,0)}}{3L(t)}\frac{\partial}{\partial
x}U(t,0)=q_0(t) ,\quad  U(t,1)=0,
\end{equation}
while the speed equation \eqref{new_speed} takes the following form:
\begin{equation}\label{new_speed_U}
\frac{d }{dt}L^2=\frac{2}{3} U_0(t),\quad \quad t>0.
\end{equation}

The system of PDEs equivalent to (\ref{w_DS}) is now defined as:
\begin{equation}\label{U_DS}
\frac{d}{d t} U=\mathcal{A}_U(U,L^2),\quad \frac{d}{d t}L^2(t)=\mathcal{B}_U(U).
\end{equation}
The operator $\mathcal{A}_U$ is described by \eqref{U_system} with
boundary conditions \eqref{U_boundary}, while the second operator
$\mathcal{B}_U$ is given by \eqref{new_speed_U}. Finally, the
initial conditions are similar to those in the previous formulation
\eqref{norm_initial}:
\begin{equation}\label{U_initial}
L(0)=1,\quad U(0, x)= w_*^3( x),\quad  x \in (0,1).
\end{equation}

\subsubsection{Reformulation of the problem in proper dependent variables. Second approach}
\label{second_approach}

The aforementioned formulation of the problem in terms of the
dependent variable $U$ has  one considerable drawback. It is
well known that for different elasticity models and different
hydrofracturing regimes one has various asymptotic behaviours of the
solution near the crack tip \citep{Adachi_2002}. For example, for exact equations of
elasticity theory and the zero toughness condition ($K_{IC}=0$), the
exponent of the leading term of $w$ varies from 2/3, for the
Newtonian fluid, to 1, for the ideally plastic fluid. Thus, the same
reformulation to the type of the proper variable might be
inconvenient, or even impossible.

For this reason, we introduce another dependent variable. Although
it does not transform the asymptotic behaviour of the solution in
such a smooth manner as it has been done previously when adopting
$U$, this variable has its own advantages. Namely, let us consider
a new dependent variable $\Omega$ defined as follows:
\begin{equation}\label{omega}
\Omega(t,x)=\int_{x}^{1}w(t,\xi)d\xi.
\end{equation}
\emph{This variable is not directly related to any particular
asymptotic representation of $w(x,t)$, however it assumes that $w
\to 0$ for $x \to 1$. As a result the form of governing equations
for $\Omega$ remains the same regardless of $w(x,t)$ asymptotics,
i.e. this formulation has a general (universal) character. Note,
that in case of $U$ the optimal way of transformation for the
lubrication equation essentially depends on the exact form of
asymptotic expansion (the leading term) for $w$. \\
Another advantage of $\Omega$ comes from the fact that it has clear
physical and technological interpretation. Namely, it reflects the
crack volume measured from the crack tip.}

Asymptotics of the function  $\Omega$ near the crack tip takes the following form:
\begin{equation}\label{omega_asymp}
\begin{split}
\Omega(t,x)=\Omega_0(t)(1- x)^{\frac{4}{3}}+\Omega_1(t)(1- x)^{\alpha+1}+o((1- x)^{\alpha+1}),\quad
x \rightarrow 1,
\end{split}
\end{equation}
where the coefficients $\Omega_0(t)$ and $\Omega_1(t)$ are related
to those in (\ref{w_asym_1n}):
\begin{equation}\label{omega_coefs}
\Omega_0(t)=\frac{3}{4} w_0(t),\quad \Omega_1(t)= \frac{1}{\alpha+1}w_1(t).
\end{equation}
\emph{Thus, similarly to $U$, the new variable is smoother than the
 crack opening, $w$, and the first singular derivative of $\Omega$ is that of the second order.}

By spatial integration of  \eqref{norm_continuity} from $x$ to 1 and
substitution of \eqref{omega} one obtains:
\begin{equation}\label{omega_continuity}
\frac{\partial \Omega}{\partial t}=
-\frac{1}{L(t)}\left[\big( V (t, x)- xV(t,1)\big)\frac{\partial\Omega}{\partial x}+ V(t,1)\Omega\right]- Q_l,
\end{equation}
where the monotonicity of $L'(t)>0$ has been taken into account and
\[
Q_l(t,x)=\int_{ x}^1 q_l(t,\xi) d\xi.
\]
Here, the particle velocity \eqref{norm_speed} is  computed in the
manner:
\begin{equation}\label{omega_speed}
V(t, x)=\frac{1}{3L(t)} \frac{\partial }{\partial  x}\left(\frac{\partial \Omega}{\partial  x}\right)^3\!\!\!.
\end{equation}
By eliminating $V(x,t)$ from the equation \eqref{omega_continuity}
we derive a new formula for the lubrication equation:
\begin{equation}\label{omega_system}
\frac{\partial \Omega}{\partial t}=
-\frac{1}{L^2(t)}\left[\left(\frac{\partial \Omega}{\partial  x}\right)^{\!3} \frac{\partial^2 \Omega}{\partial x^2}+\frac{64}{81}
\Omega_0^3\left(\Omega- x\frac{\partial \Omega}{\partial x}\right)\right]-Q_l.
\end{equation}
The boundary conditions \eqref{norm_boundary} are expressed in the
following way:
\begin{equation}\label{omega_boundary}
- \frac{1}{L(t)}\left(\frac{\partial \Omega}{\partial x}\right)^3\frac{\partial^2 \Omega}{\partial  x^2}(t,0)=q_0,\quad \frac{\partial \Omega}{\partial x}(t,1)=0.
\end{equation}
Interestingly, the first boundary condition, directly substituted
into the lubrication equation \eqref{omega_continuity} can be
equivalently rewritten in the form
\begin{equation}\label{omega_boundary_1} \frac{\partial
\Omega}{\partial
t}(t,0)=-\frac{64}{81^2L(t)}\Omega(t,0)\Omega_0^3(t)+\frac{q_0(t)}{L(t)}-
Q_l(t,0),
\end{equation}
This condition, in turn, represents nothing but the local (in time)
flux balance condition. To verify this, it is enough to apply the time
derivative to the equation (\ref{balance_2}). Furthermore it appears
much easier for implementation into a numerical procedure than
(\ref{omega_boundary})$_1$ itself, but may lead to some increase of the problem stiffness, as we will show later.

It is easy to check, by using the governing equation
\eqref{omega_continuity} and limiting values of all its terms for $x
\to 1$ , that a weaker boundary condition
\begin{equation}\label{omega_boundary_2}
\Omega(t,1)=0
\end{equation}
is equivalent to the original one \eqref{omega_boundary}${}_2$.
Finally, the speed equation (\ref{new_speed}) in the $\Omega$
formulation assumes the following form:
\begin{equation}\label{omega_L2}
\frac{d}{d t}L^2(t)=\frac{128}{81} \Omega_0^3(t).
\end{equation}

In this way we obtain another system of PDEs that is composed of two
operator relations:
\begin{equation}\label{Omega_DS}
\frac{d}{d t} \Omega=\mathcal{A}_\Omega(\Omega,L^2),\quad \frac{d}{d t}L^2(t)=\mathcal{B}_\Omega(\Omega),
\end{equation}
where, as previously, $\mathcal{A}_\Omega$ is defined by \eqref{w_system} with boundary conditions \eqref{omega_boundary}$_{1,2}$ or \eqref{omega_boundary_1} and \eqref{omega_boundary_2}.
The second operator, $\mathcal{B}_\Omega$, is given by equation \eqref{omega_L2}.
Here the initial conditions are obtained from \eqref{norm_initial}:
\begin{equation}\label{Omega_initial}
L(0)=1,\quad  \Omega(0, x)=\Omega_*( x)\equiv \int_{ x}^{1} w_*(\xi)d\xi.
\end{equation}

\subsection{$\varepsilon$-regularization and the respective boundary conditions}
\label{e_reg}

In our analysis we are going to use the so-called
$\varepsilon$-regularization technique. It was originally
introduced in \cite{Linkov_1} for the system of spatial coordinates
moving with the fracture front. In \cite{MWL}, the authors efficiently adopted
the approach for the normalised coordinate system.

The reason to separate the domain from the end point $x=1$ by a
small distance of $\varepsilon$ and to introduce
$\varepsilon$-regularisation has been thoroughly described in
\cite{Linkov_1}. It consists of replacing the Dirichlet boundary
condition (\ref{U_boundary})$_2$ with an approximate one:
\begin{equation}\label{L1}
U(t,1-\varepsilon)=3\varepsilon L(t) V(t,1),
\end{equation}
emerging from deep physical arguments. The value of the crack
propagation speed $V(t,1)$ (and simultaneously the particle velocity
at a fracture tip) was suggested to be computed from the speed
equation (\ref{new_speed_U}) in its approximated form:
\begin{equation}\label{L2}
V(t,1)=-\frac{1}{3L(t)}\frac{\partial U}{\partial x}(t,1-\varepsilon).
\end{equation}
The pair of conditions \eqref{L1} -- \eqref{L2} has shown an excellent
performance in terms of solution accuracy and, as
has been proven in \cite{MWL}, reduced the stiffness of dynamic system in case of
leak-off function vanishing near the crack tip. One can check that
for such a leak-off  model numerical error introduced by using the
approximate conditions, instead of the exact ones, is of the order
$O(\varepsilon^2)$. In view of all improvements following from
utilisation of the regularized conditions (\cite{Linkov_4,MWL}) such
a strategy is fully justified and in fact inevitable.

The conditions can be written in an equivalent form. Indeed, one can merge
(\ref{L1}) and (\ref{L2}) into a single
condition of the third type:
\begin{equation}\label{L3}
U(t,1-\varepsilon)+\varepsilon\frac{\partial U}{\partial
x}(t,1-\varepsilon)=0.
\end{equation}
Interestingly, the latter condition is nothing but the consequence of a direct utilization of the information about the
leading term of asymptotics of the solution near the crack tip (compare with (\ref{U_asymp})).

Analogously, one can define the respective pairs of boundary
conditions in the regularized formulations. Considering the dependent
variable $w$ one should take (\ref{new_speed}) together with the
condition
\begin{equation}\label{L4}
w(t,1-\varepsilon)+3\varepsilon\frac{\partial w}{\partial
x}(t,1-\varepsilon)=0,
\end{equation}
while analysing the system based on the dependent variable $\Omega$,
the speed equation (\ref{omega_L2}) should be accompanied by
\begin{equation}\label{L5}
4\Omega(t,1-\varepsilon)+3\varepsilon\frac{\partial \Omega}{\partial
x}(t,1-\varepsilon)=0.
\end{equation}

To conclude this subsection, one can make a prediction that in the case
of a singular leak-off function, even when using the
$\varepsilon$-regularization technique, the accuracy of the solution
should be worse than that presented in \cite{Linkov_1},\cite{MWL}.
However, it is always possible to use information on accurate
asymptotic behaviour of the solution (employing higher order terms)
and in this way improve the accuracy of computations.

\section{Numerical solution of the dynamic systems}
\label{numerics}

In this section, three alternative systems of PDEs (\ref{w_DS}), (\ref{U_DS}) and (\ref{Omega_DS}) describing
the problem of hydrofracruing are transformed into the corresponding
non-linear dynamic systems of the first order. Then, on the basis of
respective analytical benchmarks, we analyze their stiffness
properties, the accuracy and efficiency of computations. The
benchmark solutions in question are described in Appendix C.

\subsection{Representation of the boundary conditions and the speed equation}
\label{Rbc}

Consider a spatial domain of the problem reduced in accordance with
the $\varepsilon$-regularization technique to the interval $x
\in[0,1-\varepsilon]$, where $\varepsilon$ is a small parameter. Let
the mesh, $\{x_j\}_{j=1}^{N}$, be composed of $N $ nodes with
$x_1=0$ and $x_N=1-\varepsilon$.

For each of the problem formulations, two boundary conditions should
be accounted for: one specified at the crack inlet and a regularized
boundary condition at $x=1-\varepsilon$. In the following we present
a brief description of how these conditions are introduced to the
numerical scheme.

From now on, for the dependent variables discussed above ($w(t,x)$, $U(t,x)$, $\Omega(t,x)$),
we use common notation $f(t,x)$ together with a convention
$f_k=f(t,x_k)$.

To discretize the first boundary condition (depending on the
formulation: (\ref{norm_boundary})$_1$, (\ref{U_boundary})$_1$ or
(\ref{omega_boundary})$_1$) we exploit the smooth character of the
solution near the point $x=0$. Thus, accepting a polynomial
approximation of $f(x,t)$ on the interval $x \in [x_1,x_3]$, the
respective nonlinear relation between $f_1$, $f_2$ and $f_3$ may be
derived:
\begin{equation}
\label{BC_0} A_1(f_1,t)f_1+A_2(f_1,t)f_2+A_3(f_1,t)f_3=q_0.
\end{equation}


As mentioned in \ref{e_reg}, the regularized boundary condition in
the $\varepsilon$-regularization technique proposed in
\cite{Linkov_1} is equivalent to a mixed boundary condition based on
the leading term of the asymptotic expansion (see \eqref{L3},
\eqref{L4}, \eqref{L5}). Below we propose a modification of this
approach which consists in employing two terms of the asymptotics.
We will show that such a strategy prevents the  deterioration of accuracy
when the solution is not so smooth as in the cases originally
considered in \cite{Linkov_1}, \cite{MWL}.

According to (\ref{w_asym_1n}), (\ref{U_asymp}) and
(\ref{omega_asymp}), the following asymptotics approximation
is
acceptable in the proximity of the crack tip ($x\in[x_{N-2},1]$):
\begin{equation}
\label{expan_1}
f(t,x)=e_1^{(f)}(t)(1-x)^{\alpha_1}+e_2^{(f)}(t)(1-x)^{\alpha_2}.
\end{equation}
The values of $\alpha_1$ and $\alpha_2$ are known in advance and
depend, as has been discussed above, on the chosen variable and
the behavior of the leak-off function. Assuming that the last three
points of the discrete solution $(x_{N-2},f_{N-2})$,
$(x_{N-1},f_{N-1})$ and $(x_{N},f_{N})$ lie on the solution graph
$(x,f(t,x))$, one can derive a formula combining all these values in
one equation:
\begin{equation}
\label{BC_N} f_N+
b_{N-1}^{(f)}f_{N-1}+b_{N-2}^{(f)}f_{N-2}=0,
\end{equation}
where $b_j^{(f)}=b_j^{(f)}(x_{N-2},x_{N-1},x_{N})$. Relation
\eqref{BC_N} is consequently used to represent the regularised
boundary condition at $x=1-\varepsilon$.

{\sc Remark 1}. In the authors opinion, the presented approach is a
direct generalization of that proposed in \cite{Linkov_1}. Indeed,
if one takes $e_2=0$ in the representation (\ref{expan_1}) then the
pair of the equations (\ref{L3}) and (\ref{L2}) follows immediately.
If $\alpha_2^{(f)}-\alpha_1^{(f)}=1$, which means that the leak-off
function $q_l$ is bounded near the crack tip, the second asymptotic
term provides a small correction. However, in the case of the Carter
law, when $\alpha_2^{(f)}-\alpha_1^{(f)}=1/6$, it brings an
important contribution and improves the accuracy of the computations,
as will be shown later.

Finally, coefficient $e_1$ from \eqref{expan_1} is substituted into
the pertinent form of the speed equation (\ref{new_speed}),
(\ref{new_speed_U}) or (\ref{omega_L2}) to give the ordinary
differential equations for the crack length:
\begin{equation}
\frac{d}{dt}L^2=\frac{2}{3}\left(e_1^{(w)}\right)^3,\quad \frac{d
}{dt}L^2=\frac{2}{3} e_1^{(U)},\quad
\label{new_speed_common} \frac{d}{d t}L^2(t)=\frac{128}{81}
\left(e_1^{(\Omega)}\right)^3.
\end{equation}
Note that the right-hand sides of the equations define the boundary
operators $\mathcal{B}_w$, $\mathcal{B}_U$ and $\mathcal{B}_\Omega$
from (\ref{w_DS})$_2$, (\ref{U_DS})$_2$ or (\ref{Omega_DS})$_2$,
respectively.

{\sc Remark 2}. As it follows from this analysis, the
$\varepsilon$-regularization is, in a sense, equivalent to the
introduction of a special tip element in the discrete solution.
Thus, one can see a complementarity with the approach utilised in
\citet{Kovalyshen_1}. However, and this is crucial for the analysis,
only the speed equation together with $\varepsilon$-regularization
allows to take into account both the local and global phenomena, and
to do this in the most efficient way from the numerical point of
view.

{\sc Remark 3}. In the case of the dependent variable $U$, apart from the representations (\ref{expan_1}) of the boundary condition near the crack tip in the linear form
\begin{equation}\label{expan_3}
U_N=b_1^{(U)}U_{N-1}+b_2^{(U)}U_{N-2},
\end{equation}
one can use a nonlinear one, adopting the relationship between this dependent variable and the crack opening $w$:
\begin{equation}\label{expan_4}
U_N=\left(b_1^{(w)}\sqrt[3]{U_{N-1}}+b_2^{(w)}\sqrt[3]{U_{N-2}}\right)^3.
\end{equation}
Note that the two terms representation (\ref{expan_3}) of the function $U$ is less informative than
the same representation for the functions $w$ (or $\Omega$) and thus,
using the modified condition (\ref{expan_4}), one can expect a better solver performance.

\subsection{Spatial discretization of the Reynolds equation. Corresponding dynamic systems}

Let us consider the Reynolds equations written in different dependent
variables ((\ref{w_DS}), (\ref{U_DS}) or (\ref{Omega_DS}),
respectively). By representing the
spatial derivatives in the right-hand sides of the corresponding
equation by central three point finite difference schemes, we
obtain a nonlinear system of $N-2$ ordinary differential equations
for the values $f_i(t)$ at each internal point of the spatial domain
($x_2,...,x_{N-1}$). The respective boundary conditions are embedded
into the system  through equations \eqref{BC_0} and
\eqref{BC_N}.

Supplementing the system with the pertinent form of the speed
equation \eqref{new_speed_common}, we obtain a non-linear dynamic
system of first order describing the process of hydrofracturing which
can be written in the form:
\begin{equation}\label{governing_dis}
{\bf F}' ={\bf A}^{(f)}{\bf F}+{\bf G}^{(f)},
\end{equation}
where ${\bf F}={\bf F}(t)$ is  a vector of unknown solution
$[f(t,x_1),$ $f(t,x_2),\ldots,f(t,x_N),L^2(t)]$ of dimension $N-1$.
Note that matrix ${\bf A}^{(f)}$ and vector ${\bf G}^{(f)}$ depend,
generally speaking, on the solution. Matrix $\bf A$ is the so-called
mass matrix of the system, in the case of which a tri-diagonal form
prevails (however the boundary conditions and the last equation for
$L^2(t)$ disturb the tri-diagonal structure).

{\sc Remark 4} In the case when the
boundary condition in formulation \eqref{omega_boundary_1} is in
use, the dimension of the dynamic system is $N$. Indeed, this
condition has
the form of an ODE, and thus can be substituted directly
into the system as an additional equation.

In our numerical computations two different types of spatial meshes
are used. The first one is a regular mesh with uniformly spaced
nodes,  while the second one
gives an increased nodes density when approaching
the crack tip. Both types of meshes can be described by the formula:
\begin{equation}\label{mesh_power}
 x_m(\delta)
 =1-\left(1-\left(1-\varepsilon^{\frac{1}{\delta}}\right)\frac{m}{N}\right)^{\delta}, \quad m=1,...,N.
\end{equation}
In the foregoing, the parameter $\delta$ defines the mesh type. Namely, for $\delta=1$ one has the uniform mesh (henceforth denoted as $x^{(I)}$),
since any $\delta>1$ gives the nodes concentration near the crack tip (this mesh will be referred to as $x^{(II)}(\delta)$).
Mesh $x^{(II)}(\delta)$ allows one to choose appropriate parameter
$\delta$ to suppress the stiffness of dynamic system or to increase
the solution accuracy.

The stiffness of a dynamic system
may be described by the condition number or the condition ratio
\citep{stiff_ODE} of a mass matrix ${\bf A}^{(f)}$. In general, the values
given by various measures are different (see some consequences in
\cite{MWL}). In this paper we use the condition ratio as the measure
of the system stiffness. Some rough estimation of this parameter may
help to choose an optimal value of $\delta$ from the stiffness point
of view. In the case under consideration, one can analyse the
condition ratio of a simplified variant of the system
\eqref{U_system}, where only the leading term (with the second order
derivative) is preserved and the nonlinear multiplier is substituted
by the first term of the asymptotic expansion for $U$. It turns out that
 $\delta=2$ gives the lowest possible stiffness.
Naturally, in the general case, the optimal value of $\delta$ can be
different. We have checked however that, for three alternative
problem formulations, there are three different optimal values of
the parameter, but each of them is very close to 2. Thus, in the
following section all results concerning nonuniform mesh are
presented for $\delta=2$.

\subsection{Stiffness analysis}

In our analysis, we quantify the system stiffness by a condition
ratio $\kappa_A$ defined in (\ref{kappa}).
Since the problem is nonlinear, our
investigation is to be done for the linearized form of matrix ${\bf
A}^{(f)}$. It is obvious that for all six variants of benchmark
solutions under consideration (see Appendix \ref{app:A}) one has
different values of ${\bf A}^{(f)}$. Computations are carried out
for two types of meshes (the uniform and non-uniform one). For each
of the benchmark solutions, one obtains a constant  value for the condition ratio (independent on
time), apart from the fact that the matrix
 ${\bf A}^{(f)}$ depends on time. Those values of the condition ratio $\kappa_A$ are,
generally speaking, different for
various benchmarks and chosen meshes.

Before comparing the results for various dependent variables, we
have checked that for the dynamic system based on $U$ the stiffness
is almost identical for both forms of the regularized boundary
condition at $x=1-\varepsilon$ ((\ref{L3}) and (\ref{BC_N})
respectively). Thus, for the rest of the dependent variables ($w$ and
$\Omega$) we restrict our stiffness investigation only to the
formulation (\ref{BC_N}). Remarkably, the situation changes
dramatically when one considers the accuracy of  computations, which
shall be discussed later on.

Computing the condition ratio for the next variants of the matrix
linearization, we have confirmed the following estimate valid for large values of
$N$ for all cases under investigation:
\begin{equation}
\label{kappa}
\kappa_A^{(f)}=\frac{|\lambda_{max}|}{|\lambda_{min}|}\sim \varpi^{(f)} N^2,\quad N\to\infty.
\end{equation}
Here $|\lambda_{max}|$, and $|\lambda_{min}|$ are the largest and
smallest absolute values among the ${\bf A}^{(f)}$ matrix
eigenvalues, while the constant $\varpi^{(f)}$ is to be estimated
numerically. Its values for all six benchmark cases are shown in
 Table~\ref{table_A}.

Although the qualitative character
of the stiffness behaviour ($N^2$) is rather obvious, its
quantitative measure described by $\varpi$ can be used to select the
optimal (in terms of the stiffness properties) variant of the
dynamic system.

\begin{table}[h!]
\centering
\begin{tabular}{c| c@{}|c@{}|c@{}|c@{}|c@{}|c@{}|}
\cline{2-7}
& \multicolumn{3}{c|}{$Q_l/q_0=0.9$} & \multicolumn{3}{c|}{$Q_l/q_0=0.5$} \\ \cline{2-7}
& $q_l^{(1)}$ & $q_l^{(2)}$ & $q_l^{(3)}$ & $q_l^{(1)}$ & $q_l^{(2)}$ & $q_l^{(3)}$\\ \cline{2-7}
& \multicolumn{6}{c|}{$\varpi$ estimated for the system based on variable $w$}\\ \cline{1-7}
\multicolumn{1}{|c|}{$x^{(I)}$}
&6.5e+0&6.6e+0&6.8e+0&1.8e+1&1.8e+1&1.8e+1 \\ \cline{1-7} \multicolumn{1}{|c|}{$x^{(II)}$}
&1.7e+0&1.7e+0&1.7e+0&4.6e+0&4.7e+0&4.7e+0
  \\ \cline{1-7}
& \multicolumn{6}{c|}{$\varpi$ estimated for the system based on variable $U$ }\\ \cline{1-7}
\multicolumn{1}{|c|}{$x^{(I)}$} &3.0e+0&3.0e+0&3.2e+0&6.0e+0&6.1e+0&6.2e+0 \\ \cline{1-7} \multicolumn{1}{|c|}{$x^{(II)}$}
&7.5e-1&7.7e-1&8.1e-1&1.5e+0&1.5e+0&1.6e+0
  \\ \cline{1-7}
& \multicolumn{6}{c|}{$\varpi$ estimated for $\Omega_{(1)}$ based on  condition (\ref{omega_boundary_1})}\\ \cline{1-7}
\multicolumn{1}{|c|}{$x^{(I)}$}
&4.8e+1&4.8e+1&4.9e+1&1.7e+1&1.7e+1&1.7e+1 \\ \cline{1-7} \multicolumn{1}{|c|}{$x^{(II)}$}
&1.2e+1&1.2e+1&1.3e+1&4.3e+0&4.3e+0&4.3e+0
  \\ \cline{1-7}
& \multicolumn{6}{c|}{$\varpi$ estimated for $\Omega_{(2)}$ based on condition (\ref{BC_0})}\\ \cline{1-7}
\multicolumn{1}{|c|}{$x^{(I)}$}
&2.3e+1&2.2e+1&1.9e+1&9.6e+0&1.0e+1&1.3e+1 \\ \cline{1-7} \multicolumn{1}{|c|}{$x^{(II)}$}
&5.8e+0&5.7e+0&4.7e+0&2.5e+0&2.6e+0&3.5e+0
  \\ \cline{1-7}
\end{tabular}
\caption{Values of the parameter $\varpi^{(f)}$ from the
approximation of the condition ratio (\ref{kappa}) for the different
dynamic systems (\ref{governing_dis}) and different benchmarks. The computations were provided for $\varepsilon =10^{-3}$} \label{table_A}
\end{table}

\vspace{-4mm}

The following analysis includes investigation of stiffness
sensitivity to: i) the solution (benchmark) type, ii) choice of the
dependent variable, iii) choice of the independent variable
(spatial mesh), iv) value of the regularization parameter
$\varepsilon$.

{\sc Remark 5.} As mentioned previously, in case of the variable
$\Omega$, there are two alternative ways to introduce the boundary
condition at $x=0$ to the system -formulations
(\ref{omega_boundary_1}) and (\ref{BC_0}), respectively. In this way
one can construct two alternative dynamic systems of different
dimensions ($N$ and $N-1$). The results  in Table~\ref{table_A} show
that the stiffness properties of the system corresponding to the
boundary condition (\ref{BC_0}) are slightly better than those of
the system utilizing (\ref{omega_boundary_1}). Indeed, the
respective parameter $\varpi$ is about two times smaller. One of the
possible explanations is the aforementioned difference in the
systems' sizes: $\dim {\bf A}^{(\Omega)}_{(1)}=\dim {\bf
A}^{(\Omega)}_{(2)}+1$ (see Remark 3). We have checked that the
accuracy of computations remains practically the same regardless of
the system variant. Taking this fact into account, we restrict
ourselves in the analysis only to the system which employs the
boundary condition at point $x=0$ in the form (\ref{BC_0}). Thus,
from now on all the investigated dynamic systems (for all dependent
variables) will be based on the same mechanisms for incorporation of
the boundary conditions.

\begin{figure}[h*]
\centering
        \begin{subfigure}[b]{0.5\textwidth}
                \centering
                \includegraphics[width=\textwidth]{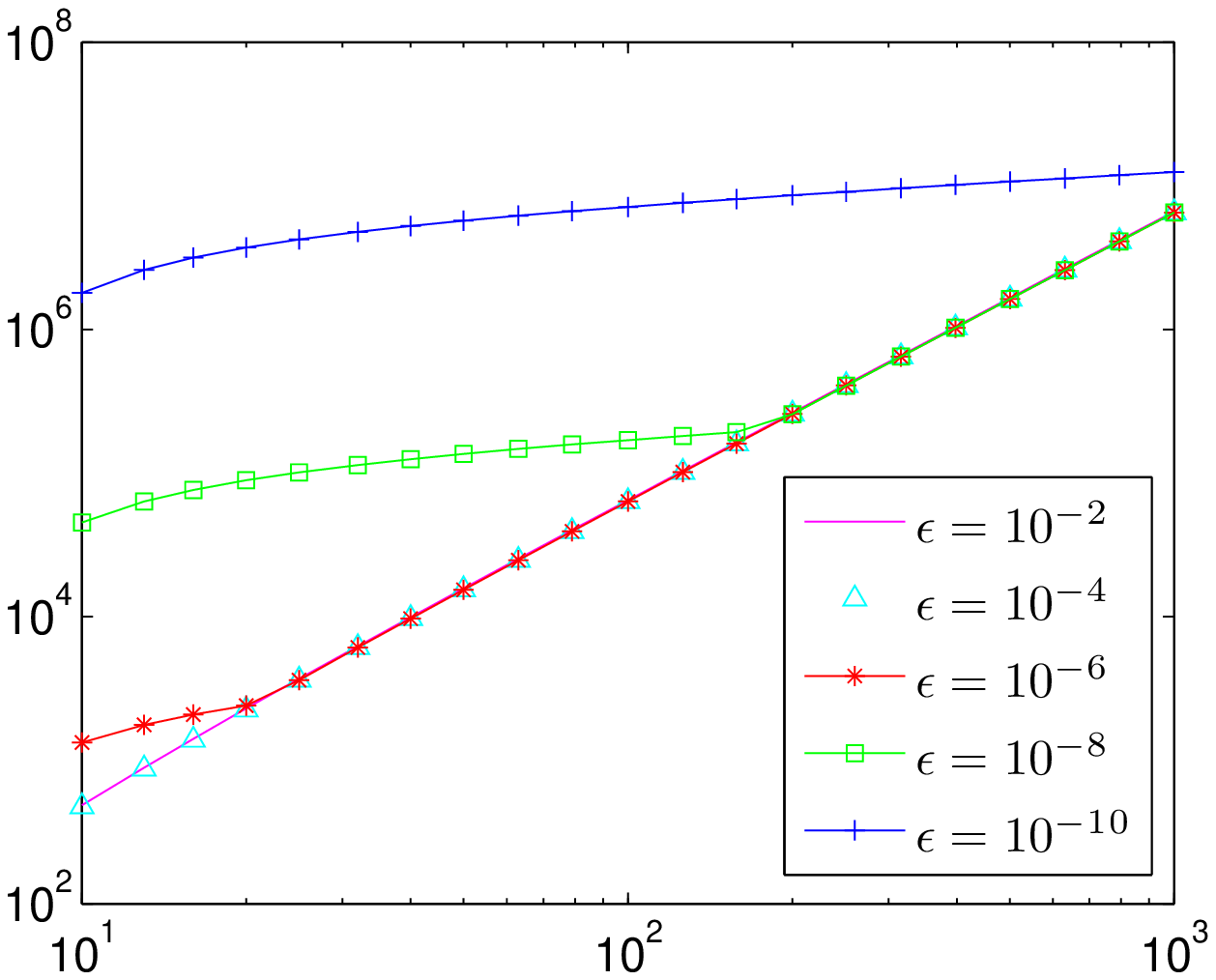}
                \begin{picture}(0,0)(85,-117)
        \put(50,-70){$Q_l/q_0=0.9$}
        \put(-8,-15){$\kappa$}
        \put(82,-110){$N$}
        \end{picture}

        \end{subfigure}

\vspace{-2mm}

\caption{Condition ratio $\kappa=\kappa^{(w)}(N)$ for the dynamic system based on $w$ variable and different values of the regularization parameter $\varepsilon$.
The case of the uniform mesh is analyzed.}
\label{fig:stiffness_b}
\end{figure}

Influence of the value of $\varepsilon$ on the condition ratio is analyzed in Fig.~\ref{fig:stiffness_b}.
As an example, we present here the benchmark $q_l^{(1)}$ for $Q_l/q_0=0.9$ (see Appendix C).
The results were obtained for the uniform mesh.
For other combinations of the benchmark solutions and different meshes the graphs have
similar character.
As anticipated, the estimation \eqref{kappa}  holds true
only for sufficiently large $N$.
The threshold value of $N$ depends
on the chosen $\varepsilon$.
Thus for a fixed number of grid
points $N$, there is a critical value of the regularization parameter
$\varepsilon_s(N)$ for which the stiffness characteristics changes its behaviour.
 By
taking $\varepsilon<\varepsilon_s(N)$ one increases appreciably the system
stiffness.

\begin{figure*}[t]
\centering
        \begin{subfigure}[b]{0.32\textwidth}
                \centering
                \includegraphics[width=\textwidth]{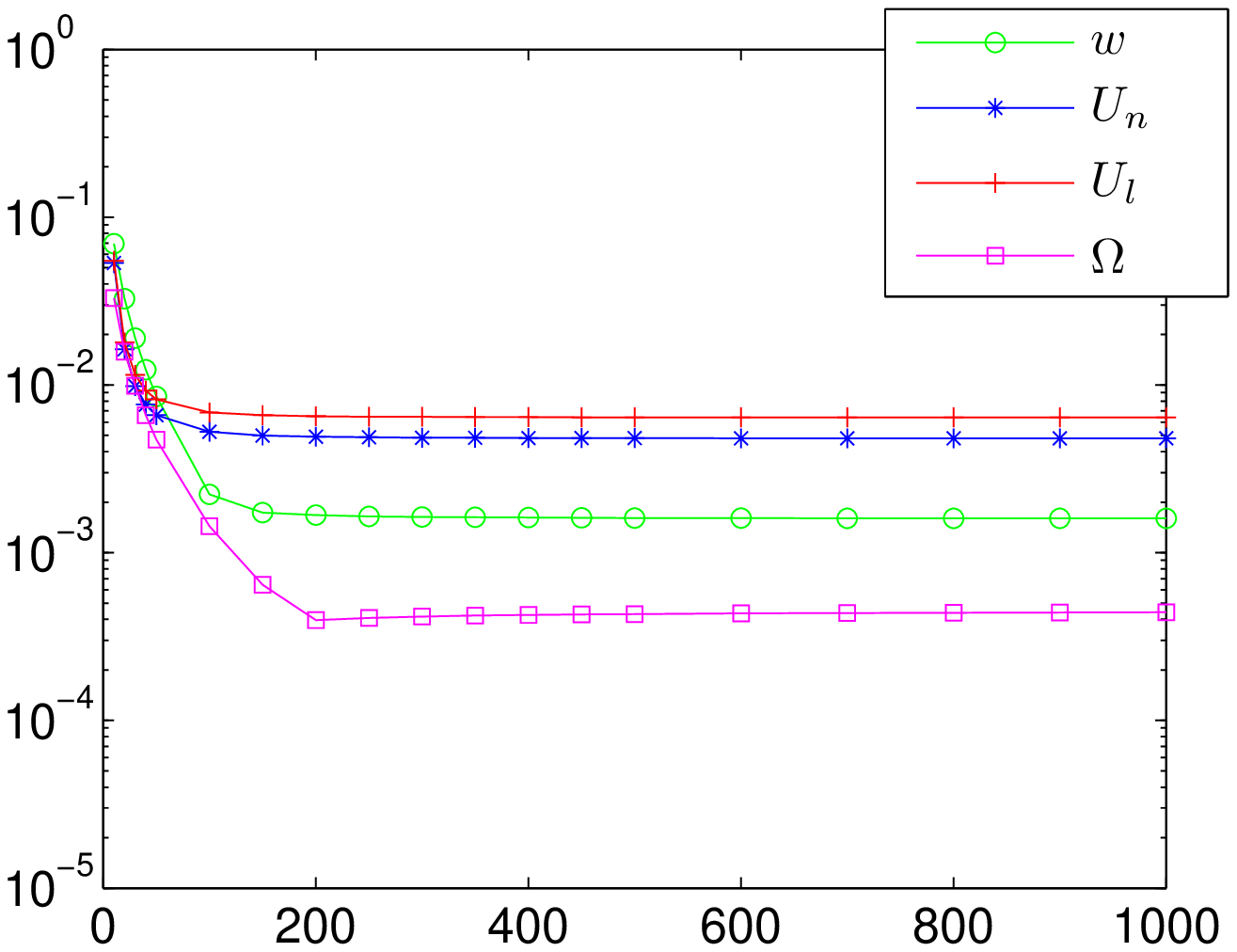}
        \end{subfigure}
        \begin{subfigure}[b]{0.32\textwidth}
                \centering
                \includegraphics[width=\textwidth]{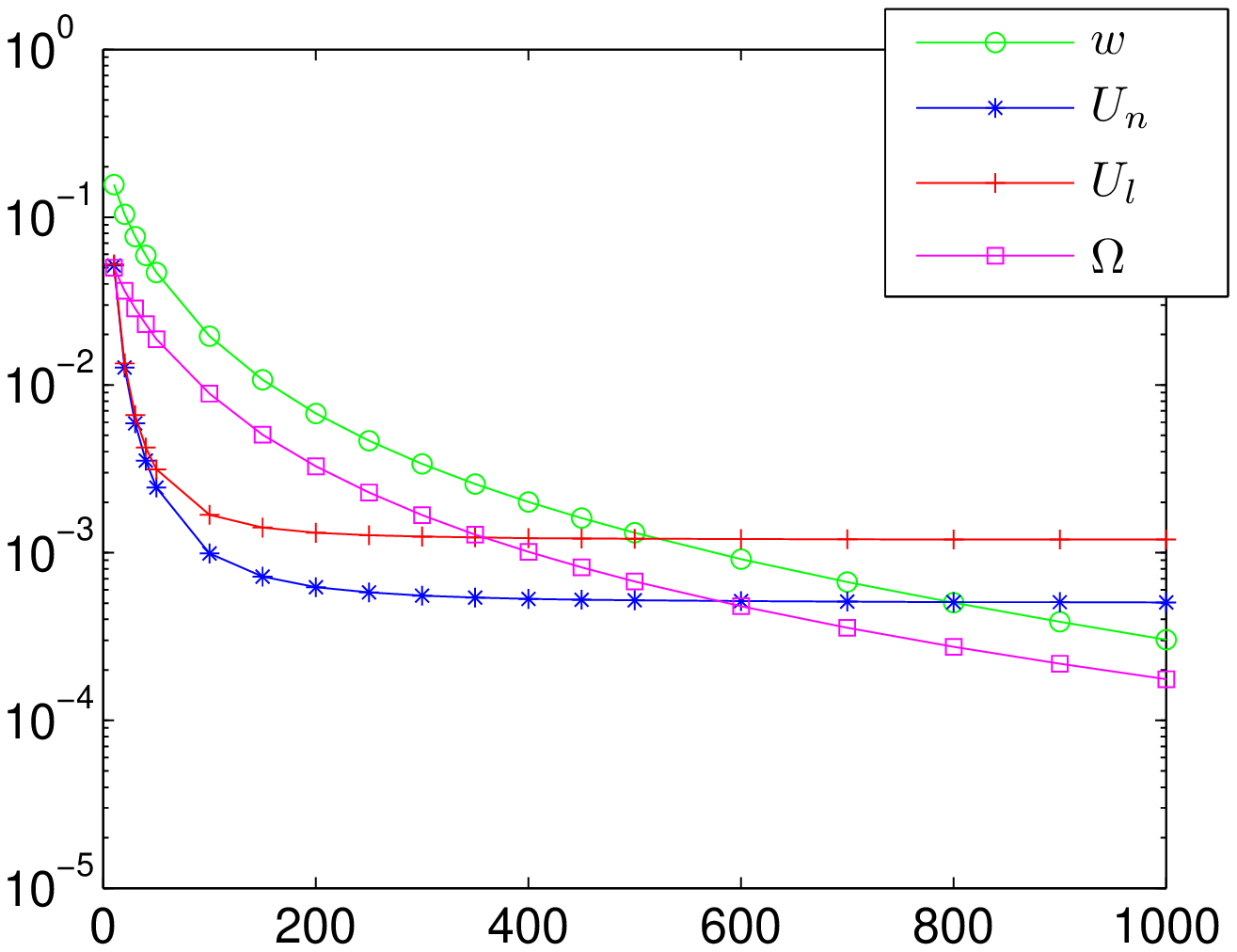}
        \end{subfigure}
        \begin{subfigure}[b]{0.32\textwidth}
                \centering
                \includegraphics[width=\textwidth]{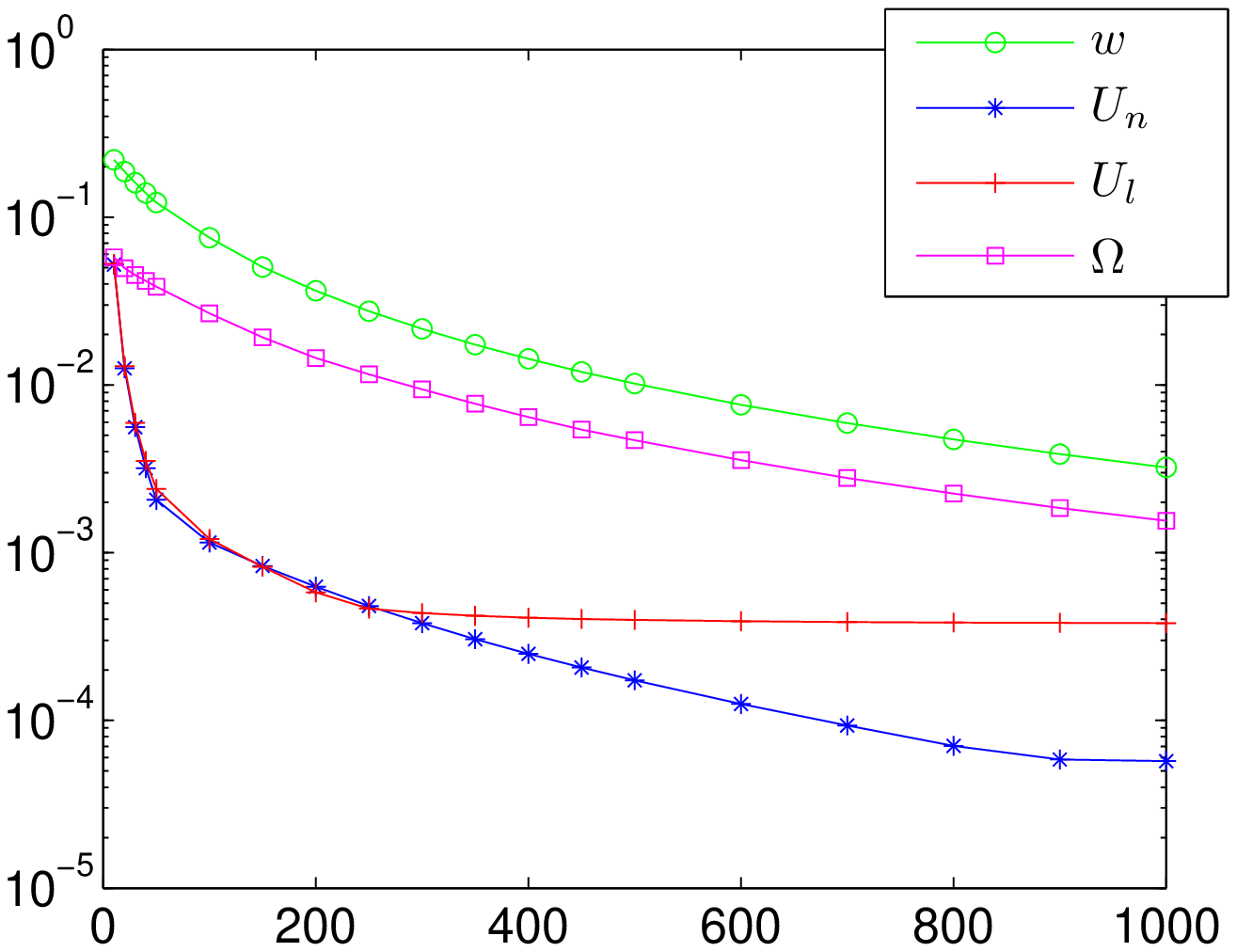}
        \end{subfigure}
        \begin{picture}(0,0)(490,-105)
        \put(85,-15){$\varepsilon=10^{-3}$}     \put(235,-15){$\varepsilon=10^{-4}$} \put(385,-15){$\varepsilon=10^{-5}$}
        \put(45,2){a)} \put(195,2){b)} \put(345,2){c)}
        \put(45,-33){$\delta f$}      \put(195,-33){$\delta f$} \put(345,-33){$\delta f$}
        \put(115,-110){$N$}    \put(265,-110){$N$} \put(413,-110){$N$}
        \end{picture}
\caption{Maximal relative errors of the solutions computed in different variables $\delta w$, $\delta U$ and $\delta \Omega$, for various number of the grid points $N$
in case of the nonuniform mesh $x^{(II)}$ ($\delta=2$). Different values of $\varepsilon$ have been considered. All computations were performed for the benchmark $q_l^{(1)}$ for
$Q_l/q_0=0.9$. Solutions $U_l$ and $U_n$ obtained by unitization of the linear and nonlinear regularized conditions (\ref{expan_3}) and (\ref{expan_4}), respectively}
\label{fig:err_N_plots}
\end{figure*}

The results of the stiffness investigation are collected in the
Table~\ref{table_A} and Fig.~\ref{fig:stiffness_b} The following
conclusions can be drawn from this data:

\begin{itemize}
\item [(i)] The nonuniform mesh reduces the stiffness approximately up to five times
regardless of the solution type (Table~\ref{table_A});

\item [(ii)] The most important parameter affecting the stiffness properties is the relation between
the injection flux rate and the leak-off to the formation $Q_l/q_0$ as can be clearly seen in
Table~\ref{table_A}. The value of this parameter is more important
than a particular distribution of the leak-off function (and its behaviour near the crack tip);

\item [(iii)] When comparing systems for various dependent variables,
the lowest condition ratio gives the system built for $U$ (one
order of magnitude lower than the others). The worst
stiffness performance takes place for the system corresponding to the
$\Omega$-variable. However, in some cases $\Omega$ may produce lower stiffness than $w$;

\item [(v)] A value of the regularization parameter $\varepsilon$
essentially affects  the stiffness of a dynamic system.
\end{itemize}

\subsection{Accuracy of the computations}

In this section, we analyze the accuracy of computations by the solvers based on different dynamic systems corresponding to the respective dependent variables.
To solve the systems, we use MATLAB ode15s subroutine utilizing a version of the Runge-Kutta method dedicated for stiff dynamic systems.

Before we compare different approaches in terms of their accuracy, let us recall two alternative ways
to define the regularized boundary condition at the end point
$x=1-\varepsilon$. The first one is based on the
$\varepsilon$-regularization technique, as it was defined in
\cite{Linkov_4} (see also \cite{MWL}).
The second approach to formulate the regularized condition is
to take into account the first two terms of asymptotics as described in
section \ref{Rbc} (compare equations: \eqref{expan_1} and
\eqref{BC_N}).
Finally, in the case of $U$, this condition may be implemented in the non-linear form  (\ref{expan_4}).
One can expect that the two term conditions would have a clear
advantage, at least in cases when the solution smoothness near the
crack tip  deteriorates due to the singularity of the leak-off
function.

The results of the computations presented in
Table~\ref{table_UUU} confirm such a prediction. We compare only conditions for $U$, as originally the $\varepsilon$-regularization technique was introduced for this variable.  Indeed, the relative errors
of the solutions $\delta U_l$ or $\delta U_n$ are at least one
order of magnitude lower than that in the case of $\delta U_*$,
corresponding to the formulation based on \eqref{L3}.\footnote{
Here and everywhere later, by $\delta f$ we understand the maximal value of the relative
error of the function $f$ over all discretized independent variables ($\delta f\equiv \|\delta f\|_\infty$).}
Surprisingly, for the variants of the non-singular leak-off function, the improvement
is even more pronounced (especially for a uniform mesh).

\begin{table}[h!]
\centering
\begin{tabular}{c c|c@{}|c@{}|c@{}|c@{}|c@{}|c@{}|}
\cline{3-8}
& & \multicolumn{6}{c|}{Comparison of conditions (\ref{L3}), (\ref{expan_3}), (\ref{expan_4})}\\ \cline{3-8}
\cline{3-8}
& & \multicolumn{3}{c|}{$q_l^{(1)} \quad Q_l/q_0=0.9$} & \multicolumn{3}{c|}{$q_l^{(3)} \quad Q_l/q_0=0.5$} \\ \cline{2-8}
& \multicolumn{1}{|c|}{$\varepsilon=$} & $10^{-2}$ & $10^{-4}$  & $10^{-6}$ &  $10^{-2}$ &  $10^{-4}$ &  $10^{-6}$ \\ \cline{1-8}
\multicolumn{1}{|c}{\multirow{2}{*}{$\delta U_*$}} & \multicolumn{1}{|c|}{$x^{(I)}$}
 &1.6e-1&1.4e-1&1.3e-1&6.1e-3&3.7e-3&3.7e-3
 \\ \cline{2-8} \multicolumn{1}{|c}{} & \multicolumn{1}{|c|}{$x^{(II)}$}
&1.4e-1&7.6e-2&6.3e-2&4.5e-3&9.3e-5&8.9e-5
 \\ \cline{1-8}
\multicolumn{1}{|c}{\multirow{2}{*}{$\delta U_l$}} & \multicolumn{1}{|c|}{$x^{(I)}$}
 &5.0e-2&1.4e-2&1.7e-2&1.2e-5&1.1e-5&1.1e-5
 \\ \cline{2-8} \multicolumn{1}{|c}{} & \multicolumn{1}{|c|}{$x^{(II)}$}
&5.0e-2&1.7e-3&2.0e-3&4.9e-5&8.2e-5&8.7e-5
 \\ \cline{1-8}
\multicolumn{1}{|c}{\multirow{2}{*}{$\delta U_n$}} & \multicolumn{1}{|c|}{$x^{(I)}$}
 &4.4e-2&1.2e-2&1.3e-2&2.2e-5&1.3e-5&1.3e-5
 \\ \cline{2-8} \multicolumn{1}{|c}{} & \multicolumn{1}{|c|}{$x^{(II)}$}
&4.4e-2&9.9e-4&1.8e-3&4.2e-5&8.2e-5&8.7e-5
 \\ \cline{1-8}
\end{tabular}
\caption{Comparison of the accuracy of the solution of dynamical system based on variable $U$.
The results depicted by $U_*$ refer to the regularized boundary condition  based on one asymptotic term, while those denoted by $U_l$ and $U_n$
correspond to two terms approximation (linear (\ref{expan_3}) and nonlinear (\ref{expan_4}), respectively).
Other problem parameters: $N=100$, $\delta=2$ for the mesh $x^{(II)}$.}
\label{table_UUU}
\end{table}


We also made the computations for three different benchmarks reported in \citet{MWL}. They correspond to the
leak-off function vanishing near the crack tip. It turned out that computational error corresponding to the modified form of the regularized conditions (based on two terms of asymptotics)
was always two orders of magnitude lower than that reported in the previous paper.

On the other hand, there is no difference observed between the solutions $\delta U_l$ but $\delta U_n$
at least for those two benchmarks and the choice of the parameters ($N=100$). However, as we will show  later,
for large numbers of nodal points, or more severe leak-off function relationship ($Q_l/q_0\sim1$),
the nonlinear formulation of the condition clearly manifests its advantage.

From now on only the regularized conditions based on two asymptotic terms \eqref{BC_N} will be utilized.
Additionally, for variable $U$, two different forms, linear
(\ref{expan_3}) and nonlinear (\ref{expan_4}), will be adopted.
For $w$ and $\Omega$ two  formulations, (\ref{L4}) and \eqref{L5}, which are equivalent
to the condition \eqref{L3} could not compete with
their more accurate analogue \eqref{BC_N} in terms of solution accuracy and will not be considered.

Graphs presented in Fig.~\ref{fig:err_N_plots} illustrate some peculiarities of the computational process.
Here, the maximal relative errors of the solutions (over the time and space) as functions of the number of mesh points $N$ ($\delta f=\delta f(N)$),
are presented for different variables. The considered benchmark assumes  $Q_l/q_0=0.9$ (see Appendix C).
Three different values of the regularization parameter $\varepsilon=10^{-3},10^{-4}$ and $10^{-5}$
were chosen.

\noindent
In Fig.~\ref{fig:err_N_plots} two basic tendencies can be observed. The first one is the
monotonous error decrease with growing $N$, up to some stabilization level. This level is different for different dependent variables and values of $\varepsilon$, and in some cases is reached for $N>1000$ (and thus cannot be identified in the figure). The second trend is discernible when comparing results for different values of $\varepsilon$. Namely, it turns out that for each dependent variable there exists an optimal  $\varepsilon$  minimizing the solution error. This value however depends on $N$.
It is not a surprise that the optimal stiffness properties and the maximal solution accuracy
are not achieved for the same values of the regularization parameter $\varepsilon$.
To increase computational accuracy one needs to decrease $\varepsilon$
and increase number of the nodal points $N$. However, both of these leads to increase of the condition ratio.

Note that the relative errors of respective dependent variables cannot be compared directly.
Indeed, even if the errors for $w$ and $U$ are interrelated via the evident relationship $\delta U=3\delta w$, their comparison with $\delta \Omega$ necessitates an additional postprocessing of the latter. This process, in turn, may introduce its own error. On the other hand, there exists a common component of the solutions, the crack length  $\delta L$,
which can be naturally used for such comparison.

\begin{table}[h!]
\centering
\begin{tabular}{c c|c@{}|c@{}|c@{}|c@{}|c@{}|c@{}|}
\cline{3-8}
& & \multicolumn{6}{c|}{Dynamic system built on the variable $w$ }\\
\cline{3-8}
& & \multicolumn{3}{c|}{$Q_l/q_0=0.9$} & \multicolumn{3}{c|}{$Q_l/q_0=0.5$} \\ \cline{3-8}
& & $q_l^{(1)}$ & $q_l^{(2)}$ & $q_l^{(3)}$ & $q_l^{(1)}$ & $q_l^{(2)}$ & $q_l^{(3)}$ \\ \cline{1-8}
\multicolumn{1}{|c}{\multirow{2}{*}{$\delta w$}} & \multicolumn{1}{|c|}{$x^{(I)}$}
  &8.5e-3&5.4e-3&5.6e-3&5.2e-3&4.0e-3&3.5e-3
 \\ \cline{2-8} \multicolumn{1}{|c}{} & \multicolumn{1}{|c|}{$x^{(II)}$}
&2.2e-3&2.6e-3&2.9e-3&1.8e-3&1.9e-3&2.0e-3
  \\ \cline{1-8} \multicolumn{1}{|c}{\multirow{2}{*}{$\Delta w$}} & \multicolumn{1}{|c|}{$x^{(I)}$}
&7.4e-3&9.1e-3&8.8e-3&4.3e-3&4.6e-3&4.7e-3
 \\ \cline{2-8} \multicolumn{1}{|c}{} & \multicolumn{1}{|c|}{$x^{(II)}$}
&2.8e-3&3.0e-3&3.2e-3&2.1e-3&2.1e-3&2.2e-3
  \\ \cline{1-8} \multicolumn{1}{|c}{\multirow{2}{*}{$\delta L$}} & \multicolumn{1}{|c|}{$x^{(I)}$}
&1.2e-3&1.3e-3&1.1e-3&5.2e-3&5.3e-3&5.2e-3
 \\ \cline{2-8} \multicolumn{1}{|c}{} & \multicolumn{1}{|c|}{$x^{(II)}$}
&4.0e-4&3.8e-4&3.1e-4&1.8e-3&1.8e-3&1.8e-3
 \\ \cline{1-8}
\end{tabular}
  \caption{Performance of the solver based on the dependent variable $w$ for number of nodal points $N=100$ and various benchmarks.
 Values of the regularized parameter are $\varepsilon=5\cdot10^{-3}$ and $\varepsilon=10^{-3}$ for the meshes for $x^{(I)}$  and $x^{(II)}$, respectively}
\label{table_w}
\end{table}

\begin{table}
\centering
\begin{tabular}{c c|c@{}|c@{}|c@{}|c@{}|c@{}|c@{}|}
\cline{3-8}
& & \multicolumn{6}{c|}{System built on $U_l$ and condition (\ref{expan_3})}\\ \cline{3-8}
\cline{3-8}
& & \multicolumn{3}{c|}{$Q_l/q_0=0.9$} & \multicolumn{3}{c|}{$Q_l/q_0=0.5$} \\ \cline{3-8}
& & $q_l^{(1)}$ & $q_l^{(2)}$ & $q_l^{(3)}$ & $q_l^{(1)}$ & $q_l^{(2)}$ & $q_l^{(3)}$ \\ \cline{1-8}
\multicolumn{1}{|c}{\multirow{2}{*}{$\delta U$}} & \multicolumn{1}{|c|}{$x^{(I)}$}
 &1.4e-2&1.0e-2&1.2e-4&2.0e-3&1.4e-3&1.1e-5
 \\ \cline{2-8} \multicolumn{1}{|c}{} & \multicolumn{1}{|c|}{$x^{(II)}$}
&1.2e-3&6.0e-4&2.5e-4&2.2e-4&1.7e-4&8.6e-5
  \\ \cline{1-8} \multicolumn{1}{|c}{\multirow{2}{*}{$\Delta U$}} & \multicolumn{1}{|c|}{$x^{(I)}$}
&7.1e-2&4.4e-2&2.0e-3&6.6e-3&4.5e-3&3.9e-4
 \\ \cline{2-8} \multicolumn{1}{|c}{} & \multicolumn{1}{|c|}{$x^{(II)}$}
&3.1e-2&2.9e-2&7.9e-3&3.5e-3&3.2e-3&7.9e-4
  \\ \cline{1-8} \multicolumn{1}{|c}{\multirow{2}{*}{$\delta L$}} & \multicolumn{1}{|c|}{$x^{(I)}$}
&4.4e-4&2.8e-4&4.4e-6&4.3e-4&2.9e-4&5.6e-6
 \\ \cline{2-8} \multicolumn{1}{|c}{} & \multicolumn{1}{|c|}{$x^{(II)}$}
&2.6e-4&2.4e-4&1.2e-4&9.5e-5&8.3e-5&4.3e-5
 \\ \cline{1-8}
\end{tabular}
 \caption{Numerical results for the system built on the dependent variable $U$ with the linear regularized condition (\ref{expan_3}) for $N=100$ and different $\varepsilon$ for the uniform and nonuniform meshes ($\varepsilon=10^{-4}$ and $\varepsilon=10^{-5}$, respectively).}
\label{table_U_l}
\end{table}

\begin{table}
\centering
\begin{tabular}{c c|c@{}|c@{}|c@{}|c@{}|c@{}|c@{}|}
\cline{3-8}
& & \multicolumn{6}{c|}{System built on $U_n$ and condition (\ref{expan_4})}\\ \cline{3-8}
\cline{3-8}
& & \multicolumn{3}{c|}{$Q_l/q_0=0.9$} & \multicolumn{3}{c|}{$Q_l/q_0=0.5$} \\ \cline{3-8}
& & $q_l^{(1)}$ & $q_l^{(2)}$ & $q_l^{(3)}$ & $q_l^{(1)}$ & $q_l^{(2)}$ & $q_l^{(3)}$ \\ \cline{1-8}
\multicolumn{1}{|c}{\multirow{2}{*}{$\delta U$}} & \multicolumn{1}{|c|}{$x^{(I)}$}
 &1.2e-2&9.2e-3&4.3e-5&1.9e-3&1.4e-3&1.3e-5
 \\ \cline{2-8} \multicolumn{1}{|c}{} & \multicolumn{1}{|c|}{$x^{(II)}$}
&1.2e-3&6.0e-4&2.5e-4&2.0e-4&1.7e-4&8.6e-5
  \\ \cline{1-8} \multicolumn{1}{|c}{\multirow{2}{*}{$\Delta U$}} & \multicolumn{1}{|c|}{$x^{(I)}$}
&6.4e-2&4.1e-2&1.7e-3&6.5e-3&4.4e-3&4.0e-4
 \\ \cline{2-8} \multicolumn{1}{|c}{} & \multicolumn{1}{|c|}{$x^{(II)}$}
&3.1e-2&2.9e-2&7.9e-3&3.5e-3&3.2e-3&7.9e-4
  \\ \cline{1-8} \multicolumn{1}{|c}{\multirow{2}{*}{$\delta L$}} & \multicolumn{1}{|c|}{$x^{(I)}$}
&4.1e-4&2.7e-4&1.5e-6&4.2e-4&2.9e-4&6.3e-6
 \\ \cline{2-8} \multicolumn{1}{|c}{} & \multicolumn{1}{|c|}{$x^{(II)}$}
&2.6e-4&2.4e-4&1.2e-4&9.5e-5&8.3e-5&4.3e-5
 \\ \cline{1-8}
\end{tabular}
 \caption{Results for the solver based on the dependent variable $U$ for nonlinear regularized condition for $N=100$ and various benchmarks.
 Values of the regularized parameter are $\varepsilon=10^{-4}$ and $\varepsilon=10^{-5}$ for the meshes for $x^{(I)}$ and $x^{(II)}$, respectively}
\label{table_U_n}
\end{table}

\begin{table}
\centering
\begin{tabular}{c c|c@{}|c@{}|c@{}|c@{}|c@{}|c@{}|}
\cline{3-8}
& & \multicolumn{6}{c|}{Dynamic system built on variable $\Omega$}\\ \cline{3-8}
\cline{3-8}
& & \multicolumn{3}{c|}{$Q_l/q_0=0.9$} & \multicolumn{3}{c|}{$Q_l/q_0=0.5$} \\ \cline{3-8}
& & $q_l^{(1)}$ & $q_l^{(2)}$ & $q_l^{(3)}$ & $q_l^{(1)}$ & $q_l^{(2)}$ & $q_l^{(3)}$ \\ \cline{1-8}
\multicolumn{1}{|c}{\multirow{2}{*}{$\delta \Omega$}} & \multicolumn{1}{|c|}{$x^{(I)}$}
 &2.5e-3&8.7e-4&3.0e-4&5.6e-4&3.3e-4&4.2e-4
 \\ \cline{2-8} \multicolumn{1}{|c}{} & \multicolumn{1}{|c|}{$x^{(II)}$}
&2.0e-3&7.3e-4&3.6e-4&4.4e-4&2.7e-4&3.1e-4
  \\ \cline{1-8} \multicolumn{1}{|c}{\multirow{2}{*}{$\Delta \Omega$}} & \multicolumn{1}{|c|}{$x^{(I)}$}
&9.4e-5&1.7e-5&5.9e-5&1.1e-4&1.3e-4&1.5e-4
 \\ \cline{2-8} \multicolumn{1}{|c}{} & \multicolumn{1}{|c|}{$x^{(II)}$}
&1.9e-4&2.1e-4&2.3e-4&1.3e-4&1.4e-4&1.4e-4
  \\ \cline{1-8} \multicolumn{1}{|c}{\multirow{2}{*}{$\delta L$}} & \multicolumn{1}{|c|}{$x^{(I)}$}
&2.7e-6&5.0e-7&1.7e-6&2.1e-5&2.5e-5&2.9e-5
 \\ \cline{2-8} \multicolumn{1}{|c}{} & \multicolumn{1}{|c|}{$x^{(II)}$}
&5.5e-6&6.2e-6&6.7e-6&2.6e-5&2.7e-5&2.8e-5
 \\ \cline{1-8}
\end{tabular}
\caption{Computation accuracy for the solver based on the dependent variable $\Omega$ for $N=100$ and $\varepsilon=10^{-2}$ for $x^{(I)}$ and $\varepsilon=5\cdot10^{-3}$ for $x^{(II)}$.}
\label{table_Omega}
\end{table}

Below we adopt the following strategy for performance test for different dynamic systems. First, we set the number of nodal points, $N$, to 100. Next, for each of the dependent variables we accept optimal (for $N=100$) values of the regularization parameter $\varepsilon$. It turned out that the optimal $\varepsilon$ differs slightly depending on the type of mesh chosen and the benchmark variant. The general trend for $\varepsilon$ can be identified for different  meshes (for $x^{(I)}$ it is always smaller than for $x^{(II)}$). However, the sensitivity to the benchmark type is low. The results of computations described by various accuracy measures are collected in Tables \ref{table_w} -- \ref{table_Omega} (the optimal values of $\varepsilon$ are specified in the captions). We present there: the relative error of solution $\delta f$, the
absolute error of solution $\Delta f$ and the relative error of the crack length $\delta L$. The following conclusions can be drawn from this data:
\begin{itemize}
\item[(i)] Similarly as in case of the stiffness properties, the solution accuracy is affected more by the value of  $Q_l/q_0$ than by the leak-off function behaviour near the crack tip.
There is a trend of simultaneous increase of the ratio $Q_l/q_0$ and the relative errors of dependent variables $\delta f$. However this tendency is not in
place (or may be even reversed) when analyzing $\delta L$.
\item [(ii)] In case of the dependent variable $U$, the way in which the regularized boundary condition is introduced (linear or non-linear) does not play an essential role for the benchmarks and ranges of the parameters under consideration in the Tables \ref{table_w} -- \ref{table_Omega}. However, there are exceptions to this rule. One of them can be seen in Fig.~\ref{fig:err_N_plots}c) for large values of $N$, where the non-linear condition proves its superiority. Another case will be presented in the end of this section.
\item[(iii)] When comparing the common accuracy parameter $\delta L$, the dynamic system for $\Omega$ gives the best results.
The dynamic system for $w$ is the worst performing scheme and comparable to the one for $U$ only in a few cases.
\item[(iv)]
Since $\Omega$ vanishes near the crack tip faster than other variables, one could expect the worst relative error in this case.
Surprisingly, even when contrasting the relative (incomparable) errors of the respective dependent variables with each other, the system for $\Omega$
seems to be the best choice.
The advantage of $\Omega$ over $w$ and $U$ is especially pronounced for the benchmarks variants with a higher ratio $Q_l/q_0$.
\item[(v)] Better solution accuracy is obtained for the non-uniform mesh in almost every case.
\end{itemize}

\begin{figure*}[h!]
\centering
        \begin{subfigure}[b]{0.32\textwidth}
                \centering
                \includegraphics[width=\textwidth]{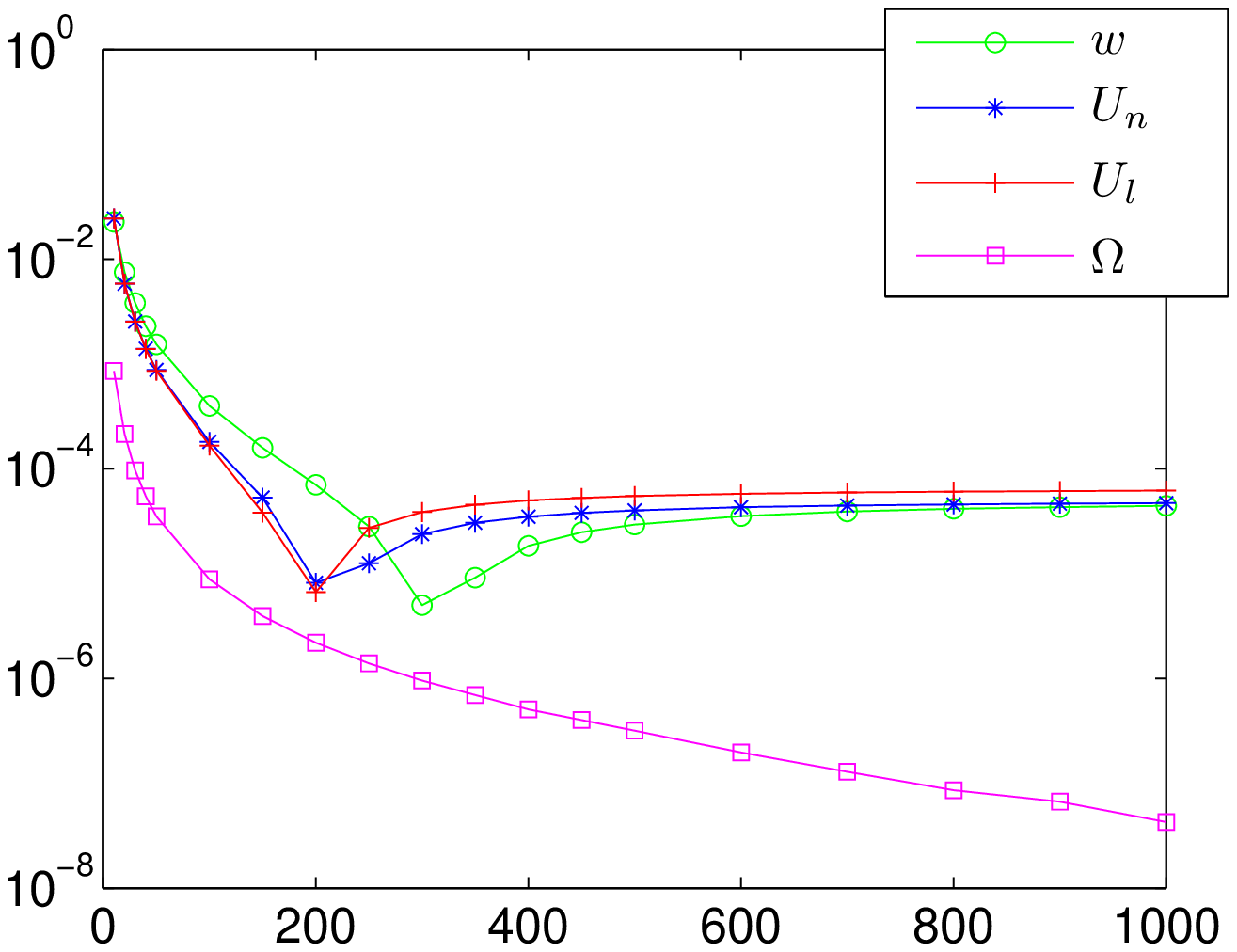}
        \end{subfigure}
        \begin{subfigure}[b]{0.32\textwidth}
                \centering
                \includegraphics[width=\textwidth]{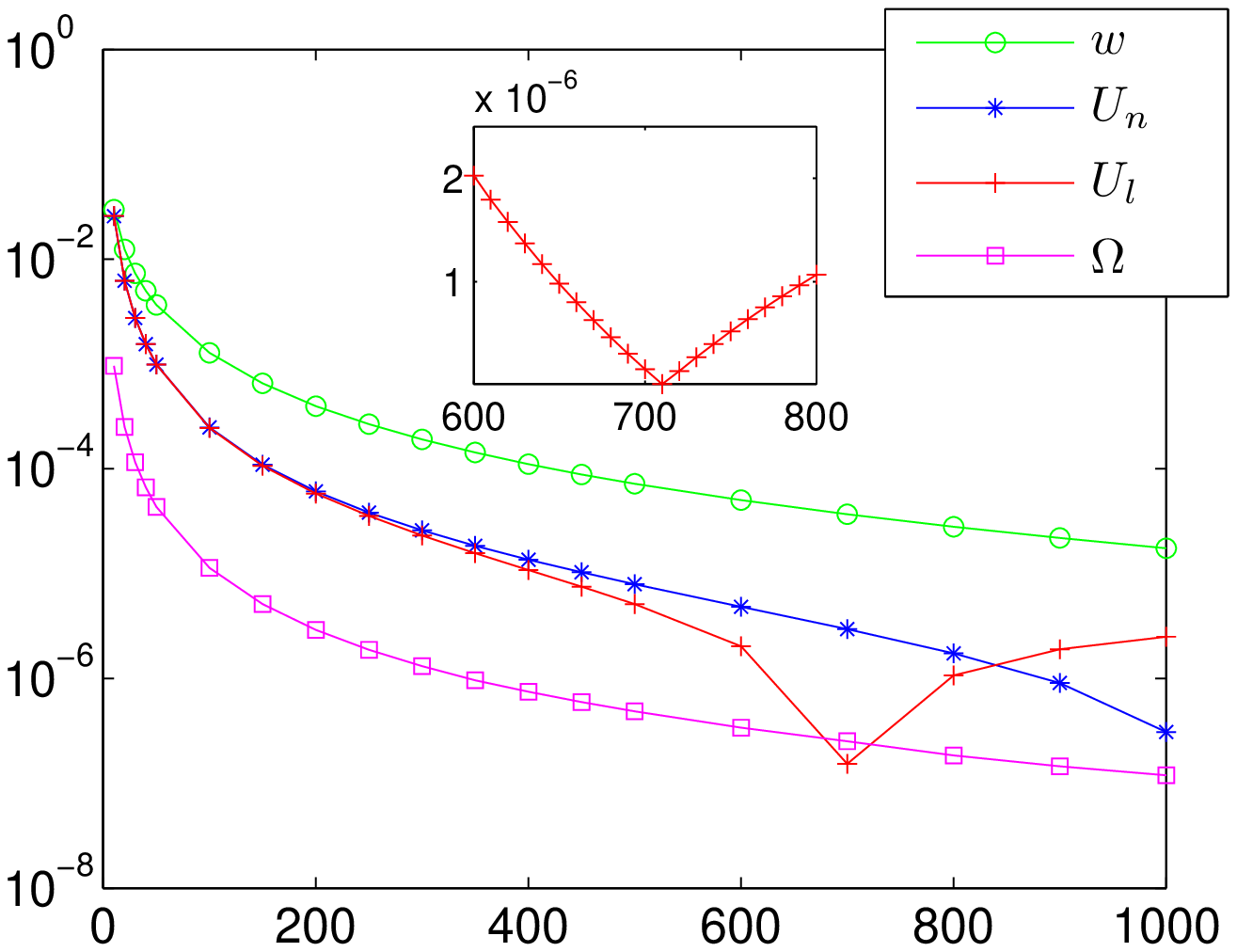}
        \end{subfigure}
        \begin{subfigure}[b]{0.32\textwidth}
                \centering
                \includegraphics[width=\textwidth]{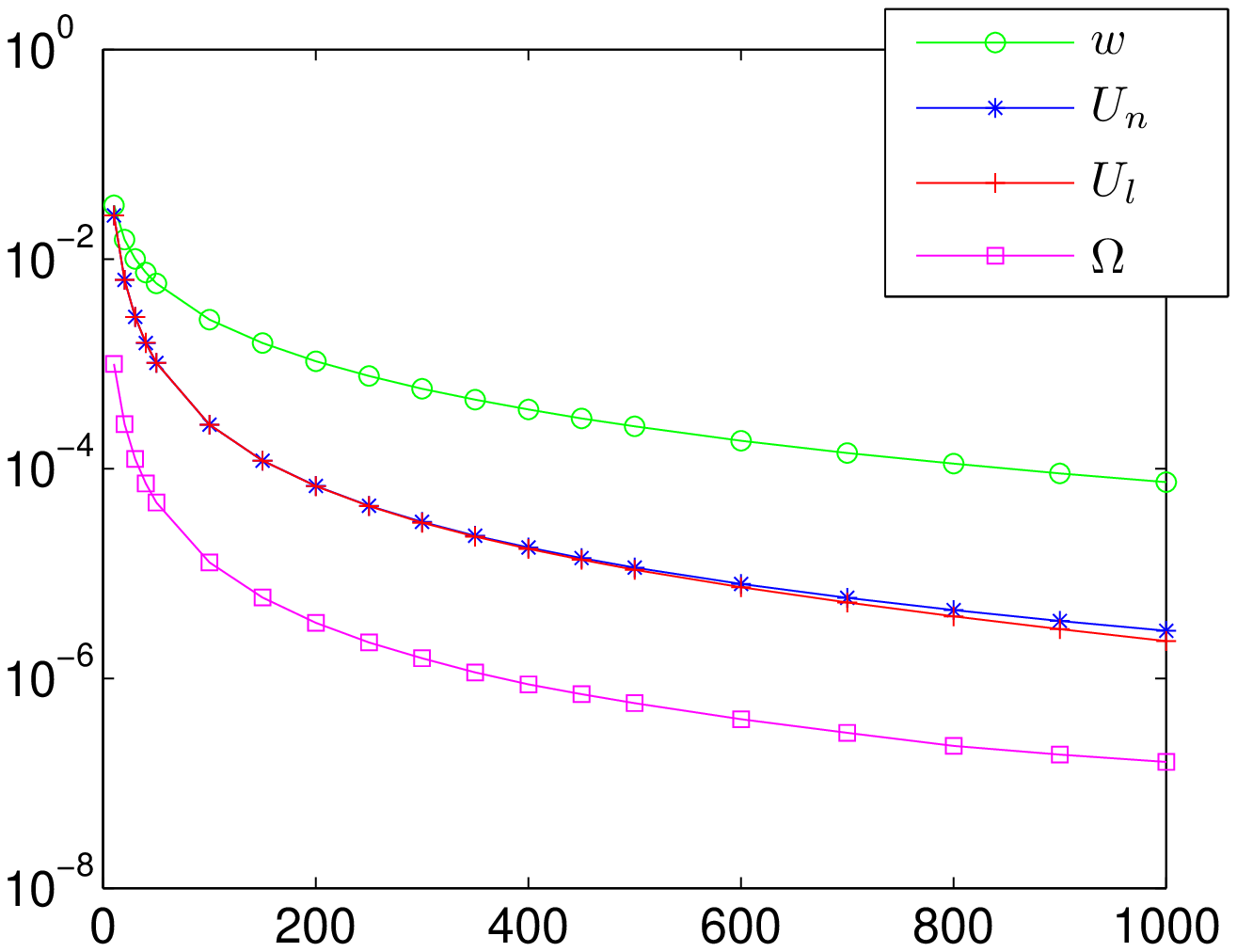}
        \end{subfigure}
         \begin{picture}(0,0)(490,-105)
        \put(80,-15){$\varepsilon=10^{-3}$}     \put(230,-85){$\varepsilon=10^{-4}$} \put(380,-15){$\varepsilon=10^{-5}$}
        \put(40,0){a)} \put(190,0){b)} \put(340,0){c)}
        \put(40,-40){$\delta L$}      \put(190,-40){$\delta L$} \put(340,-40){$\delta L$}
        \put(115,-110){$N$}    \put(265,-110){$N$} \put(413,-110){$N$}
        \end{picture}
\caption{Distribution of relative errors of the fracture length $\delta L$ computed by solvers based on different dependent variables ($w$, $U$ and $\Omega$).
When dependent variable $U$ is considered, two different regularised boundary conditions are in use: (\ref{expan_3}) for $U_l$ and (\ref{expan_4}) for $U_n$.
Other parameters are the same as in Fig~\ref{fig:err_N_plots}. Zoom picture within the Fig.~\ref{fig:err_N_plots_L} b) corresponds to the sharp minimum of $\delta L$ for the variable $U_l$.}
\label{fig:err_N_plots_L}
\end{figure*}

\begin{figure*}[h!]
        \centering
        \begin{subfigure}{0.32\textwidth}
                \centering
                \includegraphics[width=\textwidth]{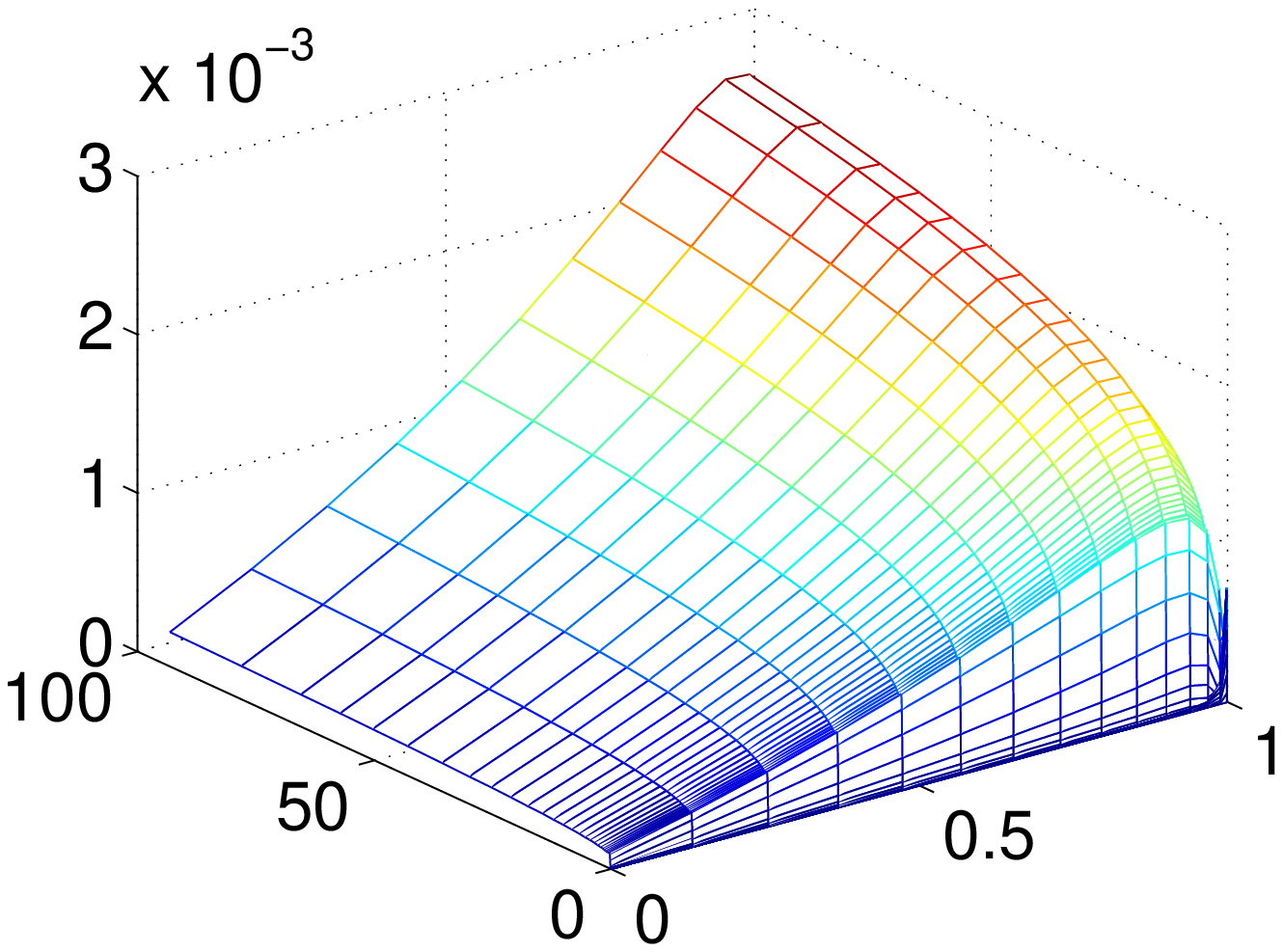}
 \end{subfigure}
 \begin{subfigure}{0.32\textwidth}
                \centering
               \includegraphics[width=\textwidth]{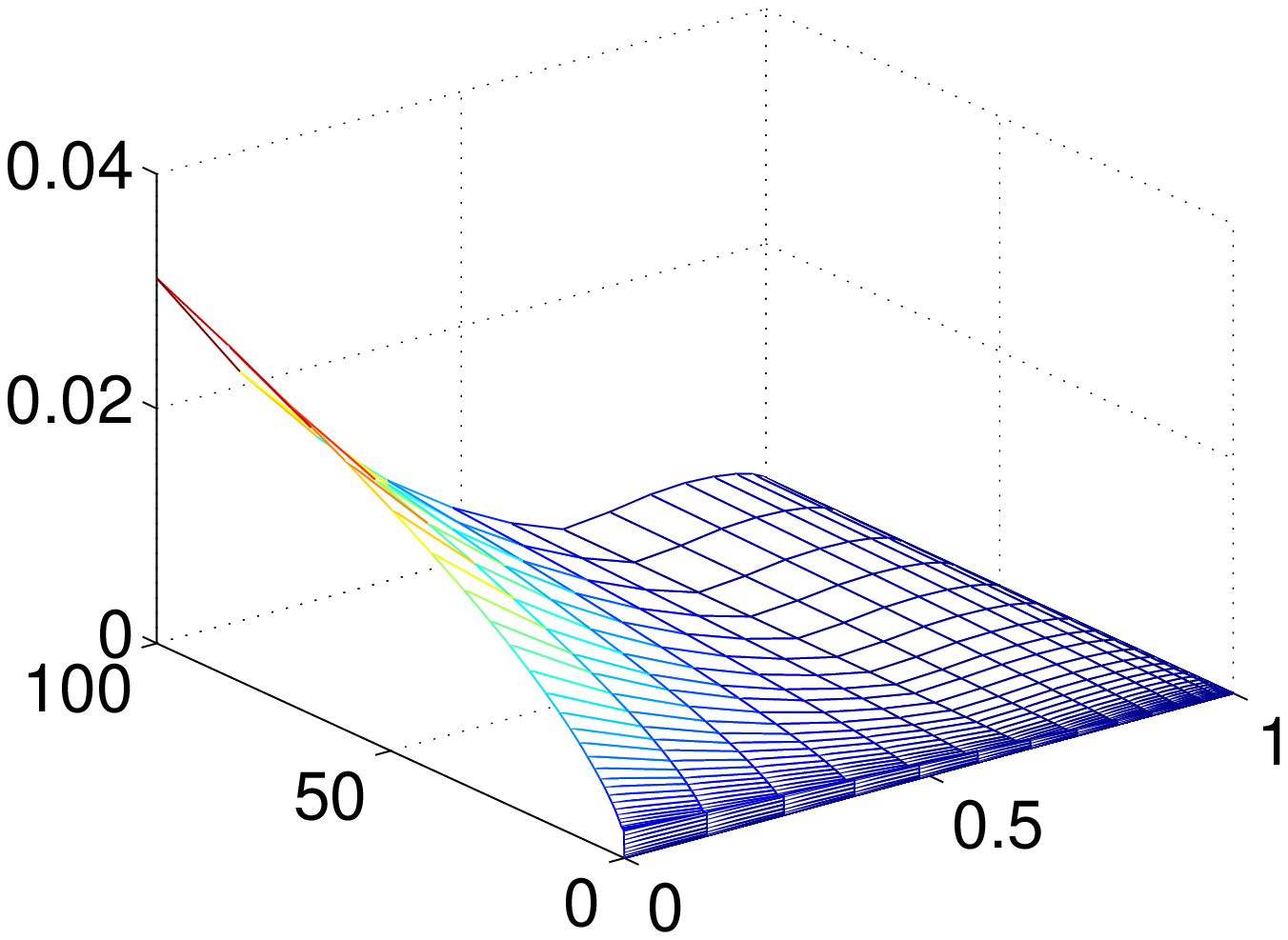}
\end{subfigure}
 \begin{subfigure}{0.32\textwidth}
                \centering
               \includegraphics[width=\textwidth]{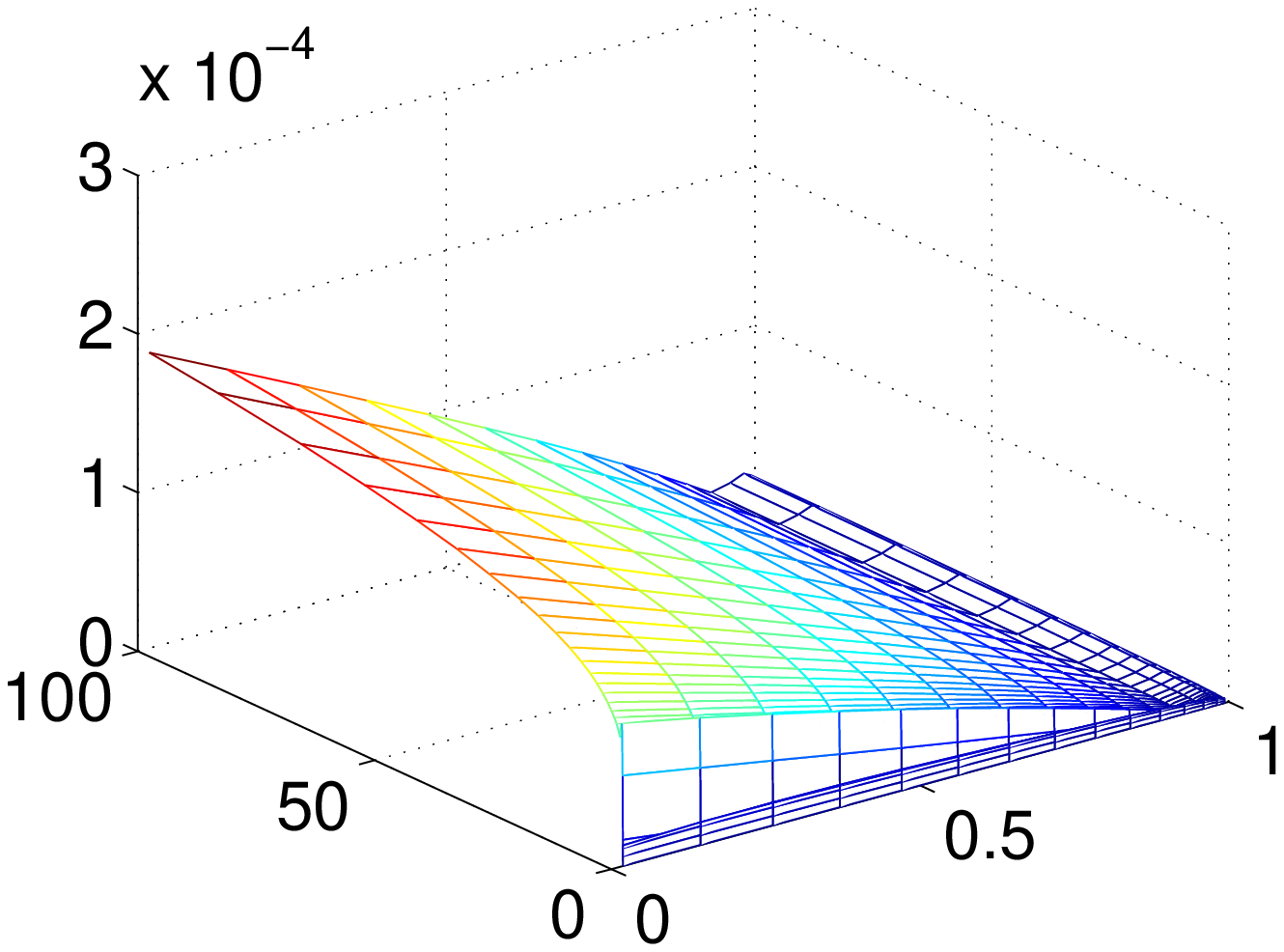}
\end{subfigure}
     \begin{picture}(0,0)(490,-50)
        \put(40,-5){a)} \put(190,-5){b)} \put(340,-5){c)}
        \put(40,-35){$\Delta w$}      \put(185,-35){$\Delta U_n$} \put(335,-35){$\Delta \Omega$}
        \put(155,-95){$x$}    \put(302,-95){$x$} \put(453,-95){$x$}
        \put(75,-95){$t$}     \put(225,-95){$t$} \put(375,-95){$t$}
        \end{picture}

  \caption{Absolute error for solutions $w$, $U_n$ and $\Omega$ computed for benchmark $q_l^{(1)}$ with ratio $Q_l/q_0=0.9$ and nonuniform mesh $x^{(II)}$ ($\delta=2$) with $N=100$ nodal points. Other parameters: $\varepsilon=10^{-3}$ for $w$, $\varepsilon=5\cdot10^{-3}$ for $\Omega$, and $\varepsilon=10^{-5}$ for $U_n$.}
    \label{distr_total_abs}
\end{figure*}

\begin{figure*}[h!]
        \centering
        \begin{subfigure}{0.32\textwidth}
                \centering
                \includegraphics[width=\textwidth]{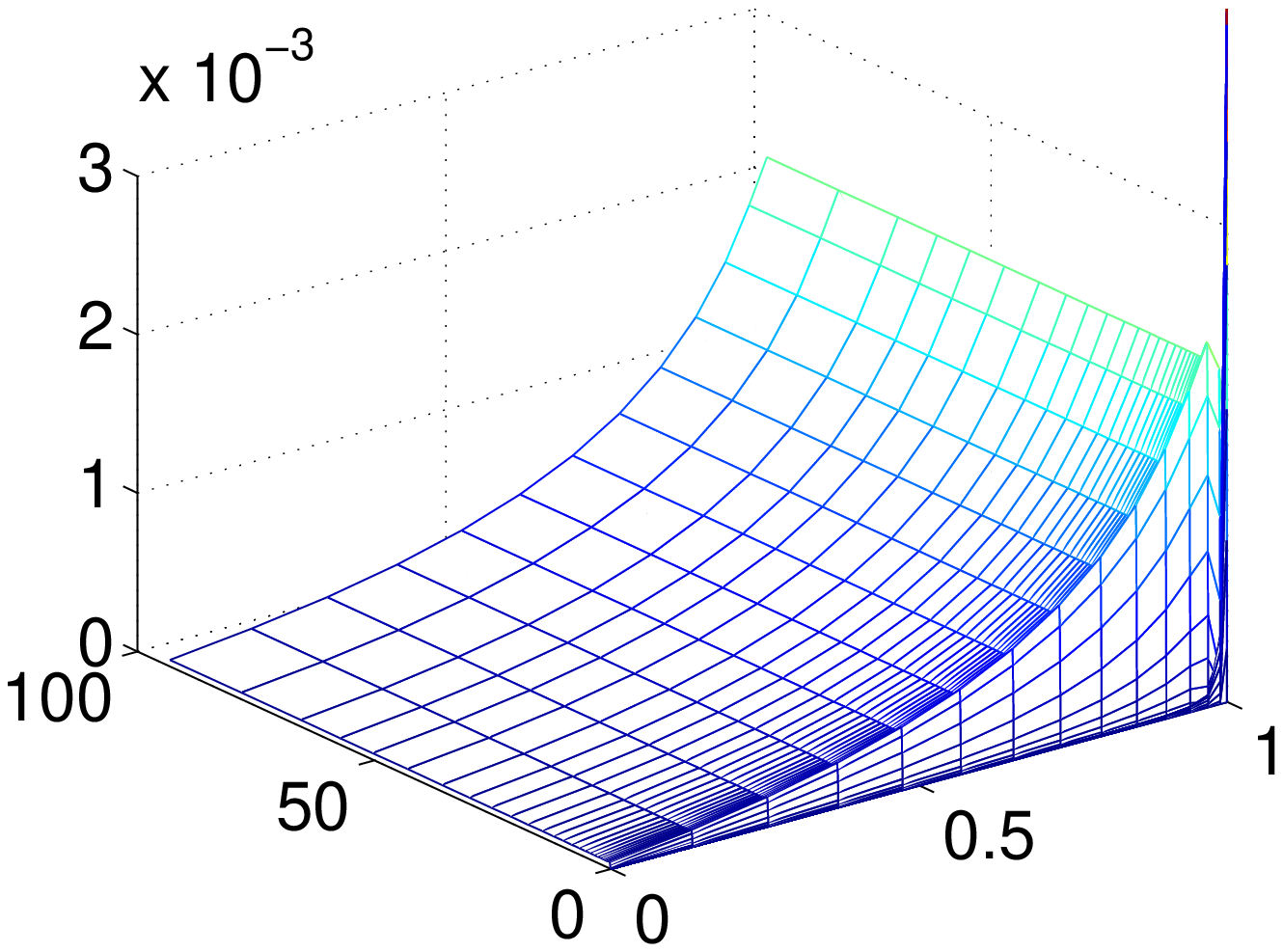}
 \end{subfigure}
 \begin{subfigure}{0.32\textwidth}
                \centering
               \includegraphics[width=\textwidth]{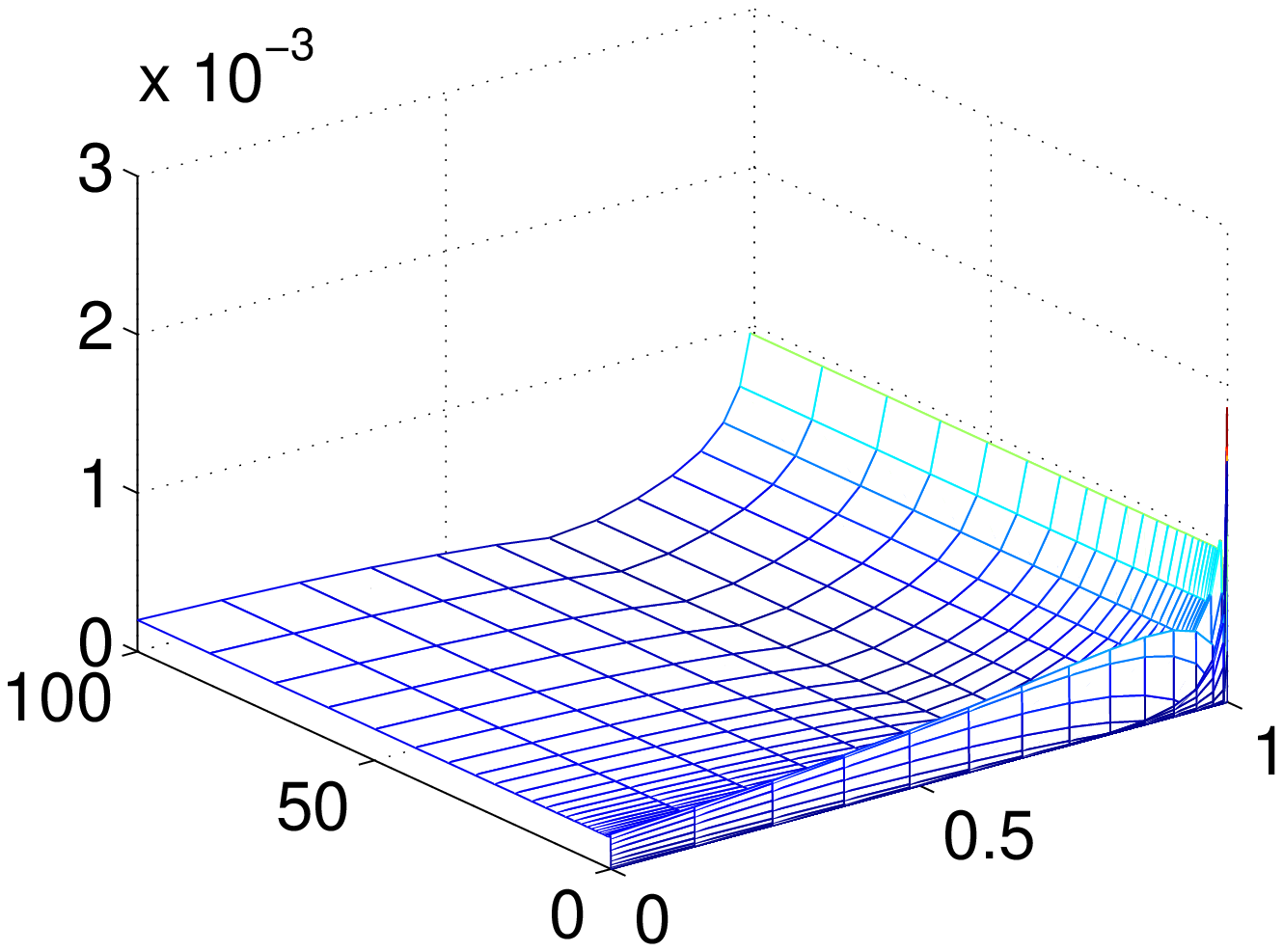}
                \end{subfigure}
 \begin{subfigure}{0.32\textwidth}
                \centering
               \includegraphics[width=\textwidth]{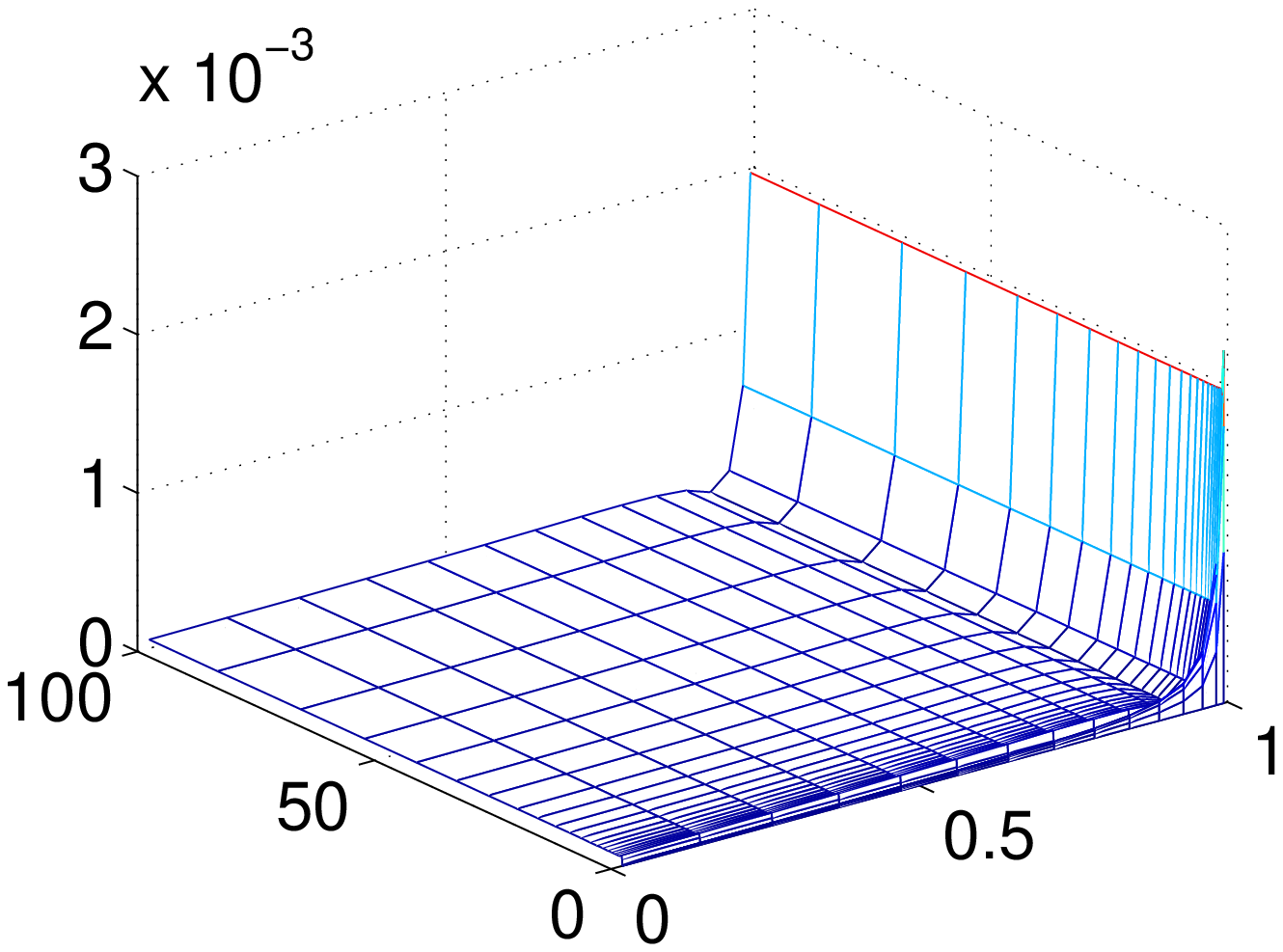}
                \end{subfigure}
                         \begin{picture}(0,0)(490,-50)
        \put(40,-5){a)} \put(190,-5){b)} \put(340,-5){c)}
                \put(40,-35){$\delta w$}      \put(185,-35){$\delta U_n$} \put(335,-35){$\delta \Omega$}
                \put(155,-95){$x$}    \put(302,-95){$x$} \put(453,-95){$x$}
        \put(75,-95){$t$}     \put(225,-95){$t$} \put(375,-95){$t$}
        \end{picture}

  \caption{
  Relative error of the solutions $w$, $U_n$ and $\Omega$ computed on the corresponding solvers for the same parameters as in  Fig.~\ref{distr_total_abs}.}
     \label{distr_total}
\end{figure*}

We have not observed any significant difference between the time step strategies chosen by the ode15s solver for the different dynamic systems.
The number of steps and the main trends were similar. For this reason we have not presented any details in the tables.

To visualize the results reported in the Tables \ref{table_w} -- \ref{table_Omega}, and to complement those presented in Fig.~\ref{fig:err_N_plots},
in Fig.~\ref{fig:err_N_plots_L} we show the relative errors of the crack length $\delta L$
computed by different dynamic systems (built on different variables).
The same benchmark and the values of all other problem parameters as previously discussed in Fig.~\ref{fig:err_N_plots} were considered.
If the trends for the different relative errors of the solution $\delta f$ and the crack length $\delta L$ in case of
 $\varepsilon=10^{-3}$ look similar, the results for the smaller value of the parameter are rather surprising. Indeed,
the relative errors $\delta U_l$ and $\delta U_n$ are smaller than $\delta w$ and $\delta \Omega$, while
the error $\delta L$ computed for $U$ no longer follows this trend.

This paradox needs an explanation. A trivial one could be that the maximal error for the solution ($\delta f$) is not situated near the crack tip but inside the computational domain.
To verify this hypothesis and to give a prospective reader a clear picture of the distribution of the solution error in time and space, we present, in  Fig.~\ref{distr_total_abs} and Fig.~\ref{distr_total},
the corresponding absolute and relative errors computed for the nonuniform mesh built on $N=100$ nodal points with the corresponding optimal regularized parameters
discussed after Fig.~\ref{fig:err_N_plots}. As it follows from Fig.~\ref{distr_total}, the maximum of the relative error is always achieved near the crack tip
($\max_{t}\delta f(t,1-\varepsilon))$). Hence, the initial guess has not been confirmed.
On the other hand, the values of the relative errors $\delta f$ and the respective $\delta L^{(f)}$
are directly interrelated. The following analysis identifies this relationship.

\begin{table*}[h1]
\centering
\begin{tabular}{c|c|c|c|c|c|c|c|c|}
\cline{2-9}
& \multicolumn{2}{c|}{$\varepsilon=10^{-2}$} & \multicolumn{2}{c|}{$\varepsilon=10^{-3}$}& \multicolumn{2}{c|}{$\varepsilon=10^{-4}$} & \multicolumn{2}{c|}{$\varepsilon=10^{-5}$}
\\ \cline{2-9}
& $x^{(I)}$ & $x^{(II)}$& $x^{(I)}$ & $x^{(II)}$& $x^{(I)}$ & $x^{(II)}$& $x^{(I)}$ & $x^{(II)}$
\\ \cline{1-9}
\multicolumn{1}{|c|}{$\delta w$}&1.6e-2&1.6e-2&3.4e-2&2.5e-3&6.8e-2&2.1e-2&--&8.2e-2
 \\ \cline{1-9}
\multicolumn{1}{|c|}{$\delta U_l$}&1.9e-1&1.9e-1&8.2e-2&8.1e-2&1.0e-1&4.3e-2&1.5e-1&2.3e-2
 \\ \cline{1-9}
\multicolumn{1}{|c|}{$\delta U_n$}&4.8e-2&4.8e-2&1.1e-2&6.3e-3&1.9e-2&3.2e-3&2.0e-2&9.1e-3
 \\ \cline{1-9}
\multicolumn{1}{|c|}{$\delta \Omega$}&2.6e-3&5.0e-3&3.0e-3&1.7e-3&3.9e-3&9.9e-3&4.0e-3&3.0e-2
 \\ \cline{1-9}
\multicolumn{1}{|c|}{$\delta L_w$}&2.0e-4&2.6e-4&2.0e-3&1.8e-4&3.4e-3&3.9e-4&--&6.2e-4
 \\ \cline{1-9}
\multicolumn{1}{|c|}{$\delta L_l$}&1.2e-3&1.1e-3&2.7e-4&1.3e-4&7.5e-4&2.8e-4&1.0e-3&3.2e-4
 \\ \cline{1-9}
\multicolumn{1}{|c|}{$\delta L_n$}&2.5e-4&1.2e-4&1.8e-4&2.6e-4&2.5e-4&3.0e-4&2.6e-4&3.2e-4
 \\ \cline{1-9}
\multicolumn{1}{|c|}{$\delta L_\Omega$}&8.1e-7&4.7e-7&1.1e-6&1.1e-6&1.2e-6&1.2e-6&1.2e-6&1.2e-6
 \\ \cline{1-9}
\end{tabular}
\caption{Accuracy parameters for the limiting (critical) variant of the benchmark solution ($Q_l/q_0=0.9857$, $\gamma_v=2.07$)
computed for different meshes composed of $N=100$ nodal points. The blank positions in the table correspond to the case when the solver ode15s could not complete the computations in a reasonable time.}
\label{table_very_bad_benchmark}
\end{table*}

Using (\ref{expan_1}), after some algebra, one has the estimate:
\begin{equation}
\label{estim}
\delta f\approx\delta e_1^{(f)}+(\delta e_2^{(f)}-\delta e_1^{(f)})\frac{e_2^{(f)}}{e_1^{(f)}}\varepsilon^{\alpha_2-\alpha_1}.
\end{equation}
For the benchmark $q_l^{(1)}$ and $Q_l/q_0=0.9$ (see Appendix C) which always provides the worst accuracy in our computations, one can conclude
\begin{equation}
\label{estim_w}
\delta w\approx\frac{1}{3}\delta U\approx\delta w_0+\frac{1}{10}(\delta w_1-\delta w_0)\sqrt[6]{\varepsilon},
\end{equation}
 \begin{equation}
\label{estim_Omega}
\delta \Omega\approx\delta w_0+\frac{8}{90}(\delta w_1-\delta w_0)\sqrt[6]{\varepsilon}.
\end{equation}
Finally, from (\ref{new_speed}) we can derive
 \begin{equation}
\label{estim_L}
\delta L\approx\frac{3}{2}\delta w_0.
\end{equation}
The last relationship has also been verified numerically by evaluating the values of the constant $w_0$ in the postprocessing procedure
using the computed solution ($w$, $U$ or $\Omega$) and the corresponding regularized boundary condition (compare (\ref{expan_1}) and (\ref{BC_N})).

It is clear from relations (\ref{estim_w}) -- \eqref{estim_Omega} that the relative errors of the respective dependent variables
also depend on the quality of approximation of the second term  in the  regularized boundary condition \eqref{expan_1}.
This explains the surprising relationship between $\delta L$ and the respective $\delta f$.

Interestingly, the results presented in Fig.~\ref{fig:err_N_plots_L} show that the value of $\varepsilon$ which provides the lowest relative error, $\delta f$, of the dependent variable $f$
does not necessarily give the best accuracy of the crack length $\delta L$. Moreover, the relation $\varepsilon=\varepsilon_L(N)$ is much more sensitive to the variation of $N$ than  $\varepsilon=\varepsilon_f(N)$.
 Indeed, one can observe sharp minima (see Fig.~\ref{fig:err_N_plots_L} a) and b)) while there is no such phenomenon in the respective graphs for $\delta f$ (see Fig.~\ref{fig:err_N_plots}).
To demonstrate that the peaks are not computational artifacts, we also include a small zoom of the corresponding area of the figure Fig.~\ref{fig:err_N_plots_L} b).

To complete the accuracy analysis, let us consider some critical
regime of  crack propagation. Namely, assume that the leak-off flux
almost entirely balances the volume of fluid injected into the
crack. Indeed, when taking the Carter type benchmark
\eqref{w_0_bench} $b_1=b_2=1$, one obtains the fluid balance ratio
$Q_l/q_0=0.9857$. This gives a very strong variation of the particle
velocity function along the crack length ($\gamma_v=2.07$ - see \eqref{gamma_v}).

In view of the previous conclusion on the  influence of the ratio $Q_l/q_0$
on the solution accuracy (which in fact confirms the observations from
\citet{MWL}), one can predict that the solution error will increase
appreciably in comparison with the figures shown in Tables
\ref{table_w} -- \ref{table_Omega}. In order to verify this
assertion the computations were made for respective dynamic systems
(the system for $U$ was analyzed again  for two forms of the
regularized boundary condition). Both types of meshes, the uniform
and the non-uniform, were utilized, each composed of 100  nodal
points ($N=100$). Four different values of the regularization
parameter $\varepsilon$ , ranging from $10^{-5}$ to $10^{-2}$, were
analyzed. The results of the computations described by respective
accuracy parameters are presented in Table
\ref{table_very_bad_benchmark}. Here, the symbols $\delta U_l$ and
$\delta U_n$ stand for the relative error of $U$ obtained for the
conditions \eqref{expan_3} and \eqref{expan_4}, respectively. The
subscript of $\delta L$ informs us which dynamic system the
corresponding result was obtained for.

The data in the table shows that the solution error increased at least one order of
magnitude, as compared to the values from Tables \ref{table_w} --
\ref{table_Omega}. The lowest deterioration of the solution accuracy
was obtained for $\Omega$, which proves the best overall performance
of the system built for this variable. Especially impressive is its
advantage when comparing the errors of crack length estimation. In
all considered cases $\delta L_{\Omega}$ is at least two orders of
magnitude lower than $\delta L$ for other dependent variables.

In this critical variant of benchmark solution, the non-linear
regularized boundary condition \eqref{expan_4} for $U$ gives, in
most cases, much better performance than its linear counterpart
\eqref{expan_3} (compare with the discussion after the Tables
\ref{table_UUU} and \ref{table_w} -- \ref{table_Omega}). Finally,
the non-uniform mesh seems to be a better choice from the point of
view of accuracy.

\begin{figure}[h!]
\centering
\begin{minipage}{.5\textwidth}
  \centering
  \includegraphics[width=.9\linewidth]{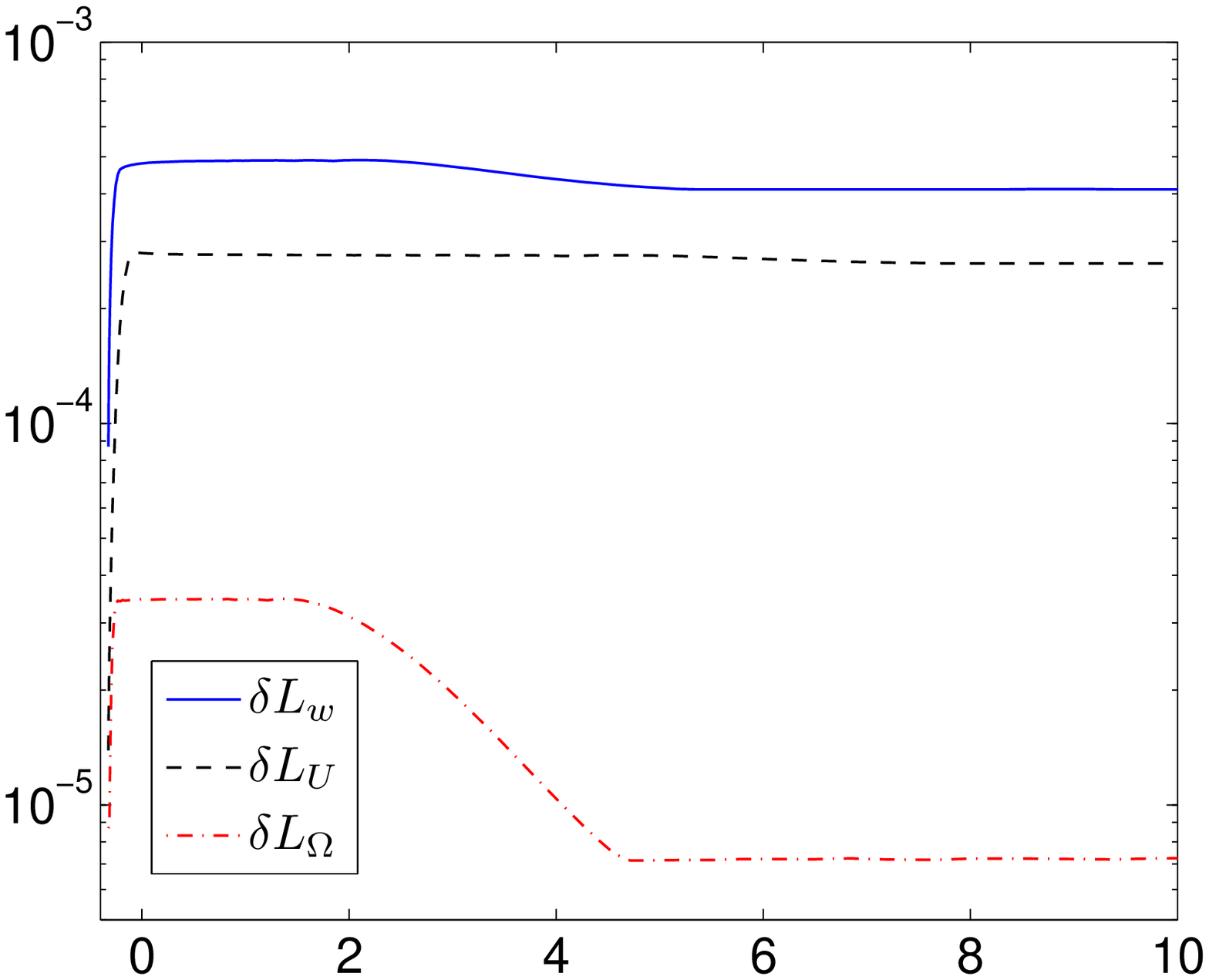}
 \put(-100,-2){$\gamma$}
  \captionof{figure}{The relative errors of the crack length\\ for different dependent variables as functions of $\gamma$.}
  \label{L_gamma}
\end{minipage}%
\begin{minipage}{.5\textwidth}
  \centering
  \includegraphics[width=.9\linewidth]{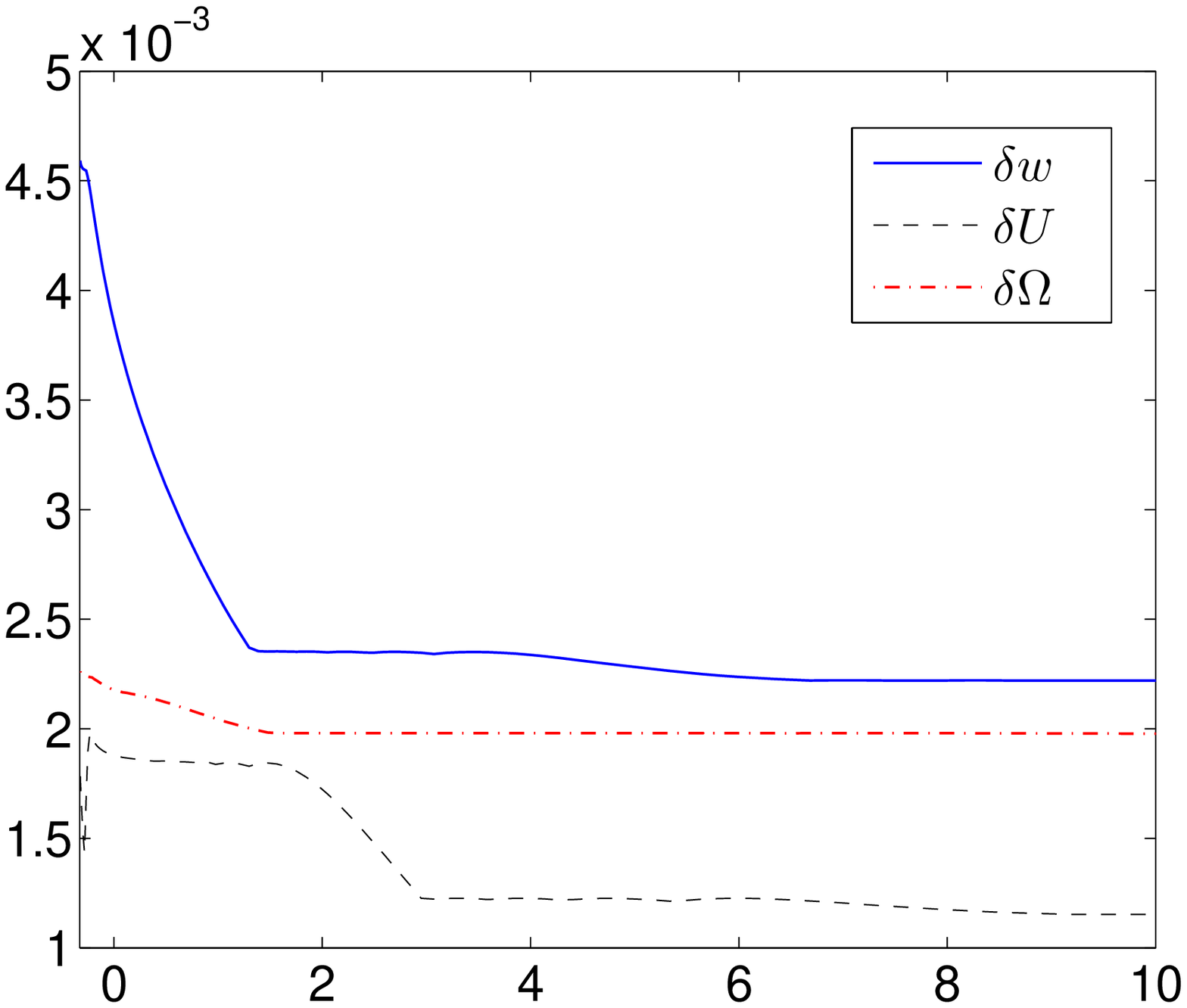}
 \put(-100,-2){$\gamma$}
  \captionof{figure}{The maximal relative errors of respective dependent variables as functions of $\gamma$.}
  \label{w_gamma}
\end{minipage}
\end{figure}

In the last test in this subsection we  discuss the sensitivity of
respective algorithms to the variation of the crack propagation
regime. To this end, consider again the benchmark solution
\eqref{w_bench} for the critical value of the ratio
$Q_l/q_0=0.9857$ ($\gamma_v=2.07$). Now, we analyze a range of parameters $\gamma
>-1/3$, motivated by the physical sense of the solution. By changing this value, one
simulates different modes of crack propagation (see Appendix
\ref{app:C}). The non-uniform mesh
composed of 100 nodes nodes was utilized. For each of the
dependent variables an optimal value of the regularization
parameter, $\varepsilon$, was taken: $\varepsilon=10^{-3}$ for $w$,
$\varepsilon=10^{-5}$ for $U$, and $\varepsilon=5\cdot10^{-3}$ for
$\Omega$. The results of the computations illustrated by the relative
errors of the crack length and the maximal relative errors of
corresponding dependent variables are shown in
Fig.~\ref{L_gamma} -- Fig.~\ref{w_gamma}, respectively.

%



As can be seen in Fig.~\ref{L_gamma}, for all dependent variables
the crack length error rapidly decreases for $\gamma \to -1/3$.
Indeed, this is the case when $L(t) \sim L_0$. For $w$ and
$U$ solvers, $\delta L$ remains very stable over most of the analyzed
interval. The solver based on $\Omega$ exhibits quite different
behaviour. For $\gamma$ greater than approximately 1.4, the error
decreases to achieve the level of its ultimate accuracy, the same as
for $\gamma \to -1/3$. Depending on the crack propagation regime,
this solver can produce up to two orders of magnitude better accuracy
of $L(t)$ than others.

Fig.~\ref{w_gamma} shows, that respective dependent variables
themselves are much less  sensitive to the changes of $\gamma$ that
the crack length. In the considered interval each solver provides a
relatively stable level of accuracy (with\-in the same order of
magnitude).

This test proves that using the solver based on $\Omega$ is
especially  beneficial when dealing with the problems of fast
propagating fractures (large values of $\gamma$).

\subsection{Comparison with known numerical results.}

Although the benchmark solutions utilized in the previous
subsections incorporate among others the leak-off term with a square root
singularity, there is no analytical solution for the Carter leak-off. In
\citet{Kovalyshen}, one can find the numerical results for such a case. This data may be utilized as a reference
solution. Unfortunately, the authors provide only some rough
estimation of the solution error. Surprisingly, they do not even
verify their numerical scheme against the early time asymptotic
model (considered as an analytical benchmark) to establish
quantitatively the accuracy of computations for the zero leak-off
case.

The numerical method used in \citet{Kovalyshen} is based on an
implicit finite volume algorithm. The data collected  in their Table 1 (p.332)
describes the normalized values of the crack length, the crack propagation speed and
the crack opening at $x=0$, at a number of times steps in the
interval $t\in [10^{-5},5\cdot 10^2]$. There is no precise
information on the utilized number of control volumes and the time
stepping strategies (the mentioned number of 10 control volumes
refers to the presented graphs, but it is not clear if the data from
table was obtained for the same parameters).

In the following we compare our numerical solution (see  Table \ref{table_Carter}) with that
by \citet{Kovalyshen}. Note that, due to different normalizations, our normalized  crack length, $L$, is two times greater than respective value in their paper. Our data was
obtained by the solver based on $U$ variable for $N=1000$ nodal points. Although
from the previous analysis it emerges that the system for $\Omega$
can provide better accuracy of $L$, we do not use it here to avoid an
additional postprocessing (numerical differentiation) when computing $w(0,t)$. On the other hand, in the light of previous
investigations, the system for $U$ for $N=1000$ can give the
accuracy of $L$ up to $10^{-6}$.

\begin{figure}[h!]
\centering
\begin{minipage}{.5\textwidth}
  \centering
  \includegraphics[width=.9\linewidth]{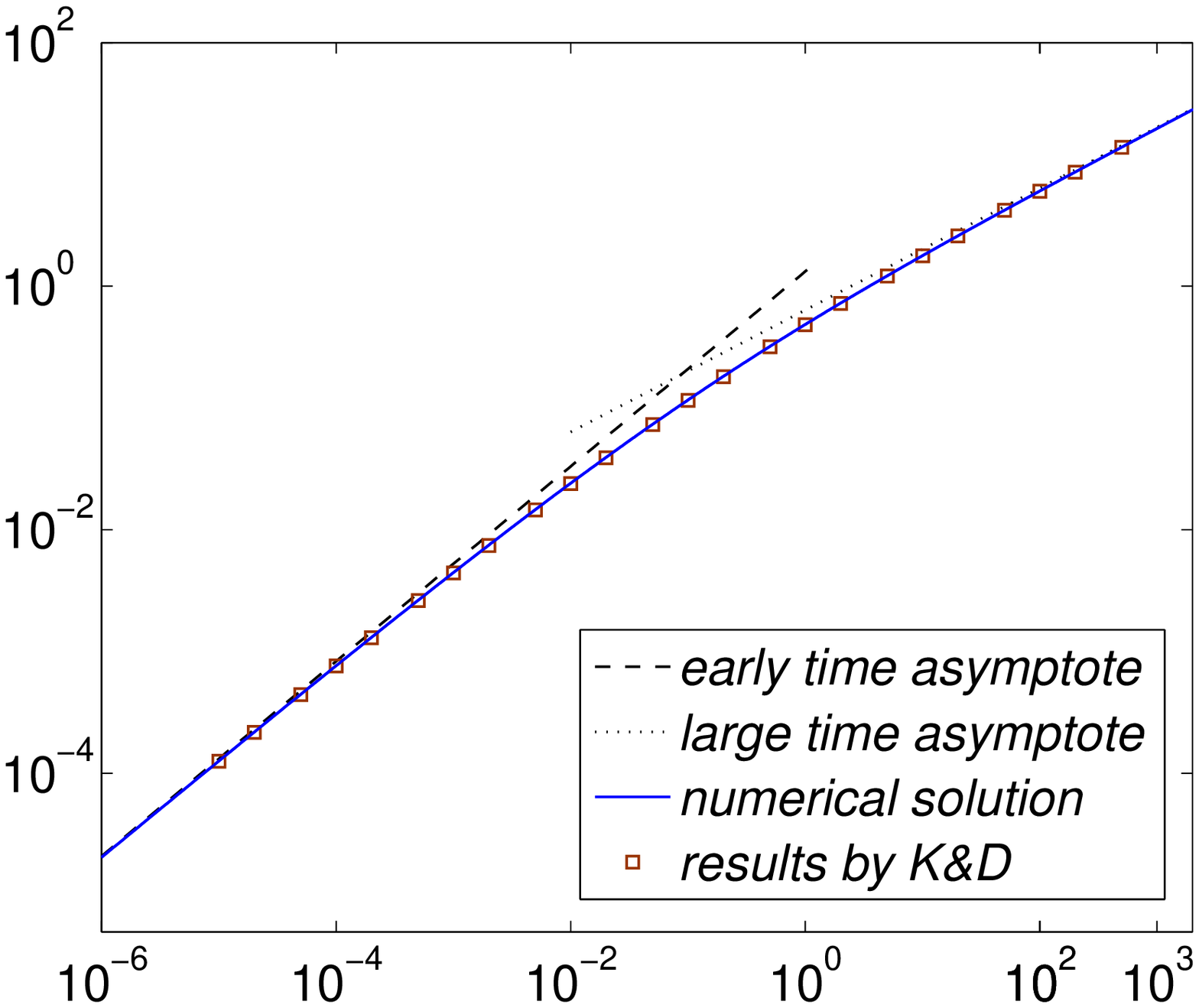}
     \put(-100,-2){$t$}
    \put(-210,80){$L(t)$}
  \captionof{figure}{The crack length evolution in time.}
  \label{K_D_Lenght}
\end{minipage}%
\begin{minipage}{.5\textwidth}
  \centering
  \includegraphics[width=.9\linewidth]{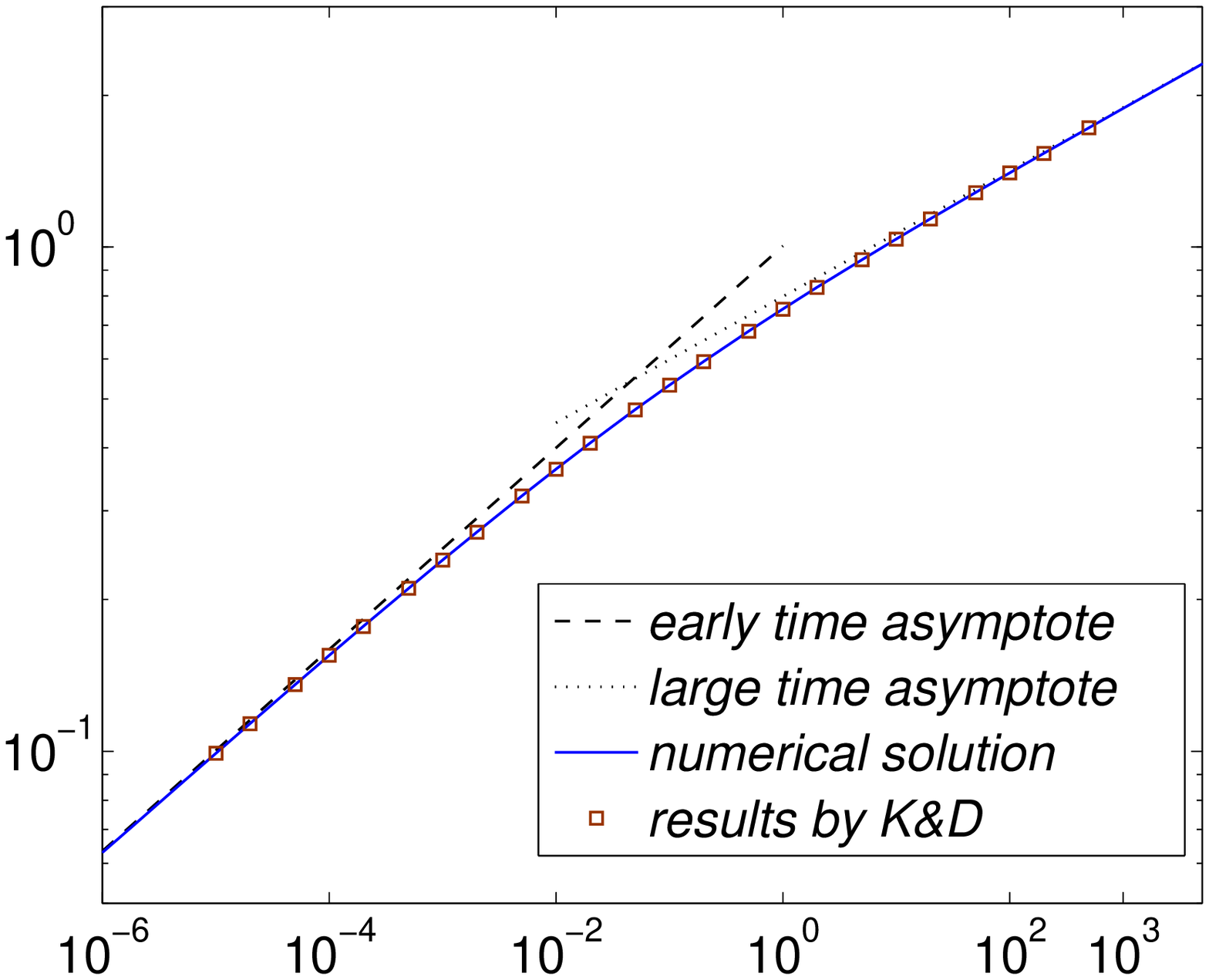}
    \put(-100,-2){$t$}
    \put(-210,80){$w(t,0)$}
  \captionof{figure}{The evolution of crack opening at zero point, $w(t,0)$.}
  \label{K_D_w_0}
\end{minipage}
\end{figure}

First, we present the graphs for evolution of the crack length, $L(t)$, -
Fig.~\ref{K_D_Lenght}, and the crack aperture at zero point, $w(t,0)$ - Fig.~\ref{K_D_w_0}. They depict the data for early time and
large time asymptotic models (respective formulae can be found also in
\citet{Kovalyshen}), and the numerical results
for a transient regime connecting these asymptotes. The solution by
\citet{Kovalyshen} is indicated by markers. A figure, equivalent to Fig.~\ref{K_D_Lenght}, has been also published in \citet{Nordgren}, however there is no data
available for comparison.

%
%

We analyze the time interval $t \in[10^{-8},10^8]$ where the initial conditions
correspond to the early time asymptote for $t=10^{-8}$. The same
initial time was taken by \citet{Kovalyshen}, but the
authors presented their data starting from $t=10^{-5}$. In order to
increase the legibility of the graphs, we have truncated the time
axis to the range $t \in[10^{-6},5\cdot10^3]$, while the complete data
is presented in Table \ref{table_Carter}.

\begin{figure}[h!]
\center
    \includegraphics [scale=0.35]{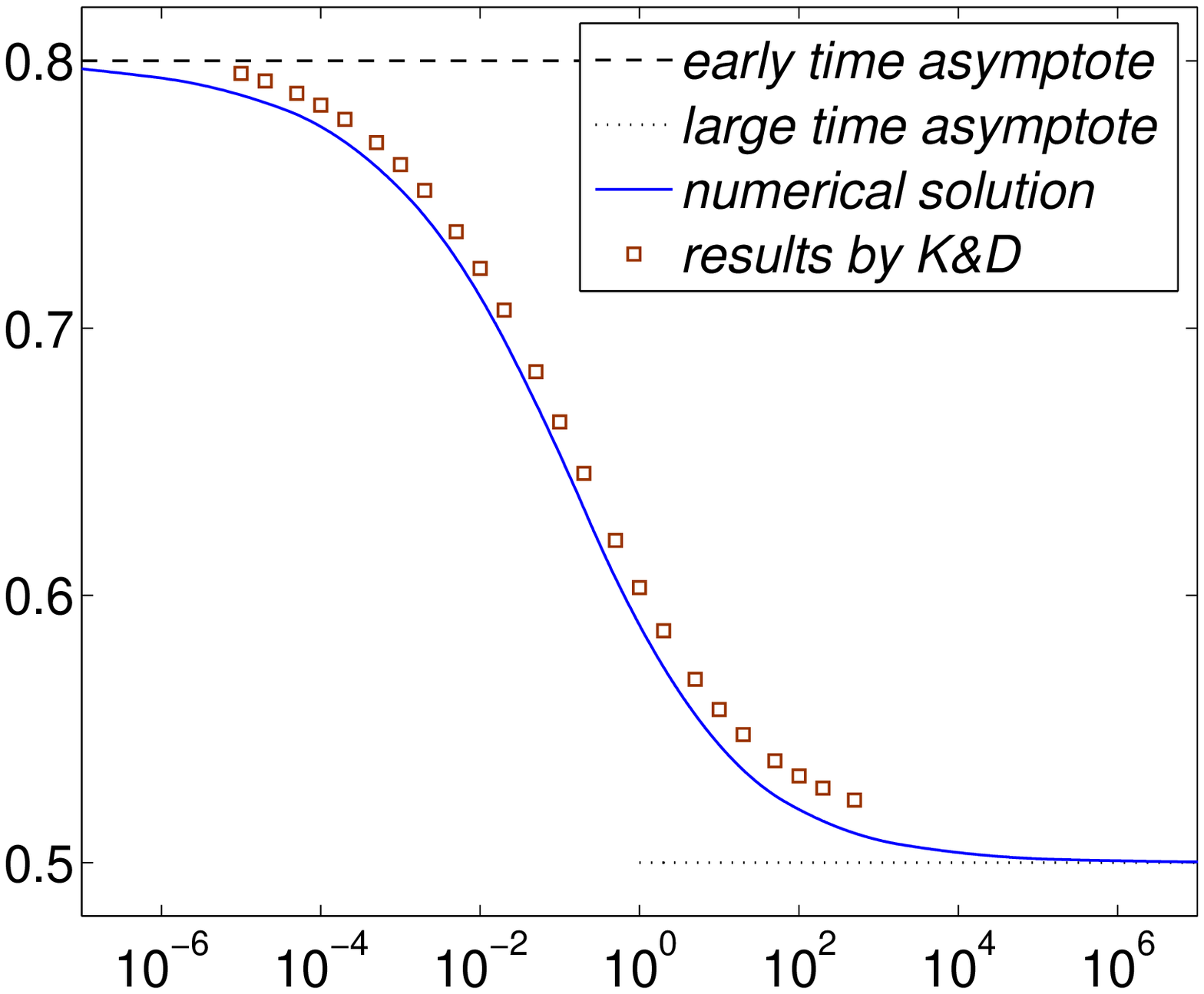}
    \put(-100,-2){$t$}
    \put(-205,80){$u(t)$}

    \caption{The evolution of the normalized crack propagation speed.}
\label{K_D_u}
\end{figure}

In Fig.~\ref{K_D_u} we show the normalized crack propagation speed, defined in a manner introduced by \citet{Kovalyshen}.

%

As can be seen in the Fig.~\ref{K_D_Lenght} -- Fig.~\ref{K_D_w_0}, in the presented scale, our solution is
undistinguishable from that by \citet{Kovalyshen} in terms of $L(t)$ and $w(t,0)$.
However, the normalized crack propagation speeds differ appreciably from each other. It shows that our solution
fits the asymptotes very well, which suggests its good quality. We cannot
examine, how the solution of \citet{Kovalyshen} approaches the asymptotic values due to the shortage of data for time intervals $t<10^{-5}$ and $t>5\cdot 10^3$ in their table.

In the analyzed case, the value of the parameter $Q_l(t)/q_0(t)$ changes continuously with time from zero to unit.
From the data presented in \citet{MWL} and in this paper, one can conclude that for $N=100$ nodal points, the relative error of the crack length
changes from $10^{-6}$ to $10^{-4}$ with the increase of the parameter $Q_l/q_0$.
On the other hand, analyzing the data from Fig.~\ref{fig:err_N_plots} -- Fig.~\ref{fig:err_N_plots_L} ($Q_l/q_0=0.9$), one can expect the achievable level
of accuracy of the order $10^{-7}$ for $N=1000$. This suggests that, in our computations, the relative error of $L$ varies between $10^{-4}$ and $10^{-6}$.

\begin{figure}[h!]
\centering
\begin{minipage}{.5\textwidth}
  \centering
  \includegraphics[width=.9\linewidth]{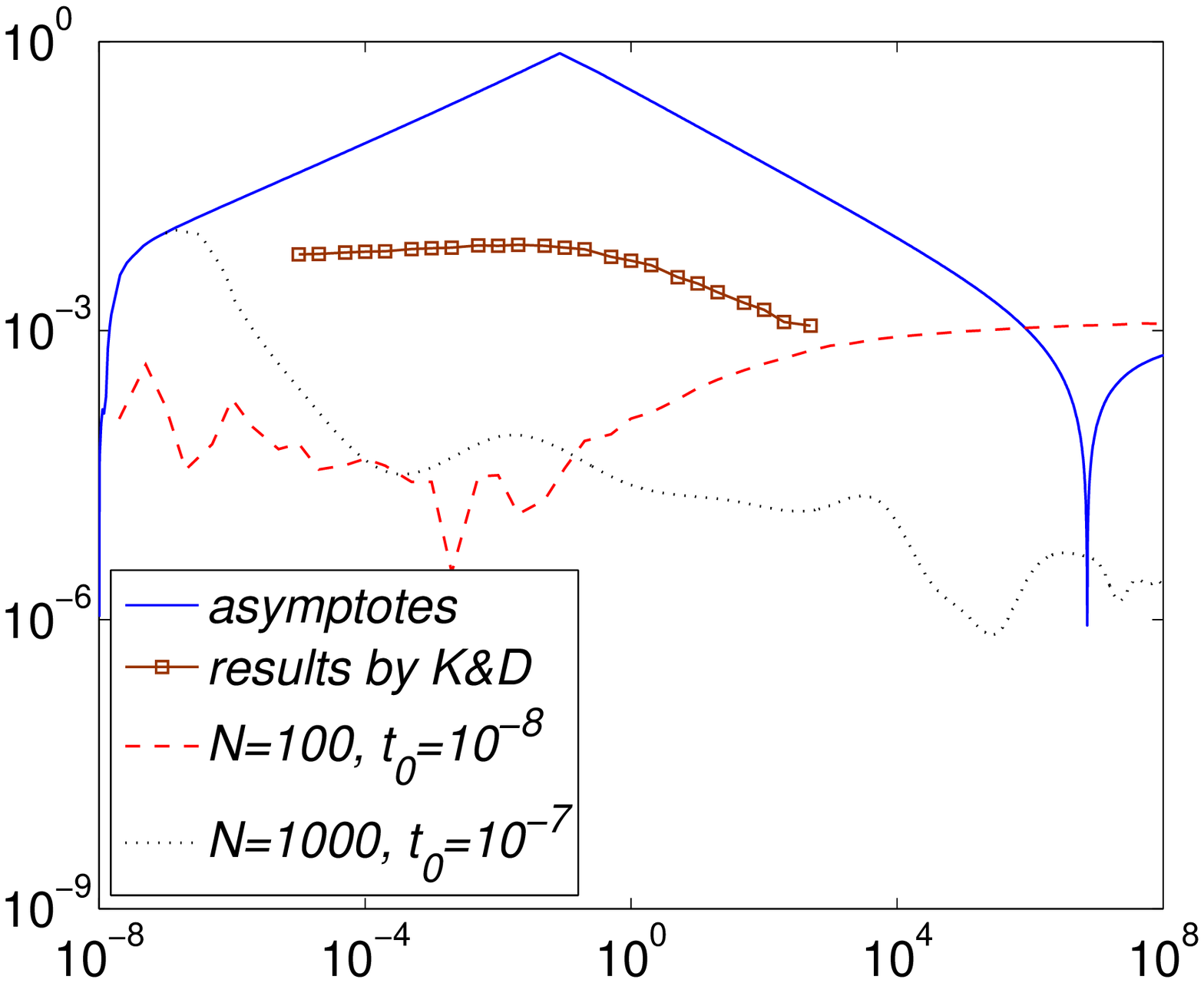}
     \put(-100,-2){$t$}
  \captionof{figure}{Relative deviations from the numerical \\solution for $L$.}
  \label{dev_L}
\end{minipage}%
\begin{minipage}{.5\textwidth}
  \centering
  \includegraphics[width=.9\linewidth]{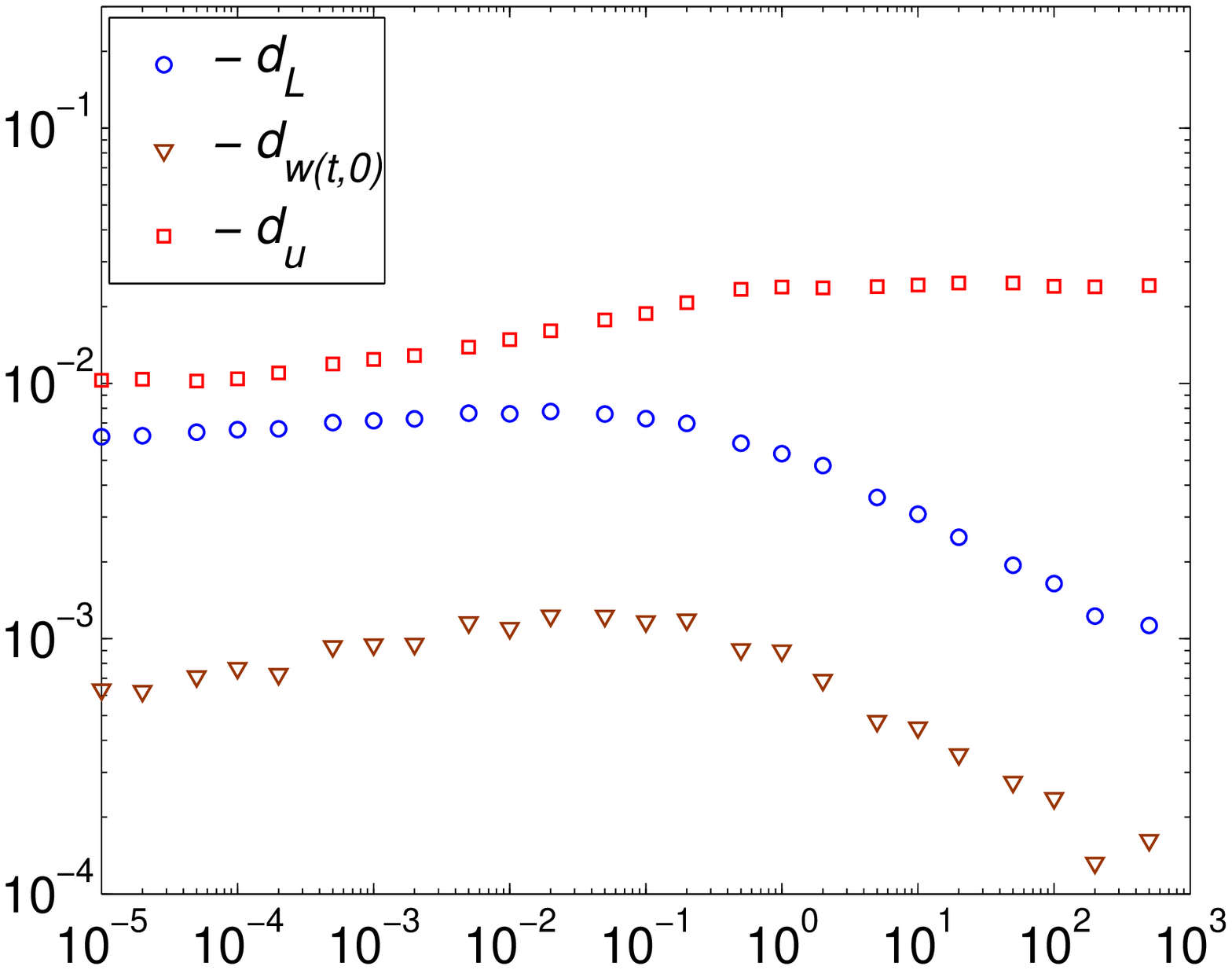}
  \put(-100,-2){$t$}
  \captionof{figure}{Relative deviations of the solution by \citet{Kovalyshen} from the results reported in Table \ref{table_Carter}.}
  \label{dev_Kov}
\end{minipage}
\end{figure}

In order to additionally asses the credibility of our solution (computed for $N=1000$ and presented in the Table \ref{table_Carter}),
we show in Fig.~\ref{dev_L} the relative deviations between it and other solutions.
Namely, we analyze the crack lengths $L$ provided by: a) early and large time asymptotes;
b) the solution by \citet{Kovalyshen};
c) the solution obtained for 100 nodal points and d) the solution obtained for 1000 nodal points at another starting point $t_0=10^{-7}$.
%

When tracing the data from Fig.~\ref{dev_L} we can see that the deviations of $L$ from the early and large time asymptotes at the ends of the considered interval are of the order $10^{-4}$. Moreover, the relative deviation of the solution obtained for 100 points is of the same order in almost entire  time range, which corresponds very well to the figures from Tab~\ref{table_very_bad_benchmark}. The discrepancy between the reference solution and the solution for $t_0=10^{-7}$ decreases rapidly with time. The last observation confirms the credibility of the reference solution.

We do not present respective graph for $w(t,0)$. However, it is worth mentioning that in this case the deviations from the asymptotes were even lower than for $L$, while the deviation of the solution for $N=100$ did not exceed the value of $10^{-4}$ on the substantial part of the interval.

The relative discrepancies between the components of our solution and the solution by \citet{Kovalyshen} are shown in Fig.~\ref{dev_Kov}. Here $d_L$, $d_{w(0,t)}$ and $d_u$ refer to the deviations of the crack length, $L$, the crack opening, $w(t,0)$ and the normalized crack velocity, $u$, respectively.


In the light of the presented results, we believe that the data collected in Table \ref{table_Carter} provide the accuracy  \textbf{at least} of the order $10^{-4}$ for both, the crack length, $L$, and the crack opening at $x=0$. Moreover, in the considerable time range ($10^{-6}< t<10^{6}$), one can expect the error lower by up to two orders of magnitude.
The normalized crack propagation speed $u$ is computed with accuracy of one to two orders of magnitude lower.

Summarizing the above discussions, the level of accuracy for the results tabulated by \citet{Kovalyshen} can be estimated as $10^{-3}\div 10^{-2}$ for $L$, $10^{-4} \div 10^{-3}$ for $w(t,0)$ and of the order $10^{-2}$ for $u$.

\begin{table}
\centering
\begin{tabular}{c|c|c|c|c|c|c|c|c|c|}
\cline{1-4}
\multicolumn{1}{|c|}{$\log(t)$}&$L(t)$&$w(t,0)$&$u(t) \times 10$
\\ \cline{1-4}
\multicolumn{1}{|c|}{$-7$}&3.283747e-6&3.988347e-2&7.9701
 \\ \cline{1-4}
\multicolumn{1}{|c|}{$-6$}&2.049209e-5&6.298786e-2&7.9355
 \\ \cline{1-4}
\multicolumn{1}{|c|}{$-5$}& 1.265786e-4&9.915967e-2&7.8716
 \\ \cline{1-4}
\multicolumn{1}{|c|}{$-4$}& 7.660018e-4&1.551088e-1&7.7536
 \\ \cline{1-4}
\multicolumn{1}{|c|}{$-3$}&4.456291e-3&2.397462e-1&7.5185
 \\ \cline{1-4}
\multicolumn{1}{|c|}{$-2$}&2.412817e-2&3.629593e-1&7.1173
 \\ \cline{1-4}
\multicolumn{1}{|c|}{$-1$}&1.163591e-1&5.326638e-1&6.5267
 \\ \cline{1-4}
\multicolumn{1}{|c|}{$0$}&4.849863e-1&7.541837e-1&5.8885
 \\ \cline{1-4}
 \multicolumn{1}{|c|}{$1$}&1.779508e0&1.037495e0&5.4408
 \\ \cline{1-4}
 \multicolumn{1}{|c|}{$2$}&6.035529e0&1.403522e0&5.1993
 \\ \cline{1-4}
  \multicolumn{1}{|c|}{$3$}&1.968511e1&1.883411e0&5.0847
 \\ \cline{1-4}
  \multicolumn{1}{|c|}{$4$}&6.308563e1&2.518338e0&5.0378
 \\ \cline{1-4}
  \multicolumn{1}{|c|}{$5$}& 2.006370e2&3.362113e0&5.0146
 \\ \cline{1-4}
  \multicolumn{1}{|c|}{$6$}&6.360179e2&4.485636e0&5.0071
 \\ \cline{1-4}
  \multicolumn{1}{|c|}{$7$}&2.013373e3&5.982935e0&5.0029
 \\ \cline{1-4}
\end{tabular}
\caption{Numerical solution of the PKN fracture.}
\label{table_Carter}
\end{table}

\subsection{Remarks on the sensitivity of the Carter leak-off model.}

It is well known that applicability of the empirical Carter law (\ref{carter})$_1$ in the vicinity of the fracture tip is questionable (see for example,
\citet{Economides2000}, \citet{Kovalyshen_1} and \citet{MKP}). Moreover, when combining Carter's leak-off with some non-local variants of elasticity models (for example, KGD model of hydrofracturing), one obtains an infinite
particle velocity at the crack tip. As a result, the speed equation (\ref{velocity_2}) cannot be applied in such a case.
One of the ways to eliminate the negative consequences of this fact is to assume that the Carter law becomes valid at some distance away from the fracture tip (see for example \citet{MKP}).

The PKN model, which does not exhibit such a drawback, gives however a unique opportunity to assess how the solution is affected by a modification of the classic Carter law in the neighborhood of the fracture tip.

To this end, let us consider two ways of modification of the law. The first one assumes that leak-off function equals zero over some distance from the crack front ($d>\varepsilon$). The second one accepts a constant value of $q_l$ in the same interval. This value is taken in such a manner to preserve the continuity of the leak-off function. Note, that both of these modifications change the volume of fluid loss to the rock formation, with respect to the original state.

\begin{figure}[h!]
\center
    \includegraphics [scale=0.35]{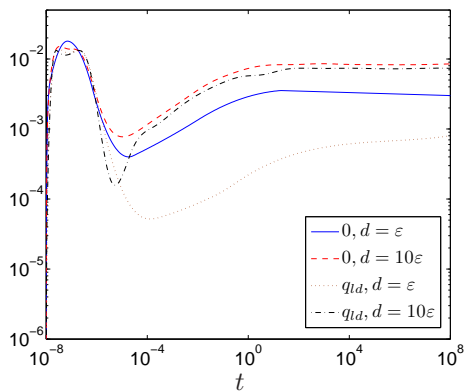}
    \put(-100,-2){$t$}

    \caption{Relative deviations of the crack lengths for different variants of truncated Carter law.}
\label{dev_carter}
\end{figure}

The relative deviations of the crack lengths for these modifications from the original one are shown in Fig.~\ref{dev_carter}. Results for two values of $d$: $d=\varepsilon$, $d=10\varepsilon$ (for $\varepsilon=10^{-5}$) are depicted. The symbol $q_{ld}$ in the legend refers to the cases when the leak-off function is complimented by the constant value over $1-d\leq x \leq 1$.

One can see from this picture that the maximal relative discrepancies (of the level of 1\%) appear at the initial and large time ranges. To explain this phenomenon we can easily compute the additional volume of fluid retained in the fracture as a result of the Carter law amendments. Taking into account (\ref{f_D}), these values are $\Delta Q_l(t)=2D(t)\sqrt{d}$ and $\Delta Q_l(t)=D(t)\sqrt{d}$, for the respective modifications (see (\ref{function_D}) for $D(t)$). Note that $D(t)=\sqrt{u(t)/t}$, which explains the level of deviation for the small time. For large time, the effect of accumulation of the difference of the fluid loss, $\int_0^t\Delta Q_l(\tau)d\tau$ = $O(\sqrt{td})$, plays a crucial role.

The above test proves that the application of the Carter law modified in the aforementioned ways, is acceptable in terms of both the stability and the accuracy of computations. This allows one to use such an approach in the cases when the law leads to unphysical results (see Mitchel et al. 2007). Note, however, that the intermediate asymptotics related to the carter law still holds true and thus should be taken into account.

On the other hand, appreciable sensitivity of the solution to the slight modifications of the Carter law calls the validity of the law near the crack tip into question.

\section{Conclusions}

In this paper, we have revisited the PKN model of hydrofracturing providing a comprehensive overview
of the known results together with:
\begin{itemize}
\item[(i)] analysis of various leak-off regimes (vanishing, boun\-ded and singular near the fracture tip) supplemented with the full asymptotic expansion of the solution for the Carter model;
\item[(ii)] introduction of a new dependent variable, $\Omega$, and the resulting problem reformulation;
\item[(iii)] implementation of a new form of the regularized boundary condition;
\item[(iv)] analysis of different aspects of application of various dependent and independent variables, including the stiffness properties, accuracy and
efficiency of the computations;
\item[(v)] comparison of the results for the Carter leak-off with the known numerical benchmark from \citet{Kovalyshen}.
\end{itemize}

The main conclusions of the paper are:

\begin{itemize}
\item
The approach proposed in \citet{Linkov_4,MWL} is
efficient even in the cases when the smoothness of the particle velocity near the
crack tip is disturbed by the singular leak-off function;
\item
For the best performance of a solver,
the regularized boundary condition should incorporate at least two leading terms of asymptotics;
\item
The stiffness properties of the dynamic systems and the accuracy of the computations can be effectively controlled by the choice of dependent and independent variables.
\item
The new independent variable, $\Omega$, leading to a slightly greater stiffness of the dynamic system, considerably improves the accuracy of computations.
\end{itemize}

The value of the aforementioned findings was demonstrated on various analytical benchmarks and for the classic Carter law.
Although, the presented analysis concerns the 1-D PKN formulation, at least some of the findings may be utilised in the more advanced cases.





\section*{Appendices}

\appendix

\section{Carter's leak-off function in the normalised formulation} \label{app:B}

Consider the transformation of the Carter law described by (\ref{carter}) when applying the normalization (\ref{dimensionless}). Assume that:
\begin{equation}
\label{f_D}
\frac{1}{\sqrt{t-\tau(x)}}=\frac{D(t)}{\sqrt{1-\tilde x}}+R(t,\tilde x),
\end{equation}
where function $D(t)$ is defined in (\ref{function_D}) while the remainder $R$ is estimated
later in (\ref{subs_2}).

To find function $D(t)$, and thus to obtain an exact form of equation (\ref{function_D}),
it is enough to compute the limit
\begin{equation}
D^2(t)=\lim_{\tilde{x}\rightarrow1}\frac{1-\tilde{x}}{t-\tau(x)}.
\label{d_t_intro}
\end{equation}
This can be done by utilising L'Hopital's rule with taking into account that $x\to L(t)$ as $\tilde{x}\rightarrow1$,
\begin{equation}
\tau(x)=\tau\left(L(t)\tilde{x}\right)=L^{-1}(L(t)\tilde{x})\label{tau_2},
\end{equation}
and that the crack length is a smooth function of time ($L\in C^1$ at least). The last fact immediately follows from the problem formulation in terms of evolution system (\ref{w_DS}).

Having the value of  $D(t)$ we can estimate the remainder $R(t,\tilde x)$
when $\tilde x\to1$, or, what it is equivalent to when $x\to l(t)$ (or $t\to \tau(x)$).
For this reason, we search for a parameter $\xi\ne0$ which guarantees that the limit
\[
A=\lim_{\tilde{x}\rightarrow1}\frac{R(t,\tilde{x})}{(1-\tilde{x})^{\xi}}=
\lim_{\tilde{x}\rightarrow1}
\frac{1}{2\xi(1-\tilde{x})^{\xi-1}}
\left(\frac{D(t)}{(1-\tilde{x})^{3/2}}-\frac{L(t)\tau'(x)}{(t-\tau(x))^{3/2}}
\right)
\]
does not turn to zero or infinity. Due to this assumption, we can write
\begin{equation}
\frac{1}{\sqrt{t-\tau(x)}}= \frac{D(t)}{\sqrt{1-\tilde{x}}}+A (1-\tilde{x})^{\xi}+o\left((1-\tilde{x})^{\xi}\right),
\label{subs_2}
\end{equation}
when $\tilde x \rightarrow 1$, or equivalently $x \rightarrow l(t)$.
Taking the last estimate into account $A$ can be expressed as:
\[
A=\lim_{\tilde{x}\rightarrow1}\frac{1}{2\xi(1-\tilde{x})^{\xi-1}}
\left(\frac{D(t)}{(1-\tilde{x})^{3/2}}-\frac{L(t)\tau'(x)}{t-\tau(x)}
\frac{D(t)}{\sqrt{1-\tilde{x}}}
\right)-
\frac{AL(t)}{2\xi}\lim_{\tilde{x}\rightarrow1}\frac{(1-\tilde{x})\tau'(x)}{t-\tau(x)}\big(1+o(1)\big).
\]
Now, on substitution of $\tau'(x)=1/L'(t)$ at $x=L(t)$ and (\ref{d_t_intro}) into the limit one has:
\[
A=\lim_{\tilde{x}\rightarrow1}\frac{D(t)}{2\xi(1-\tilde{x})^{\xi-1/2}}
\left(\frac{1}{1-\tilde{x}}-\frac{L(t)\tau'(x)}{t-\tau(x)}
\right)-\frac{AL(t)D^2(t)}{2\xi L'(t)}.
\]
Applying (\ref{d_t_intro}) and (\ref{function_D}) here gives:
\[
\frac{1+2\xi}{2\xi}A=\lim_{\tilde{x}\rightarrow1}\frac{D(t)}{2\xi(1-\tilde{x})^{\xi-1/2}}
\left(\frac{1}{1-\tilde{x}}-\frac{L(t)\tau'(x)}{\sqrt{t-\tau(x)}}
\frac{D(t)}{\sqrt{1-\tilde{x}}}
\right)
-
\frac{AD(t)L(t)}{2\xi}\lim_{\tilde{x}\rightarrow1}\frac{\tau'(x)\sqrt{1-\tilde{x}}}{\sqrt{t-\tau(x)}}.
\]
By repeating the same process one more time we have:
\[
(2+2\xi)A=\lim_{\tilde{x}\rightarrow1}\frac{D(t)}{(1-\tilde{x})^{\xi}}
\left(\frac{1}{\sqrt{1-\tilde{x}}}-\frac{L(t)\tau'(x)D(t)}{\sqrt{t-\tau(x)}}
\right).
\]
Finally by eliminating the square root with use of (\ref{subs_2}) we obtain (after some algebra)
\[
(3+2\xi)A=D(t)\lim_{\tilde{x}\rightarrow1}
\frac{1-L(t)\tau'(x)D^2(t)}{(1-\tilde{x})^{\xi+1/2}}.
\]
This relationship gives a finite value of $A$ if and only if $\xi=1/2$ and, as a result, we find:
\[
A=\frac{1}{4}D^3(t)L^2(t)\tau''(L(t))=
-\frac{1}{4}\frac{L''(t)}{L'(t)}\sqrt{\frac{L(t)}{L'(t)}}.
\]

\section{Asymptotics of the solutions for different leak-off functions} \label{app:C}

Asymptotic expansion for the crack opening and the fluid velocity near the crack tip in the normalised variables (\ref{dimensionless}) can be written in the following general forms:
\begin{equation}\label{w_asym}
w(t, x)=\sum_{j=0}^Nw_j(t)(1-x)^{\alpha_j}+O((1- x)^{\varrho_w}), \quad x\rightarrow 1,
\end{equation}
and
\begin{equation}\label{v_asym}
V(t, x)=\sum_{j=0}^NV_j(t)(1-x)^{\beta_j}+O((1- x)^{\varrho_V}),\quad  x\rightarrow 1,
\end{equation}
with $\varrho_w>\alpha_n$, $\varrho_V>\beta_n$, $\alpha_0=1/3$, $\beta_0=0$ and some increasing sequences $\alpha_0,\alpha_1,\ldots,\alpha_n$ and $\beta_0,\beta_1,\ldots,\beta_n$. Note that the asymptotics are related to each other by the speed equation \eqref{norm_speed} and thus, regardless of the chosen leak-off function, we can write
\begin{equation}\label{v_asym_1}
\sum_{j=0}^NV_j(t)(1-x)^{\beta_j}+\ldots=
\frac{1}{3L(t)}\sum_{k=0}^N\sum_{m=0}^N\sum_{j=0}^N\alpha_jw_j(t)
w_m(t)w_k(t)(1-x)^{\alpha_j+\alpha_m+\alpha_k-1}.
\end{equation}
In line with the discussion after equation (\ref{V_asym_0}),
we are interested only in the terms such that $\beta_j\le1$, restricting ourselves to the smallest $\varrho_V>1$,
since the values of $\beta_j$ are combinations of a sum of three consequent components of the exponents $\alpha_j$.
However, since $\alpha_0$ is known ($\alpha_0=1/3$), one can write (compare with (\ref{V_asym_0_coefs})):
\begin{align}
& V_{0}(t)=\frac{1}{3L(t)}w_{0}^{3}(t)\label{v_0}, \\
& V_{1}(t)=\frac{1}{L(t)}\left(\alpha_1+\frac{2}{3}\right) w_{0}^{2}(t) w_{1}(t),\quad \beta_1=\alpha_1-\frac{1}{3}.\label{v_1}
\end{align}

To continue the process one now needs to compute the value of the exponent $\alpha_1$ as it is not clear
a priori which value determining the next exponent $\beta_2=\min\{2/3+\alpha_2,1/3+2\alpha_1\}$ is larger. To do so
let us rewrite the continuity equation (\ref{norm_continuity}) in the form:
\begin{equation}
\label{norm_continuity_as}
\frac{\partial w}{\partial t}+\frac{V_0(t)}{L( t)}(1-x)\frac{\partial w}{\partial x}=\frac{1}{L( t)}\frac{\partial \big(w(V_0-V)\big)}{\partial x}-q_l(t,x).
\end{equation}
Here, the terms on the left-hand side of the equation are always bounded near the crack tip, while those on the right-hand side can behave differently depending on the chosen leak-off function.

\noindent Consider the following three cases of $q_l$ behaviour.

(\emph{i}) Assume first that
\[
q_l(t,x)=o\big(w(t,x)\big),\quad x\to1.
\]
This case naturally includes the impermeable rock formation. Analysing the leading order terms in the equation (\ref{norm_continuity_as}), it is clear that $w(V_0-V)=O((1-x)^{4/3})$, as $x\to1$. This, in turn, is only possible for $\beta_1=1$ and, therefore, $\alpha_1=4/3$. Finally, comparing the left-hand side and the right-hand side of the equation we obtain:
\begin{equation}
w_{0}'(t)= \frac{w_{0}(t)}{3L(t)} \big(V_0(t)+4V_1(t)\big),\quad V_{1}(t)=\frac{2}{L(t)}w_{0}^{2}(t) w_{1}(t).
\label{sp_1}
\end{equation}
This case has been considered in \cite{Linkov_4} and \cite{MWL}.

\vspace{2mm}

(\emph{ii})
If we assume that the leak-off function is estimated by the solution as  $O\big(w(t,x)\big)$, or equivalently;
\[
q_l(t,x)\sim\Upsilon(t)w_0(t)(1-x)^{1/3},\quad x\to1,
\]
then the previous results related to the values of $\alpha_1$ and $\beta_1$ and, therefore, the equation (\ref{sp_1})$_2$ remain the same, while the first one changes to
\begin{equation}
w_{0}'(t)= \frac{1}{3L(t)} w_{0}(t)\big(V_0(t)+4V_1(t)\big)-\Upsilon(t)w_0(t).
\label{sp_2}
\end{equation}
This case corresponds to (\ref{norm_leak_off_1})$_3$ when $C_{32}=0$ and $\Upsilon(t)=kC_{31}(t)$.

\vspace{2mm}

(\emph{iii})
The leak-off function in a general form:
\[
q_l(t,x)=\Phi(t)(1-x)^{\theta}+o((1-x)^{1/3}),\quad x\to1,
\]
where $-1/2\le \theta<1/3$. Here, one can conclude that $w(V_0-V)=O((1-x)^{1+\theta})$, as $x\to1$ or equivalently, $\beta_1=\theta+2/3$, and $\alpha_1=1+\theta$. Moreover, in this case:
\begin{equation}
(1+\theta)w_0V_1=L(t)\Phi(t), \quad V_{1}(t)=\frac{1}{L(t)}\left(\theta+\frac{4}{3}\right) w_{0}^{2}(t) w_{1}(t),
\label{sp_3}
\end{equation}
and, thus
\begin{equation}
w_{1}(t)=\frac{3L^2(t)\Phi(t)}{(4+3\theta)(1+\theta)w_{0}^{3}(t)}.
\label{sp_3a}
\end{equation}

Note, that as one would expect, the particle velocity function is not smooth in this case near the crack tip, its derivative is unbounded and exhibits the following behaviour:
\[
\frac{\partial V}{\partial x}=O\big((1-x)^{\theta-1/3}\big), \quad x\to1.
\]
To formulate the equation similar to (\ref{sp_1})$_1$ or (\ref{sp_2}), one needs to continue asymptotic analysis of the equation (\ref{norm_continuity_as}) incorporating the available information.
Apart from the fact that the analysis can be done in the general case, we restrict ourselves only to three variants used from the beginning (compare (\ref{carter})), respectively:
$\theta=0$, $\theta=1/3-1/2=-1/6$ and $\theta=-1/2$.

When $\theta=0$, $\alpha_1=1$ and $\beta_1=2/3$, returning to the equation (\ref{v_asym_1}), one concludes that $\beta_2>1$ and, therefore,
\begin{equation}
w_{0}'(t)= \frac{1}{3L(t)} w_0(t)V_0(t).
\label{sp_4}
\end{equation}
This case corresponds to (\ref{norm_leak_off_1})$_3$ when $\Phi(t)=C_3^{(2)}(t)w_0(t)$ and $C_3^{(1)}=0$.

If $\theta=-1/6$, then $\alpha_1=5/6$ and $\beta_1=1/2$. In this case the function $\Phi(t)$ can be written as $\Phi(t)=C_2 D(t)w_0(t)$ (compare to (\ref{norm_leak_off_1})$_2$)
and again equation (\ref{v_asym_1}) gives $\beta_2>1$, while equation (\ref{norm_continuity_as}) leads to
\begin{equation}
w_{0}'(t)= \frac{1}{3L(t)} \big(w_0(t)V_0(t)+4w_1(t)V_1(t)\big).
\label{sp_5}
\end{equation}
Summarizing, in both mentioned above cases, there exists a single term in asymptotics of the particle velocity which has singular derivative near the crack tip. Moreover, those terms ($w_1$ and $V_1$, respectively) are fully defined by the leak-off function $\Phi(t)$ and the coefficient $w_0$ in front of the leading term for the crack opening in (\ref{sp_3a}) and (\ref{sp_3})$_1$.

The situation changes dramatically when $\theta=-1/2$ (Carter law). We now have $\alpha_1=1/2$ and $\beta_1=1/6$ and $\Phi(t)=C_1D(t)$. In this case, however, $\beta_2<1$ and we need to continue the asymptotic analysis further to evaluate all terms of the particle velocity which exhibit non-smooth  behaviour near the crack tip. We omit the details of the derivation, presenting only the final result in a compact form.
The first six exponents in the asymptotic expansions (\ref{w_asym}) and (\ref{v_asym}), that introduce the singularity of $w_x$, are:
\[
\alpha_j=\frac{1}{2}+\frac{j}{6},\quad \beta_j=\frac{j}{6},\quad j=1,2,\ldots,6.
\]
\[
w_j(t)=\kappa_j\frac{\Phi^j(t)L^{2j}(t)}{w_0^{4j-1}(t)},\quad V_j(t)=\psi_j\frac{\Phi^j(t)L^{2j-1}(t)}{w_0^{4j-3}(t)},
\]
where $j=1,2,\ldots,5$ and
\[
\begin{array}{l}
\kappa_1=\frac{12}{7},\quad \psi_1=2,\quad \kappa_2=-\frac{270}{49},\quad \psi_2=-\frac{24}{7},
\kappa_3=\frac{9768}{343}, \,\, \psi_3=\frac{828}{49},
\\[4mm]
\kappa_4=-\frac{2097252}{12005},\,\, \psi_4=-\frac{5136}{49},
\,\, \kappa_5=\frac{1081254096}{924385},\quad \psi_5=\frac{1234512}{1715}.
\end{array}
\]

\section{Benchmark Solutions}
\label{app:A}

There are several benchmarks in the literature to be utilized for investigation of the numerical algorithms.
Benchmark solutions for impermeable rock have been constructed in \cite{Kemp,Linkov_4},
while that corresponding to the non-zero leak-off model with  $q_l$ vanishing at a crack tip has been
analyzed in \cite{MWL}.

In this paper, we introduce three different analytical benchmark
solutions cor\-res\-pond\-ing to the representations
\eqref{norm_leak_off_1}. Moreover, for each of the leak-off
functions under consideration we take two different
relationships between the injection flux rate $q_0$ and the leak-off
to formation $q_l$. In this way
 six different benchmark solutions are analyzed.

In order to formulate the benchmark solutions let us assume the
following form of the crack opening function:
\begin{equation}\label{w_bench}
w(t,x)=W_0(1+t)^\gamma h(x),\quad W_0=\sqrt[3]{\frac{3}{2}(3\gamma+1)},
\end{equation}
where $\gamma$ is an arbitrary parameter, and the function $h(x)$ ($0<x<1$) is given by:
\begin{equation}\label{w_0_bench}
h(x)=(1- x)^\frac{1}{3}+b_1(1-x)^{\lambda_1}+b_2(1-x)^{\lambda_2}.
\end{equation}
The choice of the next powers $1/3<\lambda_1<\lambda_2$ will depend on the
leak-off variant from \eqref{carter}. On consecutive
substitutions of \eqref{w_bench}-\eqref{w_0_bench} into the
relations \eqref{norm_speed}, \eqref{norm_boundary},
\eqref{w_system} and \eqref{new_speed} one obtains the remaining
benchmark quantities:
\begin{equation}\label{v_bench}
L(t)=(1+t)^{\frac{3\gamma+1}{2}},\quad V(t,x)=-W_0^3(1+t)^{\frac{3\gamma-1}{2}}h^2(x)\frac{\partial h}{\partial x}.
\end{equation}
\begin{equation}\label{q_0_bench}
q_0(t)=-W_0^4(1+t)^{\frac{5\gamma-1}{2}}\left(h^3\frac{\partial h}{\partial x}\right)|_{x=0}.
\end{equation}
\begin{equation}
\label{q_l_bench}
q_l(t,x)=W_0(1+t)^{\gamma-1}\times
\Big(\frac{3}{2}(3\gamma+1)\Big[\frac{1}{3}x\frac{\partial
h}{\partial x} +3h^2\left(\frac{\partial h}{\partial x}\right)^2
+h^3\frac{\partial^2 h}{\partial x^2}\Big]-\gamma h\Big).
\end{equation}

It can be easily checked that for $\lambda_1=1/2$ and $\lambda_2=4/3$
the leak-off function incorporates a square root singular term of
type \eqref{norm_leak_off_1}$_1$. By setting $\lambda_1=5/6$ and
$\lambda_2=4/3$ we comply with representation
\eqref{norm_leak_off_1}$_2$. Although in both of these cases
$q_{1(2)}^*$ exhibits a singular behaviour at the crack tip, it does
not detract from the applicability of our benchmarks. Finally, when
using $\lambda_1=4/3$ and $\lambda_2=7/3$, the benchmark gives a
non-singular leak-off function in the form
\eqref{norm_leak_off_1}$_3$.

\begin{figure*}[t]
        \centering
        \begin{subfigure}{0.32\textwidth}
                \centering
                \includegraphics[width=\textwidth]{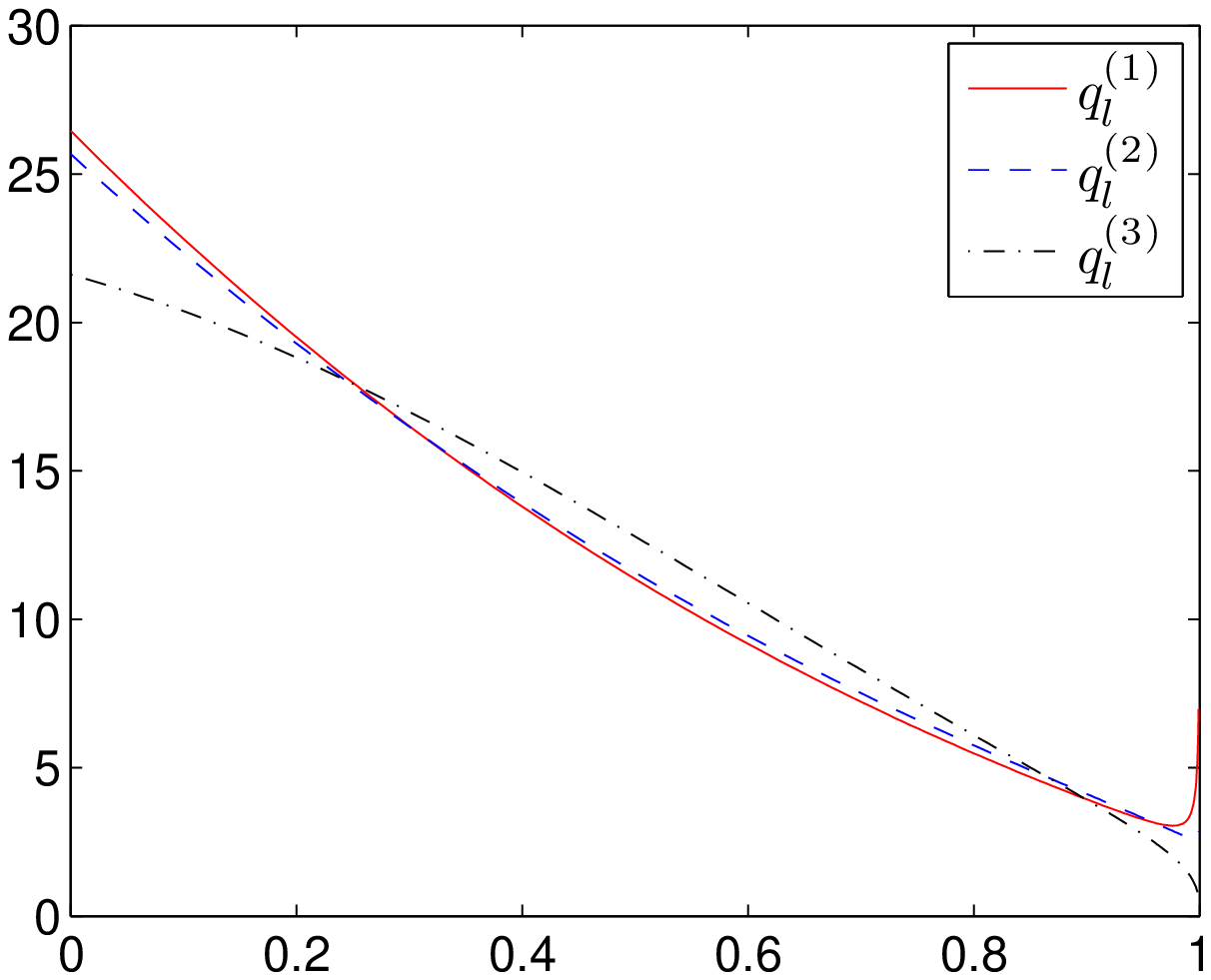}

 \end{subfigure}
 \begin{subfigure}{0.32\textwidth}
                \centering
                \includegraphics[width=\textwidth]{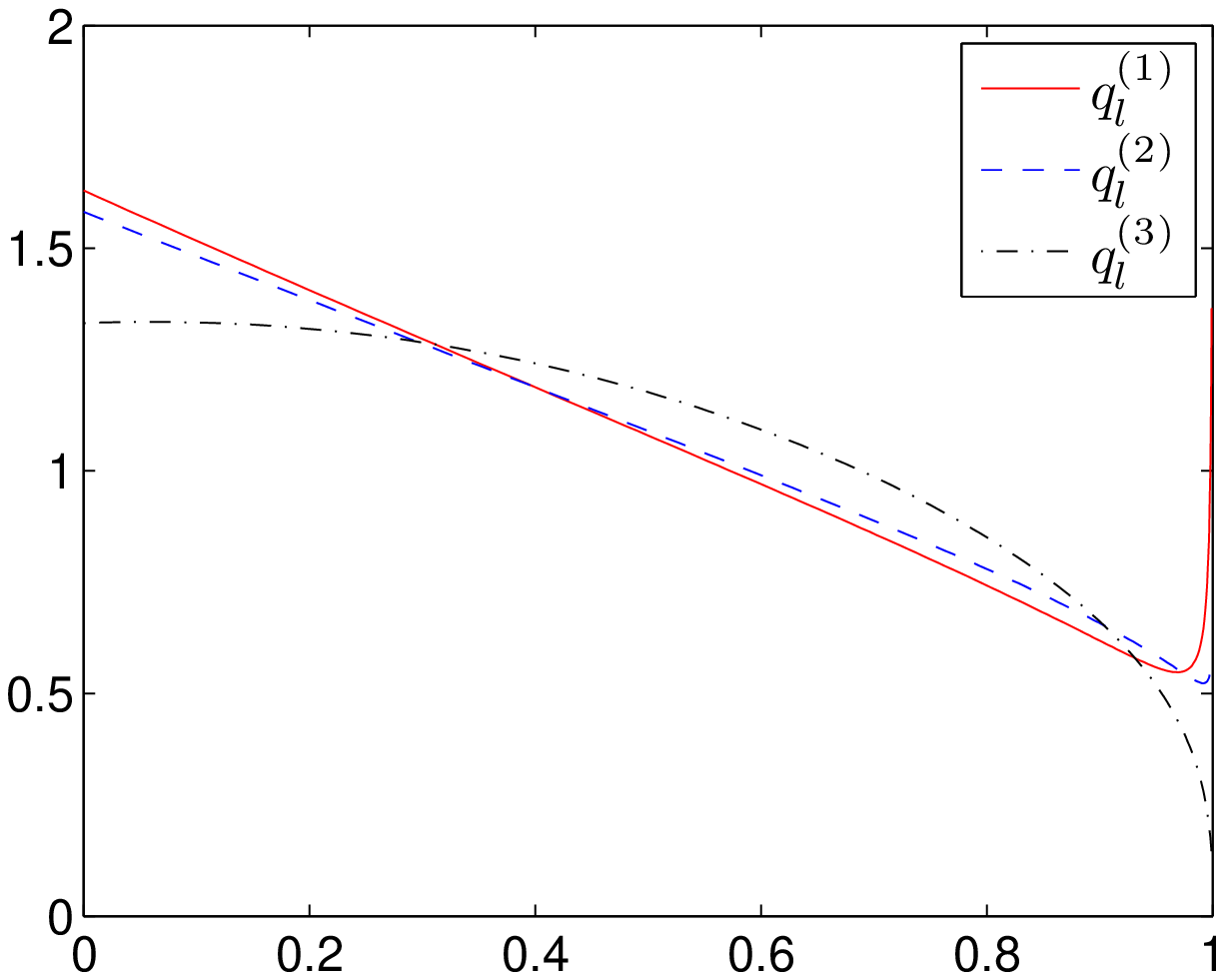}

\end{subfigure}
 \begin{subfigure}{0.32\textwidth}
                \centering
                \includegraphics[width=\textwidth]{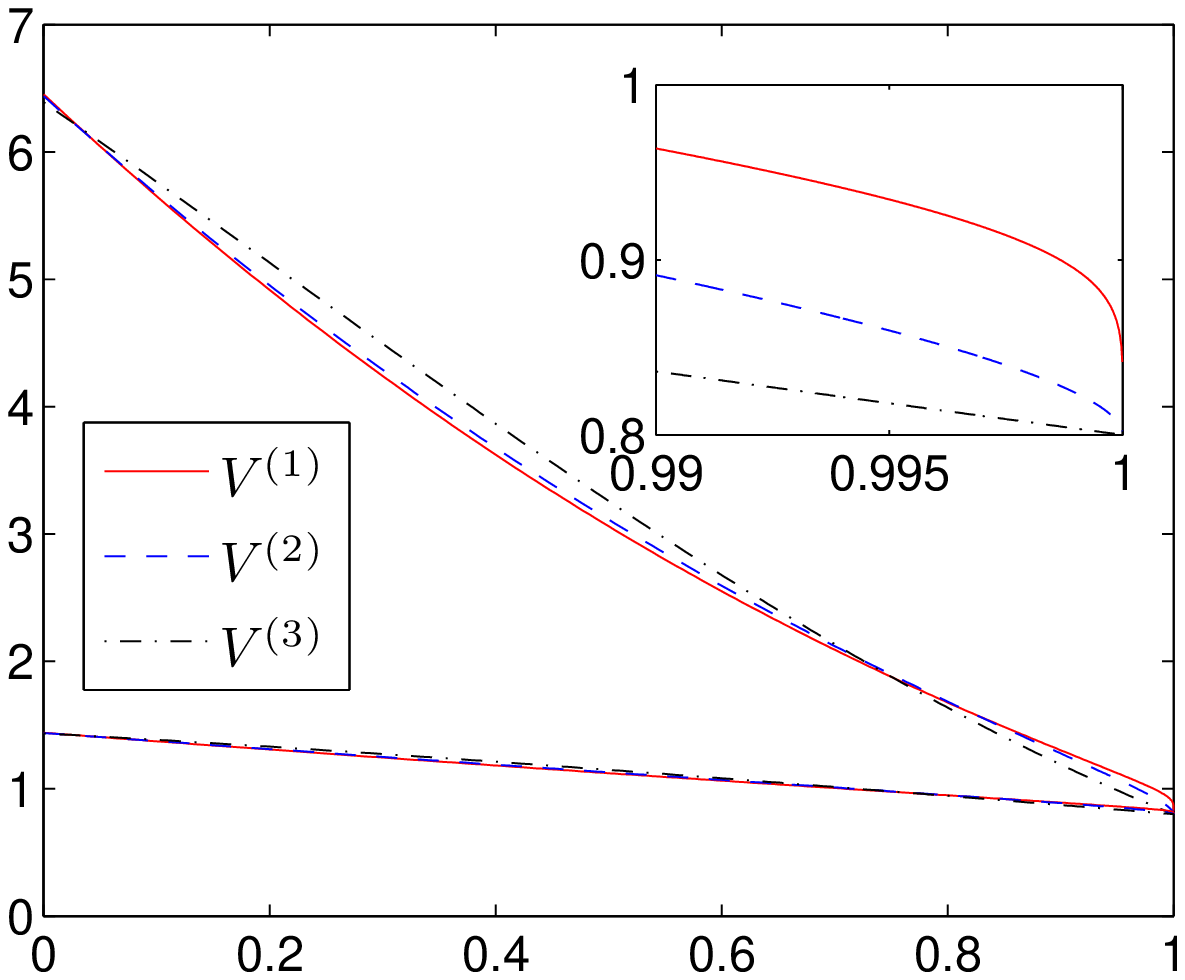}
                \begin{picture}(0,0)(0,0)
                \put(-40,35){\line(2,1){11}}
                \put(-47,29){${}_{Q_l/q_0=0.5}$}
                \put(-40,108){${}_{Q_l/q_0=0.9}$}
                \put(-40,100){\line(2,1){10}}
     \end{picture}
\begin{picture}(0,0)(405,-117)
   \put(65,-20){$Q_l/q_0=0.9$}     \put(215,-20){$Q_l/q_0=0.5$}
        \put(3,10){a)} \put(168,10){b)} \put(333,10){c)}
        \put(5,-35){$q_l$}      \put(170,-35){$q_l$} \put(332,-35){$V$}
        \put(80,-110){$x$}    \put(245,-110){$x$} \put(411,-110){$x$}
                \end{picture}
                \vspace{-2mm}
\end{subfigure}
\caption{Distributions of the leak-off functions $q_l(t,x)$ and the respective particle velocity $V(t,x)$ over $\tilde x \in(0,1)$ at initial time $t=0$.}
\label{fig:leak_offs}
\end{figure*}

Note also, that by manipulating with the value of $\gamma$ one can
simulate some very specific regimes of crack propagation. For
example $\gamma=1/5$ corresponds to the constant injection flux
rate, while $\gamma=1/3$ gives a constant crack propagation speed.
For our computations we always
set the value of $\gamma=1/5$.

Choosing appropriate values $b_1$ and $b_2$ one can change the relation between the amount of fluid loss to formation and the injection rate. This ratio can be defined by the  measure, $Q_l/q_0$, where $Q_l$ is the total volume of leak-off $\int_0^1 q_l dx$.
It is important to note that this measure decreases in time, from its maximum value to zero, for all chosen benchmarks. Thus, taking the maximal value high enough and tracing the solution accuracy in time, one can analyse performance of the algorithm for any possible value of the parameter. We consider two variants of $Q_l/q_0$,
one where fluid injection doubles the size of total fluid loss, and a second where the total fluid loss is close to injection rate.
The values of the corresponding parameters $b_1$, $b_2$ are presented in Table~\ref{tab_last}.
\begin{table}[h]
\centering
\begin{tabular}{c|c|c|c|c|c|c|}
\cline{2-7}
& \multicolumn{3}{c|}{$Q_l/q_0=0.9$ } & \multicolumn{3}{c|}{$Q_l/q_0=0.5$} \\ [.05in]\cline{2-7}
& $q_l^{(1)}$ & $q_l^{(2)}$ & $q_l^{(3)}$ & $q_l^{(1)}$ & $q_l^{(2)}$ & $q_l^{(3)}$ \\ [.05in]\cline{1-7}
\multicolumn{1}{|c|}{$b_1$} & $0.1$ & $0.19$ & $0.74$ &$0.02$& $0.03$& $0.15$ \\[.02in]\cline{1-7}
\multicolumn{1}{|c|}{$b_2$} &$0.5$ & $0.41$& $-0.13$ &$0.1$& $0.08$& $-0.02$ \\ \cline{1-7}
\multicolumn{1}{|c|}{ $\gamma_v$} &$1.69$ & $1.70$& $1.65$ &$0.55$& $0.56$& $0.55$ \\ \cline{1-7}
\end{tabular}
\caption{The values of parameters $b_1$ and $b_2$ for different benchmark solutions modelling desired leak-off to fluid injection ratios.}
\label{tab_last}
\end{table}

Additionally one can compute a parameter $\gamma_v$ defined in \cite{MWL} as a measure of the uniformity of fluid velocity distribution:
\begin{equation}
\label{gamma_v}
\gamma_v=\left[\max(V(t,x))-\min(V(t,x))\right]\left[\int_0^1V(t,\xi)d\xi\right]^{-1}.
\end{equation}
Interestingly, this measure is directly correlated with the leak-off ratio $Q_l/q_0$.

In Fig.~\ref{fig:leak_offs} the distributions  of the leak-off functions and the corresponding particle velocities for the respective benchmarks are presented.
It shows that the velocity near the crack tip depends strongly on the benchmark variant. To highlight this fact, a zoom picture is placed in the Fig.~\ref{fig:leak_offs} b).

Note that the benchmark $q_l^{(1)}$ is worse, in a sense, than the original Carter's model as it contains additional singular terms of the leak-off function.
These terms are absent in the normalised Carter's law as it follows from Appendix B.

\end{document}